\documentclass[twoside,english,3p]{elsarticle}
\usepackage[T1]{fontenc}
\usepackage{geometry}
\usepackage{amssymb}
\usepackage{amsmath}

\usepackage{amsthm}
\usepackage{stmaryrd}
\usepackage{graphicx}
\usepackage{booktabs}
\usepackage{multirow}
\usepackage{makecell}
\usepackage{xurl}
\usepackage{esint}
\usepackage{algorithm}
\usepackage{algpseudocode}
\usepackage{rotating}
\usepackage{adjustbox}
\usepackage{nomencl}
\usepackage{multicol}
\makenomenclature
\makeatletter
\theoremstyle{plain}

\theoremstyle{boldremark} 

\ifx\proof\undefined

\providecommand{\proofname}{Proof}
\fi

\renewenvironment{proof}[1][\proofname]{%
	\par\pushQED{\qed}\normalfont%
	\topsep6\p@\@plus6\p@\relax
	\trivlist\item[\hskip\labelsep\bfseries#1\@addpunct{.}]%
	\ignorespaces
}{%
	\popQED\endtrivlist\@endpefalse
}

\journal{Elsevier}

\usepackage{hyperref}
\hypersetup{colorlinks = true, allcolors = blue}

\usepackage[nameinlink]{cleveref}

\Crefname{figure}{Fig.}{Figs.}
\Crefformat{equation}{Eq.~#2(#1)#3}
\Crefformat{section}{Section~#2#1#3}
\AtBeginDocument{%
	\let\citet\cite
}

\usepackage[labelfont=bf]{caption}
\@ifundefined{showcaptionsetup}{}{%
	\PassOptionsToPackage{caption=false}{subfig}}
\usepackage{subfig}
\makeatother

\usepackage{babel}
\providecommand{\remarkname}{Remark}
\providecommand{\theoremname}{Theorem}

\begin{document}
	
	\begin{frontmatter}{}
		
		\title{Plasolver: Physics-Informed Neural Operators for Elastoplasticity}

\author[rvt]{Yizheng Wang}

\ead{wang-yz19@tsinghua.org.cn}

\author[rvt6]{Mohammad Sadegh Eshaghi}

\author[rvt]{Huadong Zhang}

\author[rvt6]{Xiaoying Zhuang}

\author[rvt3]{Timon Rabczuk}

\ead{timon.rabczuk@uni-weimar.de}

\author[rvt]{Yinghua Liu\corref{cor1}}

\ead{yhliu@mail.tsinghua.edu.cn}
\cortext[cor1]{Corresponding author}
\address[rvt]{Department of Engineering Mechanics, Tsinghua University, Beijing 100084, China}

\address[rvt3]{Institute of Structural Mechanics, Bauhaus-Universit\"{a}t Weimar, Marienstr. 15, D-99423 Weimar, Germany}

\address[rvt6]{ Institute of Photonics, Department of Mathematics and Physics, Leibniz University Hannover, Germany}

\begin{abstract}

Elastoplastic analysis is computationally demanding because its nonlinear, path-dependent constitutive behavior requires incremental loading and repeated iterative solutions. To address this challenge, we propose Plasolver, a physics-informed neural operator framework that combines the efficiency of operator learning with the accuracy and robustness of classical numerical solvers. Plasolver consists of a physics-informed pretraining stage and an optional warm-start stage. During pretraining, the neural operator is trained solely by minimizing the incremental potential energy of elastoplasticity formulated by Simo, without requiring any labeled solution data. It operates directly on unstructured point clouds by encoding spatial coordinates, loading histories, and material properties as unified point-wise prompts. This formulation provides dual invariance to spatial and loading-path discretizations, enabling consistent predictions across different spatial resolutions and different numbers of increments representing the same loading trajectory. The pretrained Plasolver achieves relative errors on the order of 1\% while providing approximately two orders of magnitude acceleration over conventional finite element simulations. In the warm-start stage, the pretrained prediction is supplied as the initial solution to a classical iterative solver, preserving its numerical accuracy, robustness, and convergence properties while substantially accelerating convergence. Numerical results show that Plasolver reduces the required number of iterations by approximately 50\% compared with conventional zero-initialized solvers and converges to solutions at any prescribed tolerance. Plasolver thus provides an efficient, accurate, and discretization-invariant computational framework for nonlinear, path-dependent elastoplastic problems.

\end{abstract}

\printnomenclature

		
\end{frontmatter}{}

\section{Introduction}

A wide range of physical phenomena are modeled using partial differential equations (PDEs) \citet{loss_is_minimum_potential_energy,rabczuk2026scientific}. Solving PDEs is essential for understanding physical phenomena and constitutes one of the central tasks in computational mathematics and physics
\citet{PINN_review,wang2024artificial}. Traditional numerical methods often face an inherent trade-off between accuracy and efficiency: higher accuracy generally requires a longer computational time \citet{kahana2023geometry}. In addition, once the geometry, material distribution, or boundary conditions change, traditional PDE solvers must solve the problem again
\citet{wang2021learning}. The finite element method is one of the most widely used conventional approaches for solving PDEs and is well known for its high accuracy and numerical robustness. However, repeatedly solving complex problems is often computationally expensive
\citet{belytschko2013nonlinear}. In many scientific and engineering applications, traditional PDE solvers must perform a new simulation whenever the geometry, material properties, or boundary conditions change \citet{hao2023gnot}. Fortunately, recent advances in neural operators provide a promising approach for substantially improving the efficiency of PDE solutions
\citet{kovachki2023neural}. In the following, we briefly review the development of neural operators.

Neural operators learn mappings between function spaces \citet{DeepOnet} and can therefore be used to learn mappings from geometries, material properties, and boundary conditions to solution spaces, as all these quantities can be mathematically represented as functions
\citet{hao2023gnot}. Once successfully trained, a neural operator can rapidly predict the target solution, typically achieving a speedup of $100$--$10{,}000$ times compared with traditional numerical methods
\citet{bi2023accurate,wang2026pretraining}. Therefore, neural operators have considerable potential for accelerating the solution of PDEs. Currently, three mainstream neural-operator architectures are DeepONet \citet{DeepOnet}, the Fourier Neural Operator (FNO) \citet{li2020fourier}, and transformer-based architectures \citet{cao2021choose}.
In recent years, transformer-based operator-learning frameworks have demonstrated substantial potential and have outperformed the Fourier Neural Operator
(FNO) and DeepONet on many problems \citet{bi2023accurate}. Because transformer architectures can directly process point-cloud inputs, they are more flexible than FNOs \citet{wang2026pfem}, particularly when dealing with complex geometries. Mathematically, a neural operator can be interpreted as using neural networks to learn integral transformations. Different forms of integral transformations therefore lead to different neural-operator architectures, such as the Fourier-transform-based FNO \citet{li2020fourier} and the Laplace-transform-based Laplace
Neural Operator (LNO) \citet{cao2024laplace}. Theoretically, the attention mechanism can itself be interpreted as a learnable integral transformation parameterized by neural networks
\citet{cao2021choose}. Consequently, transformer-based neural operators are potentially more expressive for complex problems than architectures based on fixed integral transformations, such as FNO and LNO \citet{hao2023gnot}. Considering that future foundation models for physics \citet{choi2025defining} will inevitably involve highly complex data, exploring neural operators with transformers as their backbones is particularly important. Representative transformer-based neural operators include the General Neural Operator Transformer (GNOT) proposed by Hao
et al. \citet{hao2023gnot}
and Transolver proposed by Wu et al. \citet{wu2024transolver}. GNOT directly applies transformer operations to the input information and employs linear
attention to reduce computational complexity. Subsequently, Transolver further reduced the computational complexity through physics-attention mechanisms, demonstrating considerable potential for practical applications \citet{nabian2025automotive}. The aforementioned operator-learning methods are trained entirely using data, which must typically be generated in advance using high-fidelity simulation software or physical experiments. Operator learning is therefore limited in scenarios where sufficient data are unavailable. To address this issue, Physics-Informed
Neural Operators (PINOs) have been developed \citet{li2024physics,eshaghi2025variational,wang2021learning}. PINOs integrate physical equations into operator learning, substantially reducing the amount of training data required and, in some cases, eliminating the need for labeled data entirely. Transformer-based operator learning has recently demonstrated considerable potential, particularly Transolver, which substantially reduces computational complexity. The input representation of Transolver is highly concise because it requires only a set of points, rather than the structured grid data required by FNO. Accordingly, Wang
et al. proposed the Pretrained Finite Element Method (PFEM) \citet{wang2026pfem},
which consists of two stages: pretraining and warm-start. The pretraining stage is essentially a Transolver-based PINO, whereas the warm-start stage uses the Transolver prediction as the initial solution for subsequent iterative refinement. Unlike image and text data, high-quality PDE solution data are generally scarce and expensive to generate. For many problems, such data generation is nearly infeasible, particularly for computationally intensive plasticity problems. PINOs therefore represent one of the most promising approaches for solving PDEs because they can train neural operators directly from the governing equations without requiring high-quality solution data to be generated in advance.

In this manuscript, we focus exclusively on plasticity problems because they are widespread in computational mechanics and are highly computationally demanding. Plastic deformation commonly occurs in practical applications once the deformation exceeds the elastic limit. Moreover, because plastic states are strongly history-dependent, incremental steps are required to represent the loading path. Therefore, unlike elasticity or hyperelasticity problems, which can generally be solved in a single loading step, plasticity problems require the entire loading process to be decomposed into numerous increments. Each increment involves a nonlinear solution process, making the overall computation expensive and complex. The computational cost of plasticity problems arises primarily from two aspects. First, the nonlinear equilibrium equations must be satisfied at every incremental step. Second, constitutive integration must be performed at every integration point. Therefore, developing operator-learning methods
for plasticity is of considerable importance. Generating training data for plasticity problems is particularly time-consuming because variations must be considered not only in geometries, material properties, and boundary conditions, but also in loading paths. Training neural operators using purely data-driven methods inevitably requires the generation of a large amount of data in advance. Representative studies include \citet{he2023novel} and \citet{he2024sequential}. By contrast, PINOs enable neural operators to be trained entirely using physical equations, thereby avoiding the need for preliminary data generation. The primary difficulty is that the governing equations of plasticity are highly complex. Fortunately, the incremental potential for plasticity introduced by Simo \citet{simo1998computational} provides an energy functional whose minimization is equivalent to solving the governing PDEs of plasticity. In the context of energy-based physics-informed neural networks, He
et al. have successfully employed this formulation \citet{he2023deep}. However, to the best of the authors' knowledge, no previous study has investigated PINOs
for plasticity. 
How can a neural operator for plasticity be trained without high-fidelity plasticity solution labels while retaining the ability to generalize across different problems?

\textbf{Therefore, we propose Plasolver, a PINO-based computational framework specifically designed for elastoplasticity.} Plasolver consists of two stages: a pretraining stage and a warm-start stage. The pretraining stage of Plasolver is trained exclusively using the governing physical equations and does not require any labeled data. Compared with conventional finite element simulations, it achieves a speedup of approximately $100$ times while maintaining an error of approximately $1\%$. The pretraining stage of Plasolver exhibits dual invariance to spatial and loading-path discretizations. In the warm-start stage, the prediction obtained during pretraining is used as the initial solution for a conventional iterative PDE solver, thereby retaining the ability of the classical solver to achieve arbitrarily high accuracy. Compared with a conventional numerical solver, this strategy reduces the required number of iterations by approximately $50\%$. The
main contributions of Plasolver can be summarized as follows:
\begin{itemize}
	\item \textbf{Point-cloud inputs}: We encode all point-wise information as prompts, including spatial coordinates, loading paths, and material properties. The neural operator in the pretraining stage of Plasolver processes these unified point-wise prompts.
	\item \textbf{Invariance to spatial and loading-path discretizations}: Spatial discretization invariance means that the model produces consistent predictions across different spatial resolutions. Loading-path discretization invariance means that, for the same loading trajectory, the model produces consistent predictions when different numbers of loading increments are used.
	\item \textbf{Training solely from physical equations}: We train the neural operator in the pretraining stage of Plasolver entirely using the governing physical equations, without relying on any labeled solution data.
	\item \textbf{Warm-start via classical solvers}: The solution predicted by Plasolver is supplied as the initial solution to a classical iterative solver. This approach substantially reduces the required number of iterations while enabling solutions of arbitrarily high accuracy to be obtained.
\end{itemize}
The remainder of this paper is organized as follows. \Cref{sec:Preparatory-knowledge}
introduces the classical theory of plasticity, with particular emphasis on Simo's incremental potential for plasticity. \Cref{sec:Method} describes the methodology of the proposed Plasolver framework. In
\Cref{sec:Result}, the generalization capability of Plasolver across different loading paths, geometries, and material properties is systematically evaluated. Section ~\ref{sec:Discussion}
discuss several key aspects of Plasolver. Finally, \Cref{sec:Conclusion} summarizes the principal findings and outlines directions for future research. To
the best of our knowledge, Plasolver is \textbf{the first physics-informed
	neural operator for elastoplasticity}.

\section{Preliminaries\label{sec:Preparatory-knowledge}}

The core of this work is to use physics-informed neural operators for classical elastoplasticity to improve the efficiency of plasticity simulations. Therefore, this section introduces the fundamentals of plasticity, the incremental potential for plasticity proposed by Simo \citet{simo1998computational}, and the conventional deep energy method.

\subsection{Fundamentals of plasticity theory}

A central feature of plasticity is its dependence on loading history, which is also known as path dependence. Plastic deformation is irreversible deformation caused by the motion of dislocations within crystals. Owing to this path dependence, stress and strain do not have a one-to-one correspondence in plastic deformation. This differs from elasticity, in which stress and strain have a one-to-one relationship. Therefore, only an incremental theory can describe the relationship between stress and strain increments. Specifically, for a given state and strain increment, there exists a unique corresponding stress increment. The key difference between plasticity and elasticity lies in constitutive integration, whereas their equilibrium equations are essentially identical. Constitutive integration in plasticity requires state variables, which are usually internal variables. Common internal variables include the equivalent plastic strain $\bar{\varepsilon}^{p}$, plastic strain $\boldsymbol{\varepsilon}^{p}$, and backstress $\boldsymbol{q}$.

The yield condition $f(\boldsymbol{\sigma},\boldsymbol{e})$ determines whether a material is in an elastic or plastic state. Here, $\boldsymbol{\sigma}$ and $\boldsymbol{e}$ denote the stress and internal variables, respectively, and the internal variables are determined using evolution equations. The condition $f<0$ indicates an elastic state, whereas $f=0$ indicates a plastic state. Therefore, whether the material is in the elastic or plastic regime, the yield function always satisfies
\begin{equation}
	f\leq0.
	\label{eq:yield_condition}
\end{equation}

When the material is in the plastic regime, for which $f=0$, its internal variables evolve. Plastic strain is a common internal variable, and its evolution equation is known as the plastic flow rule:
\begin{equation}
	\dot{\varepsilon}^{p}_{ij}
	=
	\dot{\gamma}r_{ij}(\boldsymbol{\sigma},\boldsymbol{e}).
	\label{eq:plastic_flow_law}
\end{equation}
Here, $\dot{\gamma}$ is the plastic multiplier. During plastic loading, the stress point must remain on the yield surface, such that $\dot{f}=0$. This requirement is known as the consistency condition and is used to determine the plastic multiplier $\dot{\gamma}$:
\begin{equation}
	\dot{f}
	=
	\frac{\partial f}{\partial\boldsymbol{\sigma}}
	:
	\dot{\boldsymbol{\sigma}}
	+
	\frac{\partial f}{\partial\boldsymbol{e}}
	:
	\dot{\boldsymbol{e}}
	=
	0.
	\label{eq:consistent_condition}
\end{equation}

In \Cref{eq:plastic_flow_law}, $r_{ij}$ describes the direction of plastic flow and is determined by the normal to the plastic potential $\psi$:
\begin{equation}
	r_{ij}(\boldsymbol{\sigma},\boldsymbol{e})
	=
	\frac{\partial\psi(\boldsymbol{\sigma},\boldsymbol{e})}
	{\partial\sigma_{ij}}.
\end{equation}
The case in which the plastic potential satisfies $\psi=f$ is commonly referred to as the associative plastic flow rule.

In numerical calculations, external loads are generally applied incrementally, and each load increment must satisfy \Cref{eq:yield_condition}. In other words, the stress state cannot lie outside the yield surface. The return-mapping method is a constitutive integration algorithm used to enforce \Cref{eq:yield_condition}. Specifically, this method first assumes that the material remains elastic during the current increment, thereby producing a trial stress state. If the trial stress does not exceed the yield surface, the current increment is an elastic step. If the trial stress lies outside the yield surface, a plastic correction is required to return the stress state to the yield surface. In general, this plastic correction is nonlinear. A Newton--Raphson iteration is therefore usually required to determine the plastic multiplier and update the internal variables. For the special case of an associative $J_{2}$ plasticity model with linear isotropic hardening, the plastic correction direction is radial in deviatoric stress space. The plastic multiplier can then be obtained analytically. Consequently, the algorithm reduces to the radial return method and requires no local Newton iteration, substantially improving the efficiency of constitutive integration.

Many models have been developed to describe the plastic behavior of materials. Examples include the well-known $J_{2}$ flow theory for metallic materials \citet{mises1913mechanik}, the Mohr--Coulomb model \citet{mohr1900umstande} and Drucker--Prager model \citet{drucker1952soil} for geomaterials, and the Gurson model \citet{gurson1977continuum} for porous plasticity. Because the associative $J_{2}$ plasticity model is among the most widely used plasticity models, the following discussion focuses on associative $J_{2}$ plasticity with mixed hardening, that is, with both isotropic and kinematic hardening.

\begin{figure}
	\begin{centering}
		\includegraphics[scale=0.50]{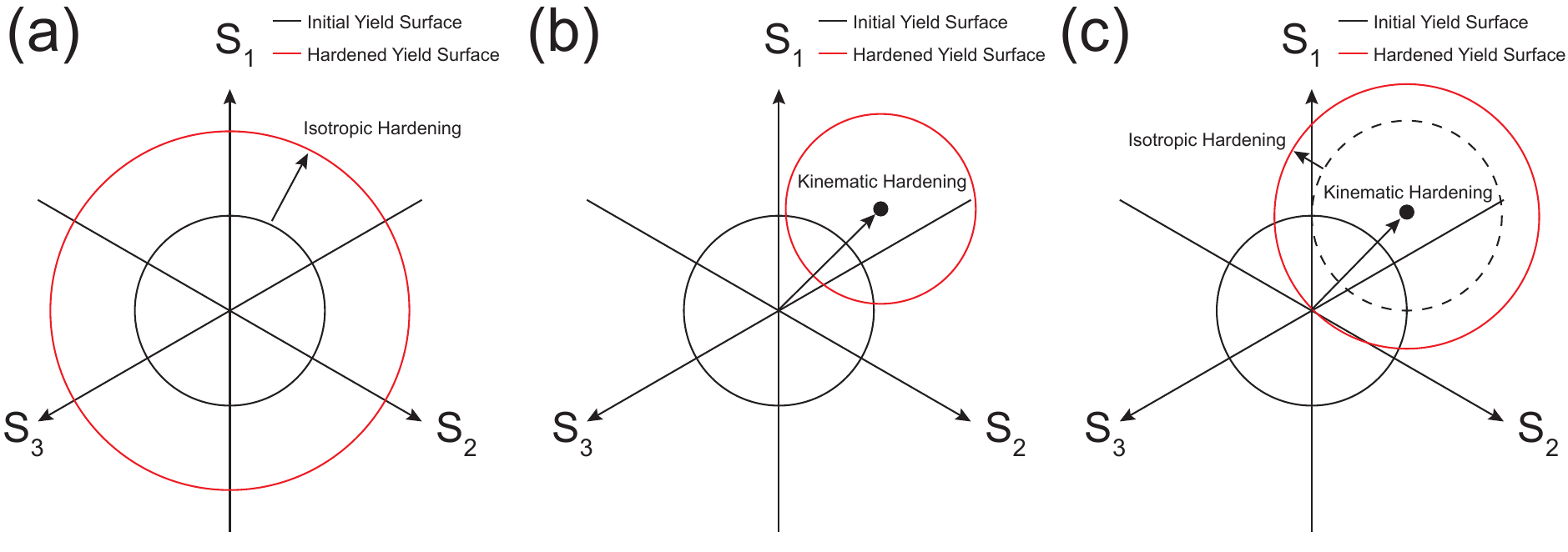}
		\par\end{centering}
	\caption{Geometric interpretation of changes in the yield surface: (a) Isotropic hardening: a change in the yield stress alters the radius of the yield surface. (b) Kinematic hardening: a change in the backstress translates the yield surface. (c) Mixed hardening: the backstress first translates the yield surface, after which a change in the yield stress alters its radius.
		\label{fig:yield-surface_geo_change}}
\end{figure}

The central idea of $J_{2}$ plasticity is to describe the yield condition using the second invariant $J_{2}$ of the shifted deviatoric stress $\boldsymbol{\eta}$:
\begin{equation}
	\begin{aligned}
		f
		&=
		\sqrt{3J_{2}}
		-
		\sigma_{y}(\bar{\varepsilon}^{p}),
		\\
		J_{2}
		&=
		\frac{1}{2}\boldsymbol{\eta}:\boldsymbol{\eta},
		\\
		\boldsymbol{\eta}
		&=
		\boldsymbol{s}-\boldsymbol{q},
		\\
		\boldsymbol{s}
		&=
		\sigma'_{ij}
		=
		\sigma_{ij}
		-
		\frac{1}{3}\sigma_{mm}\delta_{ij}.
	\end{aligned}
	\label{eq:yield_function_j2}
\end{equation}
Here, $\boldsymbol{s}$ is the deviatoric stress, $\boldsymbol{\sigma}$ is the stress, $\boldsymbol{q}$ is the backstress, $\bar{\varepsilon}^{p}$ is the equivalent plastic strain, and $\boldsymbol{\varepsilon}^{p}$ is the plastic strain, with
$d(\bar{\varepsilon}^{p})=\sqrt{2d(\boldsymbol{\varepsilon}^{p}):d(\boldsymbol{\varepsilon}^{p})/3}$.
The quantity $\bar{\sigma}=\sqrt{3J_{2}}$ is also known as the von Mises equivalent stress. The derivations of the forms of $\boldsymbol{\varepsilon}^{p}$ and $\bar{\sigma}$ are provided in \ref{sec:derivative_in_plasticity}. The quantity $\sigma_{y}$ is the yield stress, and $H=\partial\sigma_{y}/\partial\bar{\varepsilon}^{p}$ is the plastic modulus. For linear isotropic hardening,
$\sigma_{y}=\sigma^{0}_{y}+H\bar{\varepsilon}^{p}$.
It should be noted that $\boldsymbol{q}$ is deviatoric, such that $q_{mm}=0$. The quantity $\sigma^{0}_{y}$ is the initial yield stress. \Cref{fig:yield-surface_geo_change}a illustrates isotropic hardening, in which a change in the yield stress alters the radius of the yield surface. In addition, the center of the yield surface is related to the backstress $\boldsymbol{q}$, resulting in the kinematic hardening shown in \Cref{fig:yield-surface_geo_change}b. In summary, isotropic and kinematic hardening are geometrically represented by a change in the size and a translation of the yield surface, respectively, as shown in \Cref{fig:yield-surface_geo_change}c.

The evolution equations for the plastic strain and backstress are
\begin{equation}
	\begin{aligned}
		\dot{\varepsilon}^{p}_{ij}
		&=
		\dot{\gamma}r_{ij}(\boldsymbol{\sigma},\boldsymbol{e})
		=
		\dot{\gamma}\sqrt{\frac{3}{2}}
		\frac{\eta_{ij}}
		{\sqrt{\boldsymbol{\eta}:\boldsymbol{\eta}}},
		\\
		\dot{q}_{ij}
		&=
		\frac{2}{3}C\dot{\varepsilon}^{p}_{ij}
		=
		\dot{\gamma}\sqrt{\frac{2}{3}}C
		\frac{\eta_{ij}}
		{\sqrt{\boldsymbol{\eta}:\boldsymbol{\eta}}},
		\\
		r_{ij}(\boldsymbol{\sigma},\boldsymbol{e})
		&=
		\frac{\partial\psi(\boldsymbol{\sigma},\boldsymbol{e})}
		{\partial\sigma_{ij}}.
	\end{aligned}
	\label{eq:plasticity_strain_increment}
\end{equation}
Here, $\psi$ is the plastic potential, and $C$ is the kinematic hardening modulus. Because the direction in which the backstress evolves is determined solely by the direction of the plastic strain, the backstress is deviatoric. Further details can be found in the official COMSOL documentation \url{https://doc.comsol.com/6.3/doc/com.comsol.help.sme/sme_ug_theory.06.033.html}. It should be noted that the backstress follows the linear Prager hardening rule. A detailed derivation of the coefficient $2/3$ preceding $C$ in the backstress evolution equation is provided in \ref{sec:derivative_in_plasticity}. \Cref{eq:plasticity_strain_increment} adopts the associative flow rule, for which $\psi=f$. For linear isotropic hardening, the plastic multiplier $\dot{\gamma}$ can be obtained analytically:
\begin{equation}
	\Delta\gamma
	=
	\left\langle
	\frac{f_{\mathrm{trial}}}{3G+H+C}
	\right\rangle_{+}.
	\label{eq:plasticity_factor}
\end{equation}
Here, $f_{\mathrm{trial}}$ is the yield function evaluated at the elastic trial stress. \ref{sec:proof_J2_plasticity_multify} presents a detailed derivation of \Cref{eq:plasticity_factor}. The positive-part operator is defined as
$\langle s\rangle_{+}=(|s|+s)/2$.
Thus, $\langle s\rangle_{+}=s$ when $s\geq0$, whereas $\langle s\rangle_{+}=0$ when $s<0$.

After obtaining $\Delta\gamma$ from \Cref{eq:plasticity_factor}, the current state variables are updated using the evolution equations in \Cref{eq:plasticity_strain_increment}:
\begin{equation}
	\begin{aligned}
		\boldsymbol{\varepsilon}^{p(n+1)}
		&=
		\boldsymbol{\varepsilon}^{p(n)}
		+
		\Delta\gamma\sqrt{\frac{3}{2}}
		\frac{\boldsymbol{\eta}^{\mathrm{trial}}}
		{\sqrt{\boldsymbol{\eta}^{\mathrm{trial}}:
				\boldsymbol{\eta}^{\mathrm{trial}}}},
		\\
		\bar{\varepsilon}^{p(n+1)}
		&=
		\bar{\varepsilon}^{p(n)}
		+
		\Delta\gamma,
		\\
		\boldsymbol{q}^{(n+1)}
		&=
		\boldsymbol{q}^{(n)}
		+
		\Delta\gamma\sqrt{\frac{2}{3}}C
		\frac{\boldsymbol{\eta}^{\mathrm{trial}}}
		{\sqrt{\boldsymbol{\eta}^{\mathrm{trial}}:
				\boldsymbol{\eta}^{\mathrm{trial}}}},
		\\
		\sigma^{(n+1)}_{ij}
		&=
		\left[
		\lambda\delta_{ij}\delta_{kl}
		+
		G\left(
		\delta_{ik}\delta_{jl}
		+
		\delta_{il}\delta_{jk}
		\right)
		\right]
		\left[
		\varepsilon^{(n+1)}_{kl}
		-
		\varepsilon^{p(n)}_{kl}
		-
		\Delta\gamma\sqrt{\frac{3}{2}}
		\frac{\eta^{\mathrm{trial}}_{kl}}
		{\sqrt{\boldsymbol{\eta}^{\mathrm{trial}}:
				\boldsymbol{\eta}^{\mathrm{trial}}}}
		\right]
		\\
		&=
		\left[
		\lambda\delta_{ij}\delta_{kl}
		+
		G\left(
		\delta_{ik}\delta_{jl}
		+
		\delta_{il}\delta_{jk}
		\right)
		\right]
		\left[
		\frac{1}{3}\varepsilon^{(n+1)}_{mm}\delta_{kl}
		+
		\varepsilon'^{(n+1)}_{kl}
		-
		\varepsilon^{p(n)}_{kl}
		-
		\Delta\gamma\sqrt{\frac{3}{2}}
		\frac{\eta^{\mathrm{trial}}_{kl}}
		{\sqrt{\boldsymbol{\eta}^{\mathrm{trial}}:
				\boldsymbol{\eta}^{\mathrm{trial}}}}
		\right]
		\\
		&=
		\left(
		\lambda+\frac{2}{3}G
		\right)
		\varepsilon^{(n+1)}_{mm}\delta_{ij}
		+
		s^{\mathrm{trial}(n+1)}_{ij}
		-
		2G
		\left(
		\Delta\gamma\sqrt{\frac{3}{2}}
		\frac{\eta^{\mathrm{trial}}_{ij}}
		{\sqrt{\boldsymbol{\eta}^{\mathrm{trial}}:
				\boldsymbol{\eta}^{\mathrm{trial}}}}
		\right),
		\\
		\Delta\gamma
		&=
		\left\langle
		\frac{f_{\mathrm{trial}}}{3G+H+C}
		\right\rangle_{+},
		\\
		f_{\mathrm{trial}}
		&=
		\sqrt{
			\frac{3}{2}
			\boldsymbol{\eta}^{\mathrm{trial}}:
			\boldsymbol{\eta}^{\mathrm{trial}}
		}
		-
		\sigma_{y}\left(\bar{\varepsilon}^{p(n)}\right),
		\\
		\eta^{\mathrm{trial}}_{ij}
		&=
		s^{\mathrm{trial}(n+1)}_{ij}
		-
		q^{(n)}_{ij}
		\\
		&=
		\left[
		\lambda\delta_{ij}\delta_{kl}
		+
		G\left(
		\delta_{ik}\delta_{jl}
		+
		\delta_{il}\delta_{jk}
		\right)
		\right]
		\left[
		\varepsilon'^{(n+1)}_{kl}
		-
		\varepsilon^{p(n)}_{kl}
		\right]
		-
		q^{(n)}_{ij}
		\\
		&=
		2G
		\left[
		\varepsilon'^{(n+1)}_{ij}
		-
		\varepsilon^{p(n)}_{ij}
		\right]
		-
		q^{(n)}_{ij}.
	\end{aligned}
	\label{eq:update_internal_variables}
\end{equation}
Here, $\lambda$ and $G$ are the first and second Lamé parameters, respectively.

For each load increment, constitutive integration must be performed at every integration point. After constitutive integration has been completed at all integration points, the displacement correction is determined from the equilibrium equations. The resulting displacement correction produces a new strain increment for the current load step, after which constitutive integration is performed again at all integration points. This procedure is repeated until the displacement correction converges. The next load increment is then applied, and the process continues until the entire loading procedure is completed.

\subsection{Simo's incremental potential for plasticity}

The previous subsection introduced the conventional theory of elastoplasticity. Because we aim to use PINO to accelerate plasticity simulations, the construction of the loss function is crucial. For complex problems such as plasticity, strong-form loss functions are generally more difficult to optimize than energy-based loss functions, as is also the case for PINNs. When PINO uses a strong-form loss function, more hyperparameters must be selected than for an energy-based formulation, making it difficult to identify the optimal hyperparameters. Moreover, the strong form requires higher-order derivatives than the energy form. Consequently, the computational efficiency of the strong form is considerably lower than that of the energy form \citet{he2023deep}. Therefore, using the energy formulation of plasticity as the loss function is essential. Fortunately, \citet{simo1998computational} proposed the classical incremental potential theory of elastoplasticity, which can be used as the loss function for PINO.

The incremental energy formulation of the elastoplastic problem \citet{simo1998computational} is
\begin{equation}
	\begin{aligned}
		\Pi^{(n+1)}
		&=
		\int_{\Omega}
		\left[
		W^{(n+1)}_{e}
		+
		\frac{1}{2}
		\boldsymbol{v}^{(n+1)}
		\cdot
		\boldsymbol{D}^{-1}
		\cdot
		\boldsymbol{v}^{(n+1)}
		-
		\Delta\gamma f^{(n+1)}
		\right.
		\\
		&\qquad\left.
		+
		\left(
		\boldsymbol{\varepsilon}^{p(n+1)}
		-
		\boldsymbol{\varepsilon}^{p(n)}
		\right)
		:
		\boldsymbol{\sigma}^{(n+1)}
		-
		\boldsymbol{v}^{(n+1)}
		\cdot
		\boldsymbol{D}^{-1}
		\cdot
		\left(
		\boldsymbol{v}^{(n+1)}
		-
		\boldsymbol{v}^{(n)}
		\right)
		\right]
		dV
		-
		W_{\mathrm{ext}},
		\\
		W^{(n+1)}_{e}
		&=
		\frac{1}{2}
		\boldsymbol{\varepsilon}^{e(n+1)}
		:
		\boldsymbol{C}
		:
		\boldsymbol{\varepsilon}^{e(n+1)},
		\\
		W_{\mathrm{ext}}
		&=
		\int_{\Omega}
		\boldsymbol{f}\cdot\boldsymbol{u}\,dV
		+
		\int_{\Gamma^{t}}
		\bar{\boldsymbol{t}}^{(n+1)}
		\cdot
		\boldsymbol{u}\,dA,
		\\
		\textrm{s.t.}\quad
		\boldsymbol{u}^{(n+1)}
		&=
		\bar{\boldsymbol{u}}^{(n+1)},
		\qquad
		\boldsymbol{x}\in\Gamma^{u}.
	\end{aligned}
	\label{eq:simo}
\end{equation}
Here, $W^{(n+1)}_{e}$ is the elastic strain energy, $W_{\mathrm{ext}}$ is the work performed by the external forces, and $\boldsymbol{f}$ is the body force. The prescribed traction $\bar{\boldsymbol{t}}$ is applied on the traction boundary $\Gamma^{t}$. The prescribed displacement $\bar{\boldsymbol{u}}$ is imposed on the displacement boundary $\Gamma^{u}$ and must be satisfied in advance. The tensor $\boldsymbol{C}$ is the fourth-order elasticity tensor. The internal-variable vector
$\boldsymbol{v}
=
[\begin{array}{cc}
	H\bar{\varepsilon}^{p} & \boldsymbol{q}
\end{array}]$
contains the equivalent plastic strain $\bar{\varepsilon}^{p}$ and the backstress $\boldsymbol{q}$. The hardening-modulus matrix $\boldsymbol{D}$ is
\begin{equation}
	\boldsymbol{D}
	=
	\left[
	\begin{array}{cc}
		H & \boldsymbol{0}
		\\
		\boldsymbol{0} & \dfrac{2}{3}C\boldsymbol{I}
	\end{array}
	\right].
	\label{eq:matrix_harden_moduli}
\end{equation}
It should be noted that \Cref{eq:matrix_harden_moduli} assumes linear isotropic and kinematic hardening, which is the most common case. For nonlinear isotropic hardening, the internal-variable vector is generally replaced by
$\boldsymbol{v}
=
[\begin{array}{cc}
	R(\bar{\varepsilon}^{p}) & \boldsymbol{q}
\end{array}]$,
where
$\sigma_{y}
=
\sigma^{0}_{y}
+
R(\bar{\varepsilon}^{p})$.
Accordingly, $H$ in $\boldsymbol{D}$ must be replaced by
$\partial R(\bar{\varepsilon}^{p})/\partial\bar{\varepsilon}^{p}$.

When the term $\Delta\gamma f^{(n+1)}$ is included, \Cref{eq:simo} generally defines a stationarity problem rather than an extremization problem. This is caused by the Lagrange-multiplier term.

\subsection{Deep energy method}

\begin{figure}
	\begin{centering}
		\includegraphics[scale=0.57]{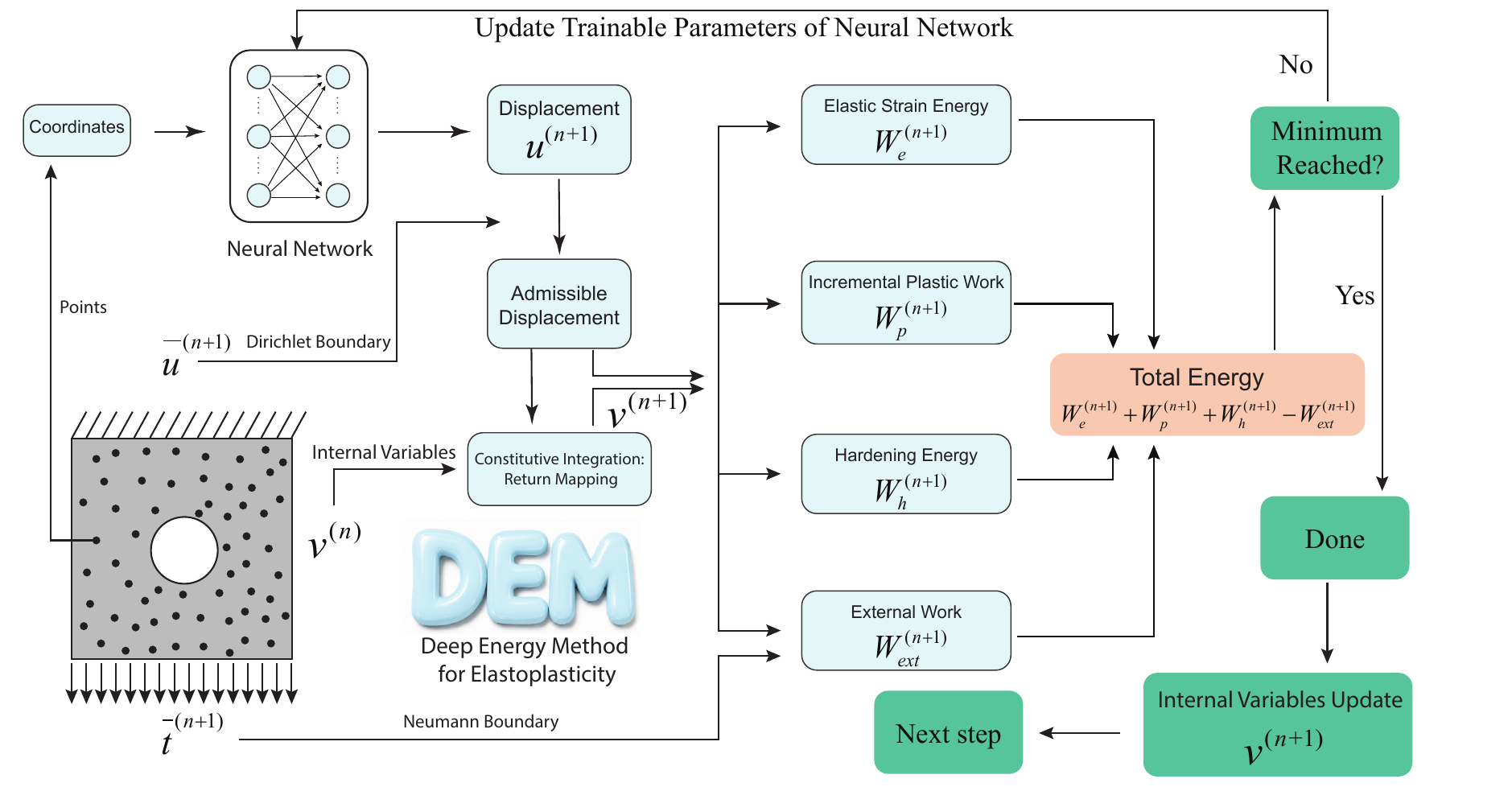}
		\par\end{centering}
	\caption{Schematic illustration of the deep energy method for elastoplasticity. Here, $\bar{\boldsymbol{u}}$ and $\bar{\boldsymbol{t}}$ denote the Dirichlet and Neumann boundary conditions, respectively, and $\boldsymbol{v}$ denotes the state variables. DEM takes the coordinates as input and produces the displacement field as output. The neural-network parameters are determined by minimizing the total energy. The total energy consists of four components: the elastic strain energy $W^{(n+1)}_{e}$, incremental plastic work $W^{(n+1)}_{p}$, hardening energy $W^{(n+1)}_{h}$, and external work $W^{(n+1)}_{\mathrm{ext}}$. After the energy minimization has converged, the internal variables are updated and the next load step begins.
		\label{fig:DEM-for-elastoplasticity}}
\end{figure}

The deep energy method (DEM) \citet{loss_is_minimum_potential_energy} is the energy-based form of PINNs, as illustrated in \Cref{fig:DEM-for-elastoplasticity}. The central idea of DEM is to use an energy functional, rather than the strong form of the PDEs, as the loss function. DEM generally achieves better computational efficiency and accuracy than strong-form methods \citet{the_comparision_of_strong_and_energy_form}. However, not all PDEs have a corresponding energy formulation \citet{wang2025physics}. Fortunately, plasticity problems possess the incremental energy formulation given in \Cref{eq:simo}. \citet{he2023deep} first applied DEM to elastoplastic problems. Although \citet{he2023deep} demonstrated that DEM still lags behind the conventional finite element method in both accuracy and efficiency, DEM can be used to verify the correctness of the energy formulation, as shown in \Cref{subsec:PINNs-to-PINOs}. Therefore, it is important to introduce the application of DEM to elastoplastic problems.

We use $\Pi^{(n+1)}$ in \Cref{eq:simo} as the loss function and minimize it:
\begin{equation}
	\begin{aligned}
		\mathcal{L}^{(n+1)}
		\left(
		\boldsymbol{u}^{(n+1)}
		(\boldsymbol{x};\boldsymbol{\theta})
		\right)
		&=
		W^{(n+1)}_{e}
		+
		W^{(n+1)}_{p}
		+
		W^{(n+1)}_{h}
		-
		W^{(n+1)}_{\mathrm{ext}},
		\\
		W^{(n+1)}_{e}
		&=
		\int_{\Omega}
		\frac{1}{2}
		\boldsymbol{\varepsilon}^{e(n+1)}
		\left(\boldsymbol{u}^{(n+1)}\right)
		:
		\boldsymbol{C}
		:
		\boldsymbol{\varepsilon}^{e(n+1)}
		\left(\boldsymbol{u}^{(n+1)}\right)
		\,dV,
		\\
		W^{(n+1)}_{p}
		&=
		\int_{\Omega}
		\left[
		\boldsymbol{\varepsilon}^{p(n+1)}
		\left(\boldsymbol{u}^{(n+1)}\right)
		-
		\boldsymbol{\varepsilon}^{p(n)}
		\right]
		:
		\boldsymbol{C}
		:
		\boldsymbol{\varepsilon}^{e(n+1)}
		\left(\boldsymbol{u}^{(n+1)}\right)
		\,dV,
		\\
		W^{(n+1)}_{h}
		&=
		\int_{\Omega}
		\left\{
		\frac{1}{2}
		\boldsymbol{v}^{(n+1)}
		\left(\boldsymbol{u}^{(n+1)}\right)
		\cdot
		\boldsymbol{D}^{-1}
		\cdot
		\boldsymbol{v}^{(n+1)}
		\left(\boldsymbol{u}^{(n+1)}\right)
		\right.
		\\
		&\qquad\left.
		-
		\boldsymbol{v}^{(n+1)}
		\left(\boldsymbol{u}^{(n+1)}\right)
		\cdot
		\boldsymbol{D}^{-1}
		\cdot
		\left[
		\boldsymbol{v}^{(n+1)}
		\left(\boldsymbol{u}^{(n+1)}\right)
		-
		\boldsymbol{v}^{(n)}
		\right]
		\right\}
		\,dV,
		\\
		W^{(n+1)}_{\mathrm{ext}}
		&=
		\int_{\Omega}
		\boldsymbol{f}^{(n+1)}
		\cdot
		\boldsymbol{u}^{(n+1)}
		\,dV
		+
		\int_{\Gamma^{t}}
		\bar{\boldsymbol{t}}^{(n+1)}
		\cdot
		\boldsymbol{u}^{(n+1)}
		\,dA,
		\\
		\boldsymbol{u}^{(n+1)}
		&=
		\mathrm{NN}(\boldsymbol{x};\boldsymbol{\theta}),
		\\
		\textrm{s.t.}\quad
		\boldsymbol{u}^{(n+1)}
		&=
		\bar{\boldsymbol{u}}^{(n+1)},
		\qquad
		\boldsymbol{x}\in\Gamma^{u}.
	\end{aligned}
	\label{eq:DEM_plasticity}
\end{equation}
Here, $\mathrm{NN}(\boldsymbol{x};\boldsymbol{\theta})$ is the approximation function represented by the neural network, and $\boldsymbol{\theta}$ denotes the trainable parameters of the neural network. The terms $W^{(n+1)}_{e}$, $W^{(n+1)}_{p}$, $W^{(n+1)}_{h}$, and $W^{(n+1)}_{\mathrm{ext}}$ denote the elastic strain energy, incremental plastic work, hardening energy, and external work, respectively. The neural network $\mathrm{NN}(\boldsymbol{x};\boldsymbol{\theta})$ takes the coordinates $\boldsymbol{x}$ as input and produces the displacement field $\boldsymbol{u}$ as output. This mapping can be expressed as
$\boldsymbol{x}\overset{\boldsymbol{\theta}}{\longrightarrow}\boldsymbol{u}$.

It should be noted that, compared with \Cref{eq:simo}, \Cref{eq:DEM_plasticity} does not contain the term $\Delta\gamma f^{(n+1)}$. This is because the yield condition is enforced at every material integration point using the return-mapping algorithm. If a material point enters the plastic regime, the yield function satisfies $f^{(n+1)}=0$. If the material point remains in the elastic regime, then $\Delta\gamma=0$. Therefore, $\Delta\gamma f^{(n+1)}=0$ in either case. This explains why the DEM energy in \Cref{eq:DEM_plasticity} does not contain the term $\Delta\gamma f^{(n+1)}$ appearing in \Cref{eq:simo}. Eliminating $\Delta\gamma f^{(n+1)}$ is important for DEM optimization because DEM is formulated as a minimization problem. Retaining this term would make the energy optimization a stationarity problem rather than an extremum problem. Moreover, if associative plasticity is adopted and the plastic modulus is positive, \Cref{eq:simo} constitutes an extremum problem, as explained in detail in \ref{sec:plasticity_variational}. Unlike conventional quasistatic problems, DEM for plasticity requires energy minimization at every load step. Therefore, DEM for plasticity involves two levels of iteration: the outer iterations correspond to the load steps, whereas the inner iterations perform energy minimization.

In summary, the optimization procedure of DEM is
\begin{equation}
	\begin{aligned}
		\boldsymbol{\theta}
		&=
		\underset{\boldsymbol{\theta}}{\arg\min}\,
		\mathcal{L}^{(n+1)}
		\left(
		\boldsymbol{u}^{(n+1)}
		(\boldsymbol{x};\boldsymbol{\theta})
		\right),
		\\
		\textrm{s.t.}\quad
		\boldsymbol{u}^{(n+1)}
		(\boldsymbol{x};\boldsymbol{\theta})
		&=
		\bar{\boldsymbol{u}}^{(n+1)}(\boldsymbol{x}),
		\qquad
		\boldsymbol{x}\in\Gamma^{u}.
	\end{aligned}
\end{equation}
Here, $\Gamma^{u}$ denotes the Dirichlet boundary. Therefore, when DEM is used to minimize the energy $\mathcal{L}^{(n+1)}$, the displacement field must satisfy the essential boundary conditions in advance. \ref{sec:DEM-for-elastoplasticity} demonstrates that DEM can effectively solve elastoplastic problems.

\section{Method\label{sec:Method}}

Although the deep energy method can effectively solve an individual plasticity problem, the problem must be solved again, as in the finite element method, whenever the loading path, geometry, or material properties change. Therefore, neural operators \citet{DeepOnet,li2020fourier,wu2024transolver} are required to accelerate the solution of plasticity problems. However, conventional operator-learning methods generally require training data. Fortunately, physics-informed neural operators (PINOs) \citet{wang2021learning,li2024physics} can be trained using only the governing partial differential equations, without requiring labeled data. In particular, the recently proposed Pretrained Finite Element Method (PFEM) for computational mechanics, which is based on physics-informed neural operators \citet{wang2026pfem}, has attracted considerable attention.

However, no previous study has applied PINO to elastoplasticity. Therefore, the main contribution of this work is the application of PINO to plasticity. The remainder of this section is divided into two parts: operator learning and Plasolver, which applies PINO to plasticity.

\subsection{Neural operator\label{subsec:Neural-Operator}}

The central objective of a neural operator is to learn mappings between functions, which has broad applications in science and engineering \citep{kovachki2023neural}. For example, the input function may represent the boundary conditions of a partial differential equation, whereas the output is the function field of interest. Although functions are mathematically infinite-dimensional, continuous functions are approximated in practice using discrete input data. Fortunately, neural operators generally possess the advantage of discretization invariance.

Conventional neural operators are trained using a purely data-driven approach, which is conceptually straightforward. Suppose that a dataset
$\{\boldsymbol{X}^{(i)},\boldsymbol{Y}^{(i)}\}_{i=1}^{N}$
is obtained for different geometries $\boldsymbol{G}$, materials $\boldsymbol{M}$, and boundary conditions $\boldsymbol{B}$. Here, $\boldsymbol{X}^{(i)}$ and $\boldsymbol{Y}^{(i)}$ are the input and output, respectively, and $N$ is the total number of samples. The optimization process for purely data-driven training is
\begin{equation}
	\begin{aligned}
		\boldsymbol{\theta}^{*}
		&=
		\arg\min_{\boldsymbol{\theta}}\mathcal{L}_{\mathrm{data}},
		\\
		\mathcal{L}_{\mathrm{data}}
		&=
		\frac{1}{N}
		\sum_{i=1}^{N}
		\left\|
		\mathcal{\boldsymbol{H}}
		\left(
		\boldsymbol{X}^{(i)};\boldsymbol{\theta}
		\right)
		-
		\boldsymbol{Y}^{(i)}
		\right\|_{2}^{2}.
	\end{aligned}
\end{equation}
Here, $\boldsymbol{\theta}$ denotes the trainable parameters of
$\mathcal{\boldsymbol{H}}(\boldsymbol{X}^{(i)};\boldsymbol{\theta})$,
and $\|\cdot\|_{2}^{2}$ denotes the squared $L_{2}$ norm. The operator
$\mathcal{\boldsymbol{H}}(\boldsymbol{X}^{(i)};\boldsymbol{\theta})$
is approximated using a neural operator and is abbreviated as
$\mathcal{\boldsymbol{H}}_{\boldsymbol{\theta}}
(\boldsymbol{x};\boldsymbol{X}^{(i)})$,
where $\boldsymbol{x}$ denotes the spatial coordinates. Both
$\boldsymbol{X}^{(i)}$ and $\boldsymbol{Y}^{(i)}$ are functions containing data at different spatial locations. The advantage of data-driven operator learning is that its training procedure is simple and straightforward. Its limitation is also evident: a high-quality dataset must be generated in advance, whereas the available data in many applications may contain considerable errors or be insufficient in quantity.

This limitation presents a major challenge for conventional operator learning in data-scarce applications. Therefore, training neural operators directly using physical equations is important. This approach is referred to as the physics-informed neural operator (PINO) \citet{li2024physics,wang2021learning,eshaghi2025variational}. Most computational-physics problems have corresponding PDEs, and the central task of computational physics is to solve these PDEs numerically. Therefore, the central idea of PINO is to incorporate PDEs into the pretraining of neural operators. The PINO loss function is generally expressed as
\begin{equation}
	\begin{aligned}
		\boldsymbol{\theta}^{*}
		&=
		\arg\min_{\boldsymbol{\theta}}\mathcal{L}_{\mathrm{pde}},
		\\
		\mathcal{L}_{\mathrm{pde}}
		&=
		\frac{1}{N}
		\sum_{i=1}^{N}
		\left\{
		\frac{1}{N_{d}}
		\sum_{j=1}^{N_{d}}
		\left\|
		\boldsymbol{P}
		\left(
		\mathcal{\boldsymbol{H}}_{\boldsymbol{\theta}}
		(\boldsymbol{x}^{(j)};\boldsymbol{X}^{(i)})
		\right)
		\right\|_{2}^{2}
		\right.
		\\
		&\qquad\left.
		+
		\frac{1}{N_{b}}
		\sum_{j=1}^{N_{b}}
		\left\|
		\boldsymbol{I}
		\left(
		\mathcal{\boldsymbol{H}}_{\boldsymbol{\theta}}
		(\boldsymbol{x}^{(j)};\boldsymbol{X}^{(i)})
		\right)
		\right\|_{2}^{2}
		\right\}.
	\end{aligned}
	\label{eq:PDEs_driven}
\end{equation}
Here, $\boldsymbol{P}$ and $\boldsymbol{I}$ denote the governing PDE operator within the domain and the boundary-condition operator, respectively. The quantities $N_{d}$ and $N_{b}$ denote the total numbers of points within the domain and on the boundary, respectively, whereas $N$ is the total number of input-function instances. For simplicity, we consider steady-state heat conduction in a two-dimensional domain with a heterogeneous thermal-conductivity field:
\begin{equation}
	\begin{cases}
		-\nabla\cdot
		\left[
		k(\boldsymbol{x})\nabla T(\boldsymbol{x})
		\right]
		=
		f(\boldsymbol{x}),
		&
		\boldsymbol{x}\in\Omega,
		\\
		T(\boldsymbol{x})
		=
		\bar{T}(\boldsymbol{x}),
		&
		\boldsymbol{x}\in\Gamma^{T},
		\\
		k(\boldsymbol{x})
		\dfrac{\partial T(\boldsymbol{x})}{\partial\boldsymbol{n}}
		=
		\bar{q}(\boldsymbol{x}),
		&
		\boldsymbol{x}\in\Gamma^{q}.
	\end{cases}
	\label{eq:poisson_equation}
\end{equation}
Here, $k(\boldsymbol{x})$ is the heterogeneous thermal-conductivity field, $T(\boldsymbol{x})$ is the temperature field, and $f(\boldsymbol{x})$ is the heat-source field. The quantities $\bar{T}(\boldsymbol{x})$ and $\bar{q}(\boldsymbol{x})$ denote the prescribed temperature on the Dirichlet boundary and the prescribed heat flux on the Neumann boundary, respectively. The operator-learning problem is defined as
\begin{equation}
\left\{
k(\boldsymbol{x}),
f(\boldsymbol{x}),
\bar{T}(\boldsymbol{x}),
\bar{q}(\boldsymbol{x})
\right\}
\overset{\mathcal{\boldsymbol{H}}_{\boldsymbol{\theta}}}
{\longrightarrow}
T(\boldsymbol{x}).
\end{equation}
The strong-form representation of \Cref{eq:poisson_equation} within the framework of \Cref{eq:PDEs_driven} is
\begin{equation}
	\begin{aligned}
		\mathcal{L}_{\mathrm{pde\text{-}s}}
		&=
		\frac{1}{N}
		\sum_{i=1}^{N}
		\left\{
		\frac{1}{N_{d}}
		\sum_{j=1}^{N_{d}}
		\left\|
		\nabla\cdot
		\left[
		k^{(i)}(\boldsymbol{x}^{(j)})
		\nabla
		\mathcal{\boldsymbol{H}}_{\boldsymbol{\theta}}
		\left(
		\boldsymbol{x}^{(j)};
		k^{(i)},f^{(i)},\bar{T}^{(i)},\bar{q}^{(i)}
		\right)
		\right]
		+
		f^{(i)}(\boldsymbol{x}^{(j)})
		\right\|_{2}^{2}
		\right.
		\\
		&\qquad
		+
		\frac{1}{N_{bT}}
		\sum_{j=1}^{N_{bT}}
		\left\|
		\mathcal{\boldsymbol{H}}_{\boldsymbol{\theta}}
		\left(
		\boldsymbol{x}^{(j)};
		k^{(i)},f^{(i)},\bar{T}^{(i)},\bar{q}^{(i)}
		\right)
		-
		\bar{T}^{(i)}(\boldsymbol{x}^{(j)})
		\right\|_{2}^{2}
		\\
		&\qquad\left.
		+
		\frac{1}{N_{bq}}
		\sum_{j=1}^{N_{bq}}
		\left\|
		k^{(i)}(\boldsymbol{x}^{(j)})
		\boldsymbol{n}\cdot\nabla
		\mathcal{\boldsymbol{H}}_{\boldsymbol{\theta}}
		\left(
		\boldsymbol{x}^{(j)};
		k^{(i)},f^{(i)},\bar{T}^{(i)},\bar{q}^{(i)}
		\right)
		-
		\bar{q}^{(i)}(\boldsymbol{x}^{(j)})
		\right\|_{2}^{2}
		\right\}.
	\end{aligned}
	\label{eq:strong_form_PDEs}
\end{equation}
Here, $N_{bT}$ and $N_{bq}$ denote the total numbers of points on the Dirichlet and Neumann boundaries, respectively. \Cref{eq:strong_form_PDEs} is a strong-form loss function used in PINO \citet{li2024physics}. Alternatively, the variational principle can be used to express the loss function in the energy form adopted by VINO \citet{eshaghi2025variational}:
\begin{equation}
	\begin{aligned}
		\mathcal{L}_{\mathrm{pde\text{-}v}}
		&=
		\frac{1}{N}
		\sum_{i=1}^{N}
		\left\{
		\int_{\Omega}
		\frac{1}{2}
		k^{(i)}(\boldsymbol{x})
		\left[
		\nabla
		\mathcal{\boldsymbol{H}}_{\boldsymbol{\theta}}
		\left(
		\boldsymbol{x};
		k^{(i)},f^{(i)},\bar{T}^{(i)},\bar{q}^{(i)}
		\right)
		\right]
		\cdot
		\left[
		\nabla
		\mathcal{\boldsymbol{H}}_{\boldsymbol{\theta}}
		\left(
		\boldsymbol{x};
		k^{(i)},f^{(i)},\bar{T}^{(i)},\bar{q}^{(i)}
		\right)
		\right]
		\,d\Omega
		\right.
		\\
		&\qquad
		-
		\int_{\Omega}
		f^{(i)}(\boldsymbol{x})
		\mathcal{\boldsymbol{H}}_{\boldsymbol{\theta}}
		\left(
		\boldsymbol{x};
		k^{(i)},f^{(i)},\bar{T}^{(i)},\bar{q}^{(i)}
		\right)
		\,d\Omega
		\\
		&\qquad\left.
		-
		\int_{\Gamma^{q}}
		\bar{q}^{(i)}(\boldsymbol{x})
		\mathcal{\boldsymbol{H}}_{\boldsymbol{\theta}}
		\left(
		\boldsymbol{x};
		k^{(i)},f^{(i)},\bar{T}^{(i)},\bar{q}^{(i)}
		\right)
		\,d\Gamma
		\right\}.
	\end{aligned}
	\label{eq:energy_form_PDEs}
\end{equation}
It should be noted that the choice between the strong-form loss
$\mathcal{L}_{\mathrm{pde\text{-}s}}$ and the energy-form loss
$\mathcal{L}_{\mathrm{pde\text{-}v}}$ depends on the properties of the governing PDEs. Corresponding energy formulations are generally available primarily for steady-state problems, and not all PDEs possess an energy form. Further details are provided in Appendix A of \citet{wang2025physics}. Empirically, VINO generally achieves better efficiency and accuracy than strong-form PINO. This advantage mainly arises because the energy formulation requires lower-order derivatives than the strong formulation.

Empirically, training a neural operator using the strong-form loss function in \Cref{eq:strong_form_PDEs} is often difficult. When the underlying problem becomes complex, as in plasticity, many hyperparameters must be adjusted. In contrast, training a neural operator using the energy-form loss function in \Cref{eq:energy_form_PDEs} is generally much easier because it involves fewer hyperparameters and lower-order derivatives. Fortunately, plasticity possesses the incremental potential given in \Cref{eq:simo}. This provides the possibility of training neural operators using an energy formulation, as discussed in detail in \Cref{subsec:PlasPINO:-Physics-informed-neura}.

It should be emphasized that PINO is essentially a framework for training neural operators using physical equations. Therefore, the specific neural-operator architecture adopted within this framework is not the central issue. Neural operators are currently implemented mainly through three types of architectures: DeepONet \citet{DeepOnet}, the Fourier neural operator (FNO), and Transformer-based neural operators. Recently, Transformer-based operator learning has demonstrated stronger performance than FNO and DeepONet in several applications. For example, Transformer-based neural operators achieve leading performance on the high-fidelity three-dimensional automotive aerodynamics problems in CarBench (\url{https://mohamedelrefaie.github.io/CarBench/}). Therefore, we focus on Transformer-based operator learning, particularly Transolver \citet{wu2024transolver}, which has recently attracted considerable attention.

\begin{figure}
	\begin{centering}
		\includegraphics[scale=0.47]{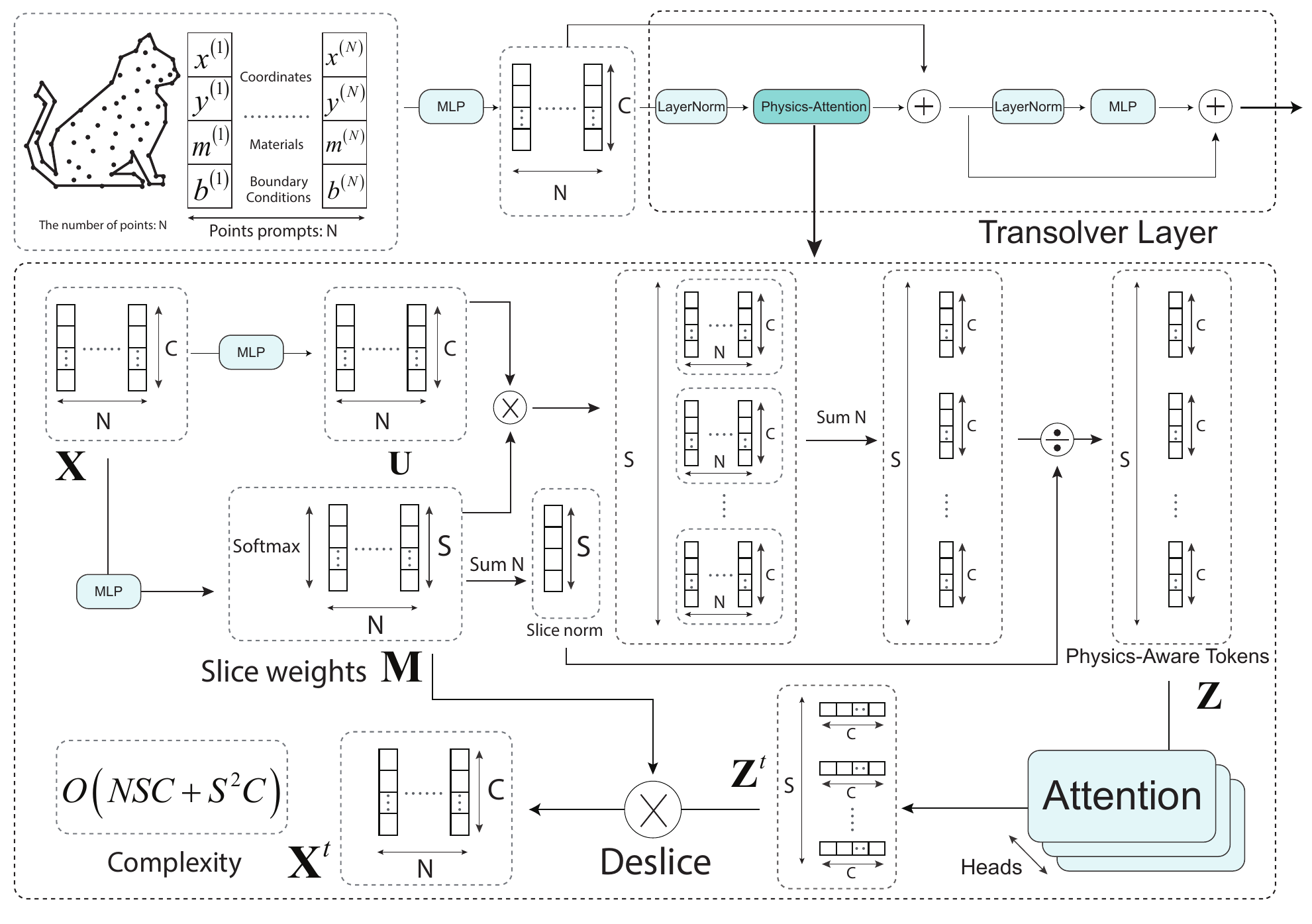}
		\par\end{centering}
	\caption{Architecture of a Transolver layer.
		\label{fig:Transolver}}
\end{figure}

The central idea of Transolver is to perform attention over a small number of physics-aware tokens, as illustrated in \Cref{fig:Transolver}. Because Transolver \citet{wu2024transolver} performs attention over physics-aware tokens, its computational complexity is reduced to $O(N+S^{2})$, where $S$ is the total number of physics-aware tokens. Since $S\ll N$, Transolver has linear complexity with respect to $N$ and therefore exhibits considerable potential for highly complex problems.

We next explain the computational procedure of Transolver. First, all points in the domain are encoded as
$\{x^{(i)},y^{(i)},m^{(i)},b^{(i)}\}_{i=1}^{N}$,
where $x^{(i)}$ and $y^{(i)}$ are the coordinates, $m^{(i)}$ denotes the material information at the corresponding location, such as the elastic modulus, and $b^{(i)}$ represents the boundary-condition information at that point. It should be noted that these points do not need to be arranged on a regular grid. The point information is first processed by a multilayer perceptron (MLP), followed by layer normalization to normalize each feature and accelerate model convergence. This produces
$\boldsymbol{X}\in\mathbb{R}^{N\times C}$,
where $C$ is the number of feature channels associated with each point. The resulting features are then processed by the physics-attention module, which is the central component of Transolver. Specifically, $\boldsymbol{X}$ is passed through two different MLPs to produce
$\boldsymbol{U}\in\mathbb{R}^{N\times C}$
and
$\boldsymbol{M}\in\mathbb{R}^{N\times S}$,
where $S$ is the number of physics-aware tokens. A softmax operation is applied to the slice weights $\boldsymbol{M}$ along the dimension associated with $S$, allowing them to be interpreted as weights. The features in $\boldsymbol{U}$ are then aggregated using $\boldsymbol{M}$ as the weights, producing the physics-aware tokens
$\boldsymbol{Z}\in\mathbb{R}^{S\times C}$:
\begin{equation}
	Z_{jk}
	=
	\frac{
		\displaystyle\sum_{i=1}^{N}M_{ij}U_{ik}
	}{
		\displaystyle\sum_{i=1}^{N}M_{ij}
	}.
\end{equation}
It should be noted that $\boldsymbol{Z}$ is considerably smaller than $\boldsymbol{X}$ because $S\ll N$. Multi-head attention is then applied along the $S$ dimension of $\boldsymbol{Z}$:
\begin{equation}
	\begin{aligned}
		\boldsymbol{Q},\boldsymbol{K},\boldsymbol{V}
		&=
		\mathrm{MLP}(\boldsymbol{Z}),
		\\
		\boldsymbol{Z}^{t}
		&=
		\mathrm{softmax}
		\left(
		\boldsymbol{Q}\boldsymbol{K}^{T}
		\right)
		\boldsymbol{V}.
	\end{aligned}
\end{equation}
After obtaining
$\boldsymbol{Z}^{t}\in\mathbb{R}^{S\times C}$,
the slice weights $\boldsymbol{M}$ are used to deslice the physics-aware tokens and obtain the final output $\boldsymbol{X}^{t}$ of the physics-attention module:
\begin{equation}
	X^{t}_{ik}
	=
	\sum_{j=1}^{S}
	M_{ij}Z^{t}_{jk}.
\end{equation}

The subsequent operations are straightforward, as shown in \Cref{fig:Transolver}, and are therefore not described further. The central component of Transolver is its physics-attention module, which reduces the quadratic complexity $O(N^{2})$ of applying a Transformer directly to all points to $O(N+S^{2})$ by operating on physics-aware tokens. Transolver also possesses strong approximation capabilities. \citet{wu2024transolver} demonstrated its potential for problems involving complex three-dimensional geometries, particularly on the ShapeNet Car dataset. Therefore, Transolver has considerable research value and potential. Multiple Transolver layers can be stacked, similarly to the Fourier layers in FNO. Increasing the number of layers can improve the generalization capability of the model, although it also increases the computational cost. The number of Transolver layers and the value of $S$ can be adjusted according to the complexity of the problem. In general, more complex problems require more Transolver layers and a larger value of $S$. The physics-aware tokens convert the original pointwise representation into a regional representation of the physical domain. The central idea of Transolver is that physical problems generally exhibit spatial continuity and similarity, making the information carried by individual points partially redundant.

\subsection{Plasolver}

\subsubsection{Pretraining stage of Plasolver: A physics-informed neural operator for elastoplasticity\label{subsec:PlasPINO:-Physics-informed-neura}}

The previous section introduced neural operators and described how they can be trained using physical equations. The central objective of this section is to apply physics-informed neural operators to plasticity, leading to the proposed Plasolver framework. Empirically, training an operator using the energy formulation in \Cref{eq:energy_form_PDEs} is more stable and efficient than using the strong formulation \citet{wang2026pfem}. Therefore, the key to constructing a PINO for elastoplasticity lies in the selection of an appropriate loss function.

Fortunately, plasticity possesses a corresponding energy formulation, namely the incremental potential in \Cref{eq:simo}. Therefore, the energy functional of the deep energy method in \Cref{eq:DEM_plasticity} can be used as the PINO loss function for training the neural operator. Based on the energy components in \Cref{eq:DEM_plasticity}, the information associated with the current load step of the plasticity problem is used as the input to the Transolver neural operator. Mathematically, the operator is expected to learn the mapping
\begin{equation}
\boldsymbol{G},
\boldsymbol{M},
\boldsymbol{B}^{(n+1)},
\boldsymbol{I}^{(n)}
\overset{\mathcal{\boldsymbol{H}}}{\longrightarrow}
\boldsymbol{U}^{(n+1)},
\end{equation}
where $\boldsymbol{G}$, $\boldsymbol{M}$, $\boldsymbol{B}$, $\boldsymbol{I}$, and $\boldsymbol{U}$ denote the geometry, material properties, boundary conditions, internal variables, and displacement field, respectively. Here, $n$ denotes the current time step, and $n+1$ denotes the next time step. Therefore, constructing a PINO for elastoplasticity is transformed into training the neural operator
\begin{equation}
\mathcal{\boldsymbol{H}}
\left(
\boldsymbol{G}^{(i)},
\boldsymbol{M}^{(i)},
\boldsymbol{B}^{(i)(n+1)},
\boldsymbol{I}^{(i)(n)};
\boldsymbol{\theta}
\right),
\end{equation}
where $i$ denotes the input instance and $\boldsymbol{\theta}$ denotes the trainable parameters of the neural operator $\mathcal{\boldsymbol{H}}$. The complete dataset can be represented as
\begin{equation}
\left\{
\boldsymbol{G}^{(i)},
\boldsymbol{M}^{(i)},
\boldsymbol{B}^{(i)(n+1)},
\boldsymbol{I}^{(i)(n)},
\boldsymbol{U}^{(i)(n+1)}
\right\}_{i=1}^{N},
\qquad
1\leq n\leq\tau^{(i)},
\end{equation}
where $\tau^{(i)}$ denotes the total number of load steps in the $i$th loading trajectory.

Because Plasolver trains the neural operator using physical equations, only
\begin{equation}
\left\{
\boldsymbol{G}^{(i)},
\boldsymbol{M}^{(i)},
\boldsymbol{B}^{(i)(n+1)},
\boldsymbol{I}^{(i)(n)}
\right\}_{i=1}^{N}
\end{equation}
are used, without the labeled displacement field $\boldsymbol{U}^{(i)(n+1)}$. The neural operator $\mathcal{\boldsymbol{H}}$ first predicts the displacement field
$\hat{\boldsymbol{u}}^{(i)(n+1)}$ at the next step. A distance function $\boldsymbol{Dis}$ is then used to impose the displacement boundary conditions and construct a kinematically admissible displacement field:
\begin{equation}
	\begin{aligned}\boldsymbol{u}^{(i)(n+1)} & =\boldsymbol{Dis}(\boldsymbol{x};\boldsymbol{B}^{(i)(n+1)})*\hat{\boldsymbol{u}}^{(i)(n+1)}+\boldsymbol{u}^{(n+1)}_{p}(\boldsymbol{x};\boldsymbol{B}^{(i)(n+1)})\\
		\boldsymbol{Dis}(\boldsymbol{x};\boldsymbol{B}^{(i)(n+1)}) & =\min_{y\in\Gamma^{u}}\sqrt{(\boldsymbol{x}-\boldsymbol{y})\cdot(\boldsymbol{x}-\boldsymbol{y})}\\
		\hat{\boldsymbol{u}}^{(i)(n+1)} & =\mathcal{\boldsymbol{H}}(\boldsymbol{G}^{(i)},\boldsymbol{M}^{(i)},\boldsymbol{B}^{(i)(n+1)},\boldsymbol{I}^{(i)(n)};\boldsymbol{\theta})\\
		\boldsymbol{u}^{(n+1)}_{p}(\boldsymbol{x};\boldsymbol{B}^{(i)(n+1)}) & =\bar{\boldsymbol{u}}^{(n+1)},\boldsymbol{x}\in\Gamma^{u}
	\end{aligned}
	\label{eq:admissible_displacement}
\end{equation}
It can be readily verified that $\boldsymbol{u}^{(i)(n+1)}$ satisfies the essential boundary conditions.

The displacement field $\boldsymbol{u}^{(i)(n+1)}$ in \Cref{eq:admissible_displacement} is substituted into the incremental potential in \Cref{eq:simo} to calculate the loss function in \Cref{eq:Plasolver_plasticity}:
\begin{equation}
	\begin{aligned}
		\mathcal{L}
		\left(
		\boldsymbol{u}^{(i)(n+1)}
		\right)
		&=
		W^{(i)(n+1)}_{e}
		+
		W^{(i)(n+1)}_{p}
		+
		W^{(i)(n+1)}_{h}
		-
		W^{(i)(n+1)}_{\mathrm{ext}},
		\\
		W^{(i)(n+1)}_{e}
		&=
		\int_{\Omega}
		\frac{1}{2}
		\boldsymbol{\varepsilon}^{e(i)(n+1)}
		\left(
		\boldsymbol{u}^{(i)(n+1)}
		\right)
		:
		\boldsymbol{C}^{(i)}
		:
		\boldsymbol{\varepsilon}^{e(i)(n+1)}
		\left(
		\boldsymbol{u}^{(i)(n+1)}
		\right)
		\,dV,
		\\
		W^{(i)(n+1)}_{p}
		&=
		\int_{\Omega}
		\left[
		\boldsymbol{\varepsilon}^{p(i)(n+1)}
		\left(
		\boldsymbol{u}^{(i)(n+1)}
		\right)
		-
		\boldsymbol{\varepsilon}^{p(i)(n)}
		\right]
		:
		\boldsymbol{C}^{(i)}
		:
		\boldsymbol{\varepsilon}^{e(i)(n+1)}
		\left(
		\boldsymbol{u}^{(i)(n+1)}
		\right)
		\,dV,
		\\
		W^{(i)(n+1)}_{h}
		&=
		\int_{\Omega}
		\left\{
		\frac{1}{2}
		\boldsymbol{v}^{(i)(n+1)}
		\left(
		\boldsymbol{u}^{(i)(n+1)}
		\right)
		\cdot
		\left(
		\boldsymbol{D}^{-1}
		\right)^{(i)}
		\cdot
		\boldsymbol{v}^{(i)(n+1)}
		\left(
		\boldsymbol{u}^{(i)(n+1)}
		\right)
		\right.
		\\
		&\qquad\left.
		-
		\boldsymbol{v}^{(i)(n+1)}
		\left(
		\boldsymbol{u}^{(i)(n+1)}
		\right)
		\cdot
		\left(
		\boldsymbol{D}^{-1}
		\right)^{(i)}
		\cdot
		\left[
		\boldsymbol{v}^{(i)(n+1)}
		\left(
		\boldsymbol{u}^{(i)(n+1)}
		\right)
		-
		\boldsymbol{v}^{(i)(n)}
		\right]
		\right\}
		\,dV,
		\\
		W^{(i)(n+1)}_{\mathrm{ext}}
		&=
		\int_{\Omega}
		\boldsymbol{f}^{(i)(n+1)}
		\cdot
		\boldsymbol{u}^{(i)(n+1)}
		\,dV
		+
		\int_{\Gamma^{t}}
		\bar{\boldsymbol{t}}^{(i)(n+1)}
		\cdot
		\boldsymbol{u}^{(i)(n+1)}
		\,dA.
	\end{aligned}
	\label{eq:Plasolver_plasticity}
\end{equation}
The optimization objective of Plasolver is
\begin{equation}
	\boldsymbol{\theta}^{*}
	=
	\arg\min_{\boldsymbol{\theta}}
	\sum_{i=1}^{N}
	\sum_{n=1}^{\tau^{(i)}}
	\mathcal{L}
	\left(
	\boldsymbol{u}^{(i)(n+1)}
	\right).
\end{equation}
In practice, because the available GPU memory is limited, a batch of size $B$ is generally selected to calculate the gradient:
\begin{equation}
	\nabla_{\boldsymbol{\theta}}
	\sum_{i=1}^{B}
	\sum_{n=1}^{\tau^{(i)}}
	\mathcal{L}
	\left(
	\boldsymbol{u}^{(i)(n+1)}
	\right).
\end{equation}
After the gradient has been obtained, the converged parameters $\boldsymbol{\theta}$ are determined iteratively. Plasolver thereby learns the operator mapping
\begin{equation}
\boldsymbol{G},
\boldsymbol{M},
\boldsymbol{B}^{(n+1)},
\boldsymbol{I}^{(n)}
\overset{\mathcal{\boldsymbol{H}}}{\longrightarrow}
\boldsymbol{U}^{(n+1)}.
\end{equation}
It should be noted that the derivatives in the Plasolver energy loss are evaluated explicitly using shape functions. Compared with automatic differentiation (AD), this approach can substantially reduce the training cost, as detailed in Appendix A of \citet{wang2026pfem}. Because the derivatives are obtained using shape functions, the essential boundary conditions could theoretically be imposed strongly using the conventional finite element procedure, without employing a distance function. This issue is discussed in detail in \Cref{subsec:Shape_function_not_need_distance}. The distance function is used here because it improves the stability of Plasolver training. In addition, the Plasolver loss is evaluated over an entire loading trajectory and is then used to train the neural operator. The neural operator is not trained separately at each load step. Optimizing over the complete trajectory accounts for every load step simultaneously and identifies a suitable overall optimization direction. Training the neural operator separately at individual load steps can cause gradient cancellation, resulting in inefficient or even unsuccessful training.

\begin{figure}
	\begin{centering}
		\includegraphics[scale=0.57]{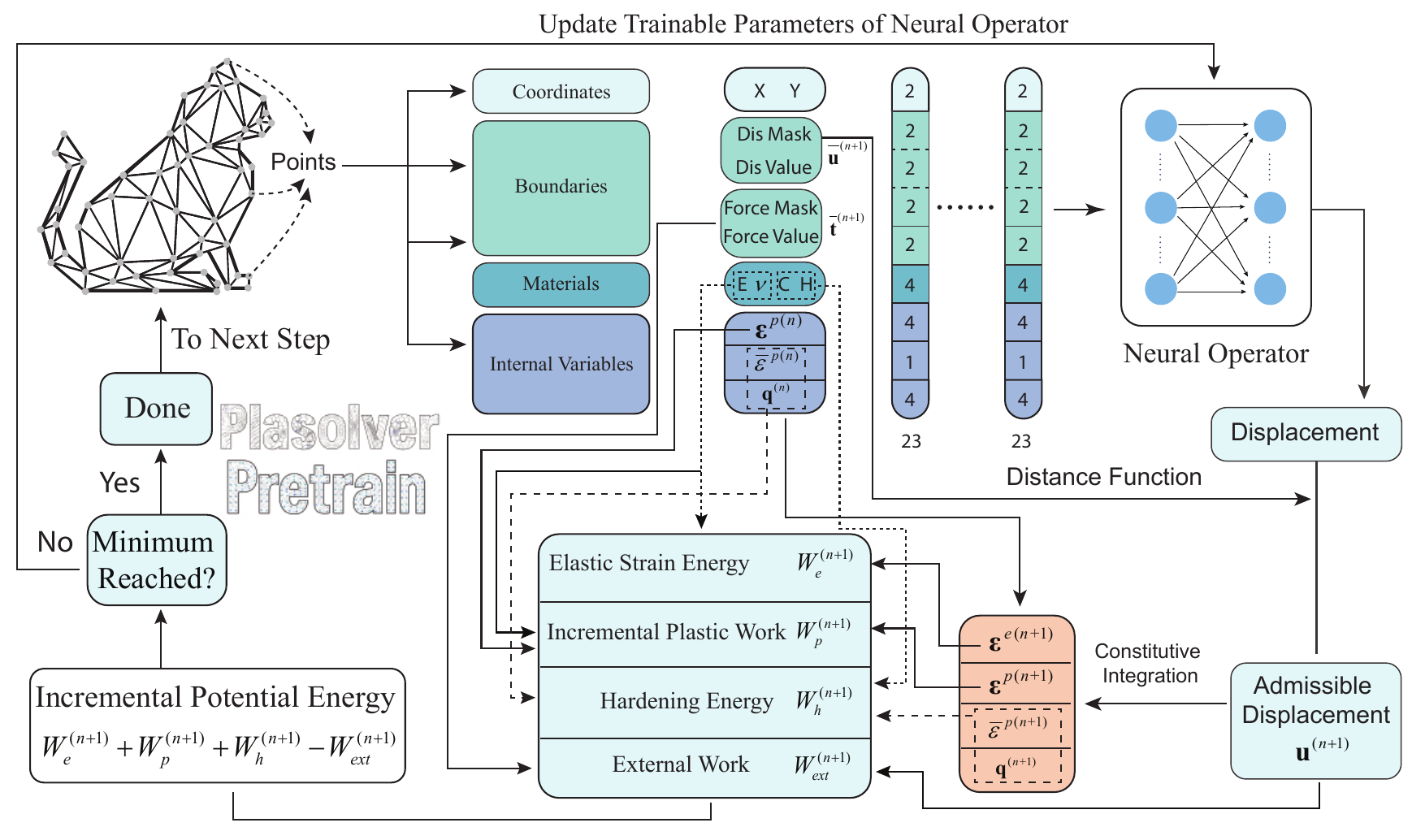}
		\par\end{centering}
	\caption{Schematic illustration of the pretraining stage of Plasolver. The points representing the problem are provided as input to the neural operator, which outputs the corresponding displacement field. The Dirichlet boundary conditions are imposed on the displacement field to construct a kinematically admissible displacement. The predicted displacement field is processed through constitutive integration to satisfy the yield condition. Based on Simo's incremental potential, the inputs and outputs are used to construct the elastic strain energy $W^{(n+1)}_{e}$, incremental plastic work $W^{(n+1)}_{p}$, hardening energy $W^{(n+1)}_{h}$, and external work $W^{(n+1)}_{\mathrm{ext}}$. These four energy components constitute the incremental potential energy, which is used as the loss function for optimizing the trainable parameters of the neural operator. Plasolver has four categories of inputs: coordinates, boundary conditions, material properties, and internal variables.
		\label{fig:Plasolver_pretrain}}
\end{figure}

We use a two-dimensional plane-strain problem to illustrate Plasolver, as shown in \Cref{fig:Plasolver_pretrain}. The inputs include the coordinates $x$ and $y$ of the two-dimensional structure. The coordinate point cloud
$\{x^{(m)},y^{(m)}\}_{m=1}^{N_{p}}$
represents the geometry $\boldsymbol{G}$, where $N_{p}$ is the total number of spatial points. The boundary-condition information consists of two parts: the displacement boundary conditions and the traction boundary conditions. Each component is represented by a mask and a corresponding prescribed value. For example, a value of zero in a displacement-boundary mask indicates that the corresponding displacement component is not prescribed at that point, whereas a value of one indicates that it is prescribed. The same representation is used for the traction boundary conditions. When a mask has a value of one, the corresponding prescribed displacement $\bar{\boldsymbol{u}}^{(n+1)}$ or traction $\bar{\boldsymbol{t}}^{(n+1)}$ is provided. Therefore, the boundary-condition input $\boldsymbol{B}$ has eight dimensions for a two-dimensional problem. For mixed hardening with linear isotropic and kinematic hardening, four parameters describe the material properties $\boldsymbol{M}$: the elastic modulus $E$, Poisson's ratio $\nu$, the plastic modulus $H$, and the kinematic hardening modulus $C$. The central feature of a plasticity problem is its internal variables, including the plastic strain $\boldsymbol{\varepsilon}^{p(n)}$, equivalent plastic strain $\bar{\varepsilon}^{p(n)}$, and backstress $\boldsymbol{q}^{(n)}$. A total of nine dimensions are used to represent the internal variables $\boldsymbol{I}$.

Consequently, each point contains 23 input features. All points are provided as input to the Transolver neural operator, which outputs the displacement field $\boldsymbol{u}^{(n+1)}$ at step $n+1$. The corresponding energy loss is then calculated using \Cref{eq:Plasolver_plasticity}, and Plasolver is trained through gradient-based minimization of this energy loss. It should be noted that elastoplastic problems are strongly dependent on the loading path. In practice, the same loading trajectory may be discretized using different numbers of load steps. Plasolver must therefore possess path-discretization invariance. Path-discretization invariance means that Plasolver should produce similar predictions for the same loading path when different numbers of load steps are used, thereby providing robust predictions. The Plasolver framework shown in \Cref{fig:Plasolver_pretrain} learns an incremental operator and does not explicitly encode a fixed number of load steps. Therefore, Plasolver learns local increments along the loading path, allowing it to produce similar predictions when the same path is discretized using different numbers of steps. This property is experimentally verified in \Cref{fig:Plasolver_path_discretization_invariance}, which shows that Plasolver possesses good path-discretization invariance. Plasolver also possesses spatial-discretization invariance, meaning that it provides robust predictions when different spatial point discretizations are used. \Cref{fig:Plasolver_space_discretization_invariance} demonstrates the spatial-discretization invariance of Plasolver.

\Cref{sec:Result} systematically evaluates the generalization capability of Plasolver with respect to the geometry $\boldsymbol{G}$, material properties $\boldsymbol{M}$, and boundary conditions $\boldsymbol{B}$. Its spatial- and path-discretization invariance are also evaluated.

The neural operator employed in the current implementation of Plasolver is primarily Transolver \citet{wu2024transolver}, which has recently received considerable attention. As neural-operator architectures continue to develop, Transolver can be replaced by more powerful neural operators in future implementations. One example is AB-UPT \citet{alkin2025ab}, which has recently demonstrated strong performance in automotive flow-field prediction.

\subsubsection{Warm-start phase of Plasolver}

The pretrained Plasolver model can generally achieve extremely high simulation speeds, providing an approximately 100-fold speedup over conventional algorithms while maintaining errors within $5\%$. If this level of accuracy is insufficient, the warm-start phase of Plasolver can be used to further improve the accuracy.

\begin{figure}
	\begin{centering}
		\includegraphics[scale=0.57]{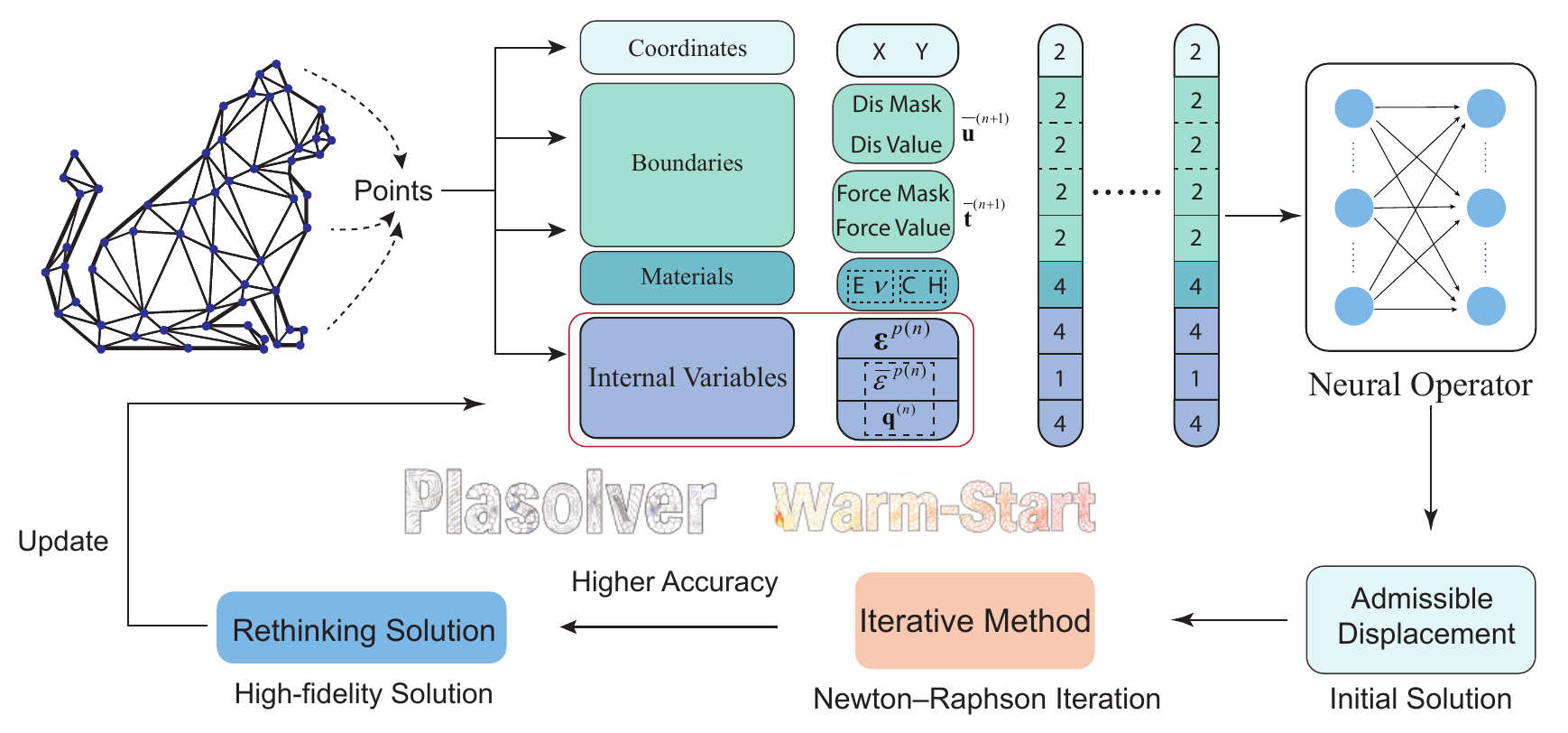}
		\par\end{centering}
	\caption{Schematic illustration of the Plasolver warm-start phase. The prediction obtained during the pretraining phase is used as the initial solution of the iterative algorithm.
		\label{fig:Plasolver_warmstart}}
\end{figure}

The central idea of the Plasolver warm-start phase is to use the output of the pretrained Plasolver model as the initial solution of an iterative solver, as illustrated in \Cref{fig:Plasolver_warmstart}. After the physics-based pretraining of Plasolver has been completed, researchers from different fields can submit a specific problem to the pretrained Plasolver model, which rapidly provides an initial solution. It should be noted that pretraining Plasolver generally requires a large number of problem instances and is therefore typically performed on a supercomputer. After training has been completed, the pretrained Plasolver model can be deployed in the cloud. Individual users can then employ their local computational resources to refine the initial solution through warm-start iterations, thereby obtaining high-accuracy numerical solutions more efficiently. This ``cloud pretraining + local warm start'' paradigm can substantially improve the solution efficiency.

The technical basis of the Plasolver warm-start phase is an iterative numerical algorithm. We focus here on the iterative finite element algorithm. Because plasticity problems are nonlinear, a nonlinear finite element iterative algorithm is employed. The central idea of the Plasolver warm start is to use the pretrained prediction
$\boldsymbol{u}^{(n+1)}_{\mathrm{no}}$
at the current load step $n$, where the subscript $\mathrm{no}$ denotes the trained neural operator, as the initial solution $\boldsymbol{U}^{(n+1,0)}$ of the finite element iteration:
\begin{equation}
	\begin{aligned}
		\boldsymbol{U}^{(n+1,k+1)}
		&=
		\boldsymbol{\phi}
		\left(
		\boldsymbol{U}^{(n+1,k)};
		\boldsymbol{K}^{(k)},
		\boldsymbol{F}^{(k)}
		\right),
		\\
		\boldsymbol{U}^{(n+1,0)}
		&=
		\boldsymbol{u}^{(n+1)}_{\mathrm{no}}.
	\end{aligned}
	\label{eq:iteraitive_linear_TINO}
\end{equation}
Here, $\boldsymbol{U}$ denotes the nodal values of the field to be solved, such as the displacement field, and $k$ denotes the current iteration. The vector $\boldsymbol{F}$ is the global nodal force vector determined by the boundary conditions. The matrix $\boldsymbol{K}$ is the global stiffness matrix determined by the geometry and material properties, and $\boldsymbol{\phi}$ denotes the iterative algorithm. The iterations terminate when the prescribed convergence tolerance is satisfied:
\begin{equation}
	\left\|
	\boldsymbol{r}
	\left(
	\boldsymbol{U}^{(n+1,k+1)}
	\right)
	\right\|_{2}
	<
	\mathrm{tol}.
	\label{eq:tol_TINO}
\end{equation}
Here, $\boldsymbol{r}$ is the residual used to evaluate convergence:
\begin{equation}
\boldsymbol{r}(\boldsymbol{U})
=
\boldsymbol{f}^{\mathrm{int}}(\boldsymbol{U})
-
\boldsymbol{f}^{\mathrm{ext}}(\boldsymbol{U}),
\end{equation}
where $\boldsymbol{f}^{\mathrm{int}}$ and $\boldsymbol{f}^{\mathrm{ext}}$ denote the global internal- and external-force vectors, respectively. The quantity $\mathrm{tol}$ is the convergence threshold. After the displacement field at the current load step has converged, the internal variables obtained from the converged finite element solution are used to update the internal-variable input of Plasolver. The algorithm then proceeds to the next load step, $n+1$. This process is repeated until all load steps have been completed.

Because the pretrained Plasolver model can provide predictions at arbitrary spatial locations, predictions at the nodes of the warm-start mesh can be readily obtained by supplying their coordinates to Plasolver. The prediction from the pretraining phase is then used as the initial solution of the iterative algorithm. Conceptually, the pretraining and warm-start phases of Plasolver are similar to the pretraining and reinforcement-learning stages of large language models \citet{guo2025deepseek}. Because the pretraining phase of Plasolver provides a high-quality initial solution, the warm-start phase can require significantly fewer iterations than conventional methods initialized using random vectors. Fewer iterations lead directly to a reduction in computational time. Theoretically, the solution refined through the Plasolver warm-start phase can converge to the corresponding discrete finite element solution, provided that the convergence threshold $\mathrm{tol}$ of the iterative algorithm is sufficiently small. It should be noted that the warm-start phase is optional and can be applied according to the accuracy requirements of a specific problem. To determine whether the warm-start phase is necessary, we propose substituting the initial solution provided by the pretrained Plasolver model into \Cref{eq:tol_TINO} and evaluating its residual $\boldsymbol{r}$. Because $\boldsymbol{r}$ reflects the accuracy of the initial solution, the warm-start phase is considered unnecessary if
\begin{equation}
\|\boldsymbol{r}\|_{2}<\mathrm{tol}_{\mathrm{fine}},
\end{equation}
where $\mathrm{tol}_{\mathrm{fine}}$ is the threshold used to determine whether warm-start refinement should be performed.

If a moderate level of error is acceptable, the pretrained Plasolver model can be used directly. If a high-accuracy solution is required, the warm-start phase can be employed at the cost of additional computational resources. It should be noted that Plasolver follows the same central idea as PFEM \citet{wang2026pfem}. Specifically, Plasolver extends the PFEM framework to plasticity problems.

\section{Results\label{sec:Result}}

The main objective of this section is to systematically evaluate the generalization capability of Plasolver with respect to the geometry $\boldsymbol{G}$, material properties $\boldsymbol{M}$, and boundary conditions $\boldsymbol{B}$, as well as its spatial- and path-discretization invariance. All experimental results are summarized in \Cref{tab:Plasolver_diff_problem}, and the performance of Plasolver is described in detail below. All results are evaluated against finite element reference solutions, and convergence studies have been performed for all finite element results detailed in \ref{sec:FEM_convergence_proof}. Therefore, the finite element solutions are considered sufficiently reliable. The computations were performed using an NVIDIA A100 GPU with 80 GB of memory and 128 GB of CPU memory on the rented ``Kaituo 1000'' GPU-accelerated HPC cluster at Tsinghua University.

\begin{table}
	\caption{
		Accuracy and computational efficiency of Plasolver. 
		``Layer'' and ``Slice'' denote the Transolver architecture, and ``Learning rate'' is the learning rate used during training.
		The reported relative errors and standard deviations are averaged over all intermediate loading steps.
		The ``P: Cartoon'' case contains 80 loading steps, whereas all other cases contain 40 loading steps.
		``Time'' denotes the total computational time required to complete all loading steps, where the Plasolver time only includes the pretrained inference stage.
		``Iterative Steps'' compares the number of Newton iterations required by the conventional FEM and the Plasolver warm-start strategy.
		For the first four experiments (generalization to different loading paths), the displacement convergence tolerance is set to $10^{-5}$.
		For the fifth experiment (generalization to different geometries and loading paths) and the last two experiments (generalization to different materials, geometries, and loading paths), the displacement convergence tolerance is set to $10^{-4}$.
		All experiments are conducted on a mesh with a resolution of $100\times100$.
		``P'' denotes different loading paths, ``G'' denotes different geometries, and ``M'' denotes different materials.
		``NP'', ``UL'', and ``CY'' represent non-proportional loading, loading--unloading, and cyclic loading, respectively.
		``Cartoon'' denotes loading paths defined by cartoon-shaped contours.
		``L'' and ``N'' denote linear isotropic hardening and nonlinear isotropic hardening, respectively, where the constitutive integration is performed using the radial return algorithm and the return-mapping algorithm.  “LR” represent the learning rate.
		All reported relative errors and standard deviations correspond to the converged models.
		\label{tab:Plasolver_diff_problem}
	}
	
	\centering
	\begin{adjustbox}{max width=\textwidth}
\begin{tabular}{ccccccc}
	\toprule 
	Problem & Layer/Slice/LR & Relative Error: $|\boldsymbol{u}|$,$\sigma_{mise}$,$\bar{\varepsilon}^{p}$,$q_{mise}$ & Std: $|\boldsymbol{u}|$,$\sigma_{mise}$,$\bar{\varepsilon}^{p}$,$q_{mise}$ & Time: s (FEM/Plasolver) & Iterative Steps (FEM/Plasolver) & Optimizer\\
	\midrule 
	P: NP & 6/64/0.0002 & $0.0075,0.0154,0.0417,0.0416$ & $0.0028,0.0026,0.0191,0.0185$ & $38.89/0.474=82.05$ & $0.00001:135.24/88.72=1.52$ & Adam\\
	P: UL & 6/64/0.0002 & $0.0051,0.0132,0.0211,0.0213$ & $0.0015,0.0039,0.0060,0.0047$ & $37.42/0.475=78.78$ & $0.00001:136.16/84.63=1.60$ & Adam\\
	P: CY & 6/64/0.0002 & $0.008,0.0196,0.0216,0.0426$ & $0.00392,0.0102,0.0066,0.0285$ & $35.17/0.477=73.73$ & $0.00001:130.12/77.79=1.67$ & Adam\\
	P:Cartoon & 6/64/0.0002 & $0.0032,0.0059,0.00569,0.0115$ & $0.00173,0.0023,0.0014,0.0031$ & $89.31/0.942=95.10$ & $0.00001:147.29/71.51=2.06$ & Adam\\
	G+P: CY & 6/64/0.0005 & $0.009,0.0504,0.0698,0.0855$ & $0.00481,0.0104,0.0144,0.0276$ & $37.12/0.472=78.64$ & $0.0001:120.04/84.44=1.42$ & Adam\\
	M+G+P: L & 6/64/0.0005 & $0.0154,0.077,0.0592,0.1002$ & $0.009,0.0170,0.01535,0.032$ & $37.82/0.473=79.95$ & $0.0001:121.79/97.63=1.24$ & Adam\\
	M+G+P: N & 6/64/0.0005 & $0.0158,0.058,0.0852,0.1096$ & $0.008,0.0163,0.01880,0.038$ & $52.70/0.733=71.89$ & $0.0001:146.51/92.14=1.59$ & Adam\\
	\bottomrule
\end{tabular}
\end{adjustbox}
\end{table}
\subsection{Different loading paths\label{subsec:path}}

\begin{figure}
	\begin{centering}
		\includegraphics[scale=0.47]{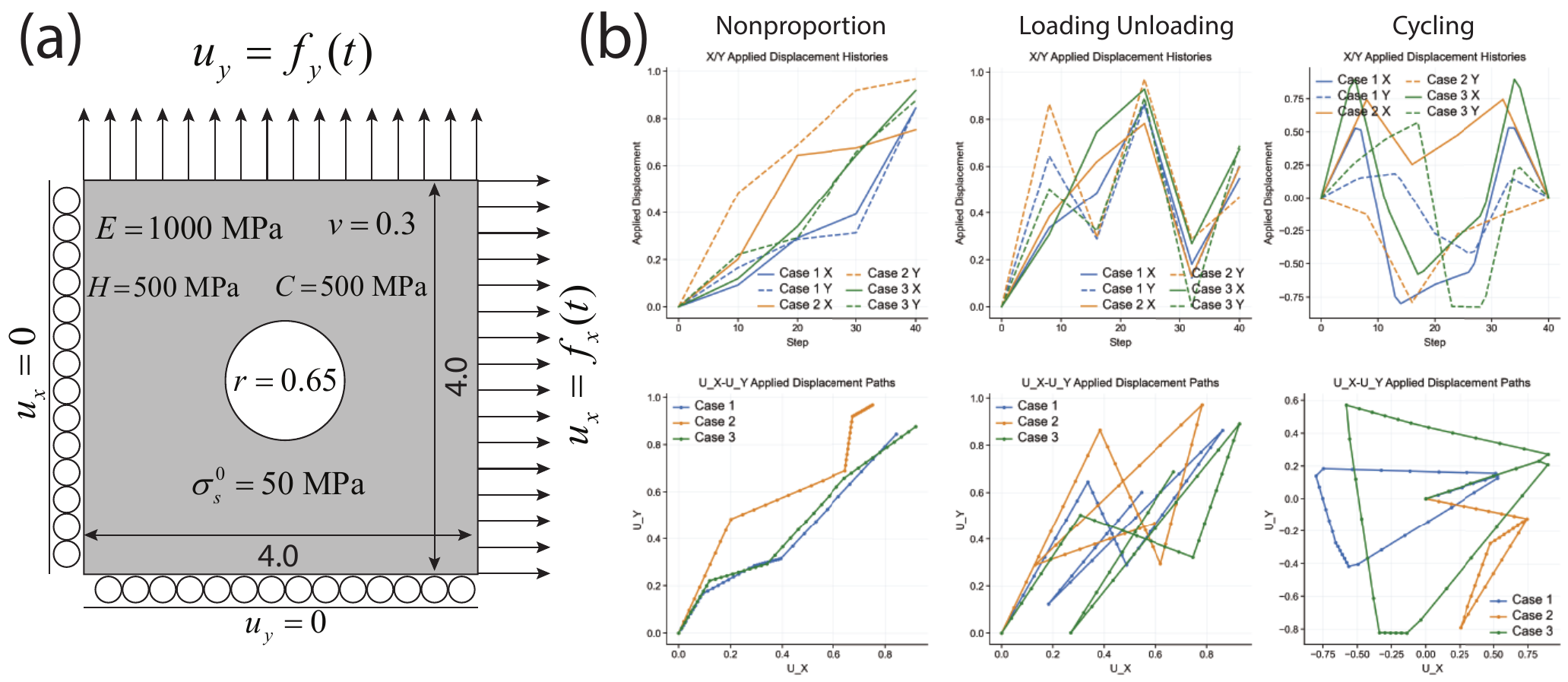}
		\par\end{centering}
	\caption{Elastoplastic problems under different loading paths. (a) Geometry, material properties, and boundary conditions of the elastoplastic problem. The geometry is a square plate with a length and width of $4$ mm and a circular hole with a radius of $0.65$ mm. The material parameters are the elastic modulus $E=1000$ MPa, Poisson's ratio $\nu=0.3$, plastic modulus $H=500$ MPa, and kinematic hardening modulus $C=500$ MPa. The initial yield stress is $\sigma^{0}_{y}=500$ MPa. The displacement in the $x$ direction is constrained on the left boundary, and the displacement in the $y$ direction is constrained on the bottom boundary. Displacement loading $u_{x}=f_{x}(t)$ is applied on the right boundary, and displacement loading $u_{y}=f_{y}(t)$ is applied on the top boundary. Both $f_{x}(t)$ and $f_{y}(t)$ are spatially uniform. (b) Three loading conditions: non-proportional loading, loading--unloading, and cyclic loading. Three randomly generated loading paths are presented for each condition. The first row shows the variations in $f_{x}(t)$ and $f_{y}(t)$ over the load steps, whereas the second row shows the corresponding trajectories in the $u_{x}$--$u_{y}$ loading space.
		\label{fig:Plasovler_path_intro}}
\end{figure}

Path dependence is a central feature of plasticity. This subsection focuses on evaluating the performance of Plasolver under different loading paths. To isolate this effect, the geometry and material properties are fixed, and only the loading paths are varied, as illustrated in \Cref{fig:Plasovler_path_intro}a. Specifically, three loading conditions are considered: non-proportional loading, loading--unloading, and cyclic loading, as shown in \Cref{fig:Plasovler_path_intro}b. These three loading conditions are investigated separately below. The displacement boundary conditions are imposed using distance functions:
\begin{equation}
	\begin{aligned}
		u^{(n+1)}
		&=
		u^{(n+1)}_{p}
		+
		\mathcal{D}_{x}
		\mathcal{H}_{x}
		\left(
		\boldsymbol{x};
		\boldsymbol{G},
		\boldsymbol{M},
		\boldsymbol{B}^{(n+1)},
		\boldsymbol{I}^{(n)},
		\boldsymbol{\theta}
		\right),
		\\
		v^{(n+1)}
		&=
		v^{(n+1)}_{p}
		+
		\mathcal{D}_{y}
		\mathcal{H}_{y}
		\left(
		\boldsymbol{x};
		\boldsymbol{G},
		\boldsymbol{M},
		\boldsymbol{B}^{(n+1)},
		\boldsymbol{I}^{(n)},
		\boldsymbol{\theta}
		\right),
		\\
		u^{(n+1)}_{p}
		&=
		\frac{x}{4}f_{x}(t),
		\\
		v^{(n+1)}_{p}
		&=
		\frac{y}{4}f_{y}(t),
		\\
		\mathcal{D}_{x}
		&=
		\frac{x(4-x)}{16},
		\\
		\mathcal{D}_{y}
		&=
		\frac{y(4-y)}{16}.
	\end{aligned}
\end{equation}

\begin{figure}
	\begin{centering}
		\includegraphics[scale=0.49]{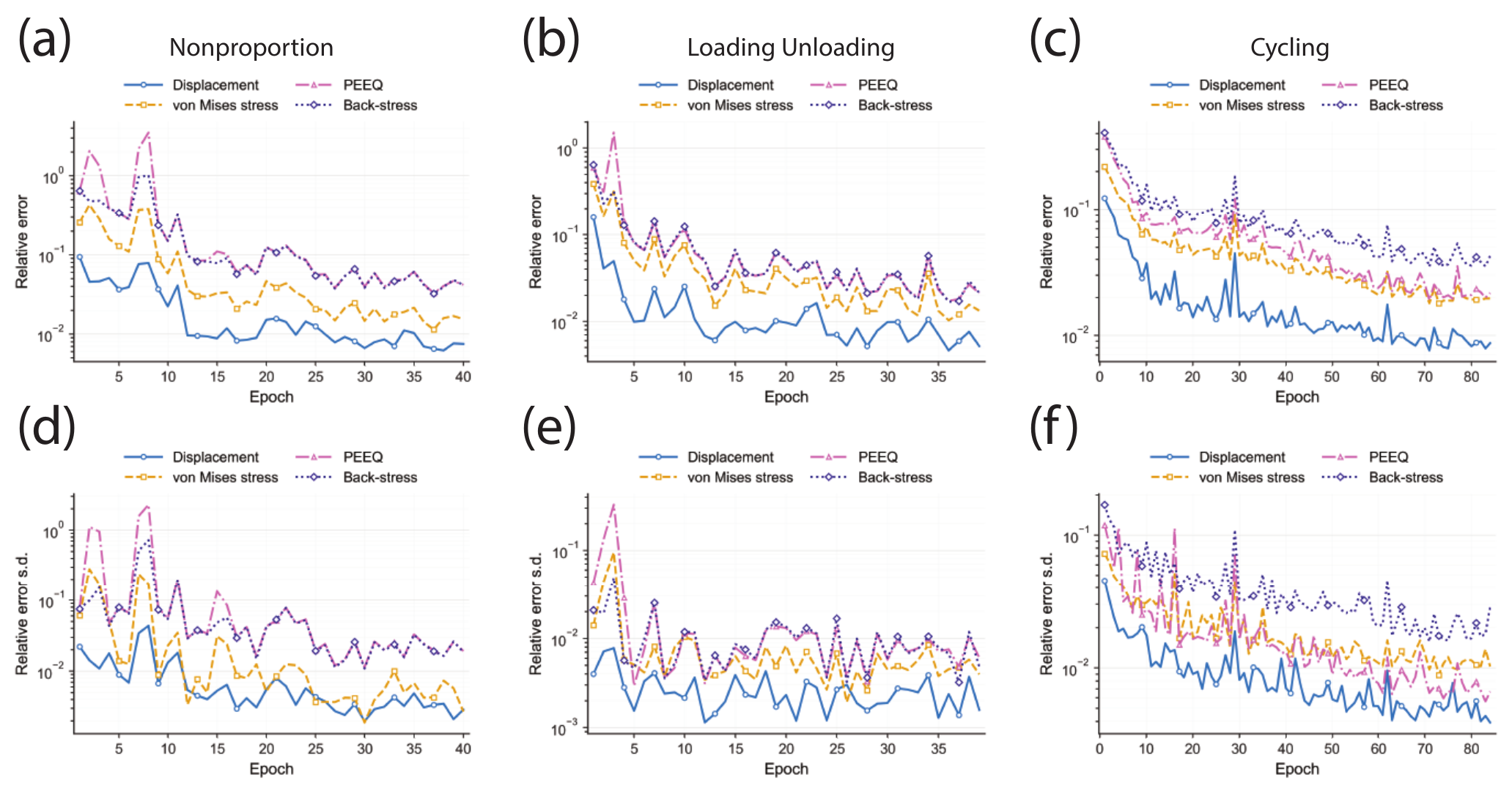}
		\par\end{centering}
	\caption{Evolution of the mean errors and standard deviations of the pretrained Plasolver model under different loading paths. The evaluated quantities include the displacement magnitude, von Mises elastic stress, equivalent plastic strain (PEEQ), and von Mises equivalent backstress. (a) Evolution of the mean relative errors under non-proportional loading. (b) Evolution of the mean relative errors under loading--unloading. (c) Evolution of the mean relative errors under cyclic loading. (d) Evolution of the standard deviations of the relative errors under non-proportional loading. (e) Evolution of the standard deviations under loading--unloading. (f) Evolution of the standard deviations under cyclic loading. The dataset sizes for non-proportional loading, loading--unloading, and cyclic loading are $38.15$ GB, $32.56$ GB, and $35.48$ GB, respectively. Each dataset contains 1000 trajectories, including 900 training trajectories and 100 test trajectories. Each trajectory contains 40 load steps and approximately 10,000 finite element nodes. For each loading condition, one training epoch requires 22 minutes. The reported errors and standard deviations are evaluated on the independent test set.
		\label{fig:Plasovler_error_evolution}}
\end{figure}

The first loading condition is non-proportional loading. First, an endpoint is uniformly sampled from $[0.6,1.0]^{2}$ in the loading space defined by $f_{x}(t)$ and $f_{y}(t)$. Three additional ratios are then sampled from a uniform distribution over $[0.18,0.82]$. Together with the origin $(0,0)$, these samples define five points and four loading segments. Appropriate random perturbations are added to the three intermediate points. Each segment is divided uniformly into ten load steps, resulting in 40 load steps in total, as illustrated in the ``Non-proportional'' column of \Cref{fig:Plasovler_path_intro}b. Under the same perforated-plate geometry and material distribution, 1000 non-proportional loading trajectories are generated, of which 900 are used for training and 100 are reserved for testing. The complete dataset occupies $38.15$ GB. The neural operator is the Transolver described in \Cref{subsec:Neural-Operator}. Details of the network architecture and training parameters are provided in the first row, ``P: NP'', of \Cref{tab:Plasolver_diff_problem}. The results show that the pretrained Plasolver model effectively handles unseen non-proportional loading paths, with relative errors below $5\%$ and stable predictions whose standard deviations are approximately $1\%$. In terms of efficiency, Plasolver is nearly 100 times faster than the reference finite element solver. \Cref{fig:Plasovler_error_evolution} shows the convergence histories of the errors and standard deviations. The errors in the backstress and equivalent plastic strain exhibit similar trends. This is because the loading trajectory of $\boldsymbol{\eta}$ evolves in an approximately consistent direction under the considered non-proportional loading paths. Consequently, the backstress updates in \Cref{eq:update_internal_variables} also occur in approximately the same direction. Additional load--displacement curves under non-proportional loading are presented in \Cref{fig:Plasovler_nonproportional_loading_morepath}. The finite element reference model contains 9977 nodes and 9776 four-node bilinear quadrilateral elements, with $2\times2$ Gauss integration.

\begin{figure}
	\begin{centering}
		\includegraphics[scale=0.47]{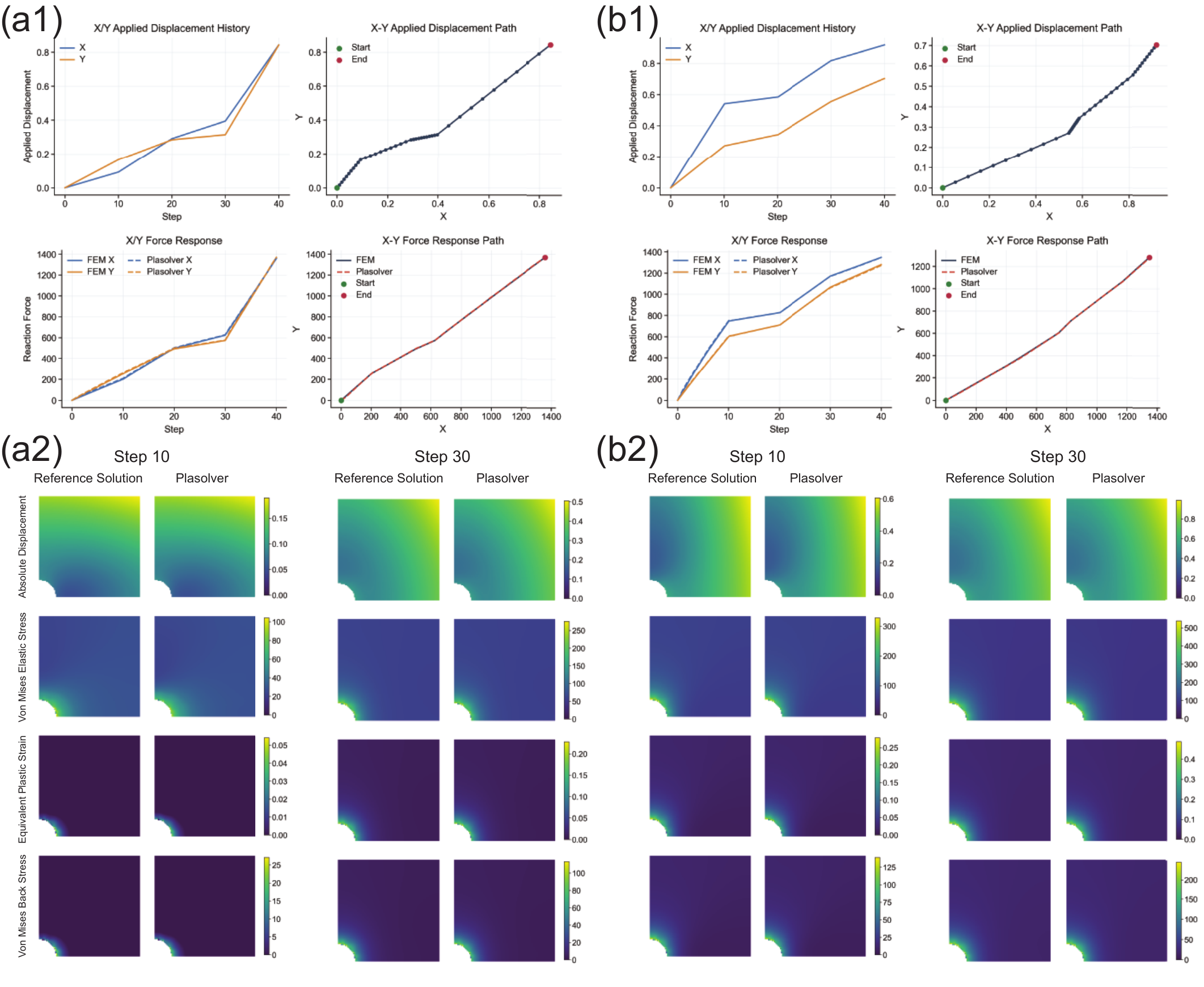}
		\par\end{centering}
	\caption{Performance of the pretrained Plasolver model under non-proportional loading. In (a1) and (b1), the first row shows the prescribed displacement-loading paths, and the second row compares the reaction-force predictions obtained using the pretrained Plasolver model and the finite element reference solution. Panels (a2) and (b2) compare the contour fields predicted by Plasolver and FEM. From the first to the fourth row, the fields are the displacement magnitude, von Mises elastic stress, equivalent plastic strain, and von Mises equivalent backstress. The loading paths in (a1) and (b1) correspond to the contour fields in (a2) and (b2), respectively.
		\label{fig:Plasovler_nonproportional_loading_path}}
\end{figure}

\begin{figure}
	\begin{centering}
		\includegraphics[scale=0.47]{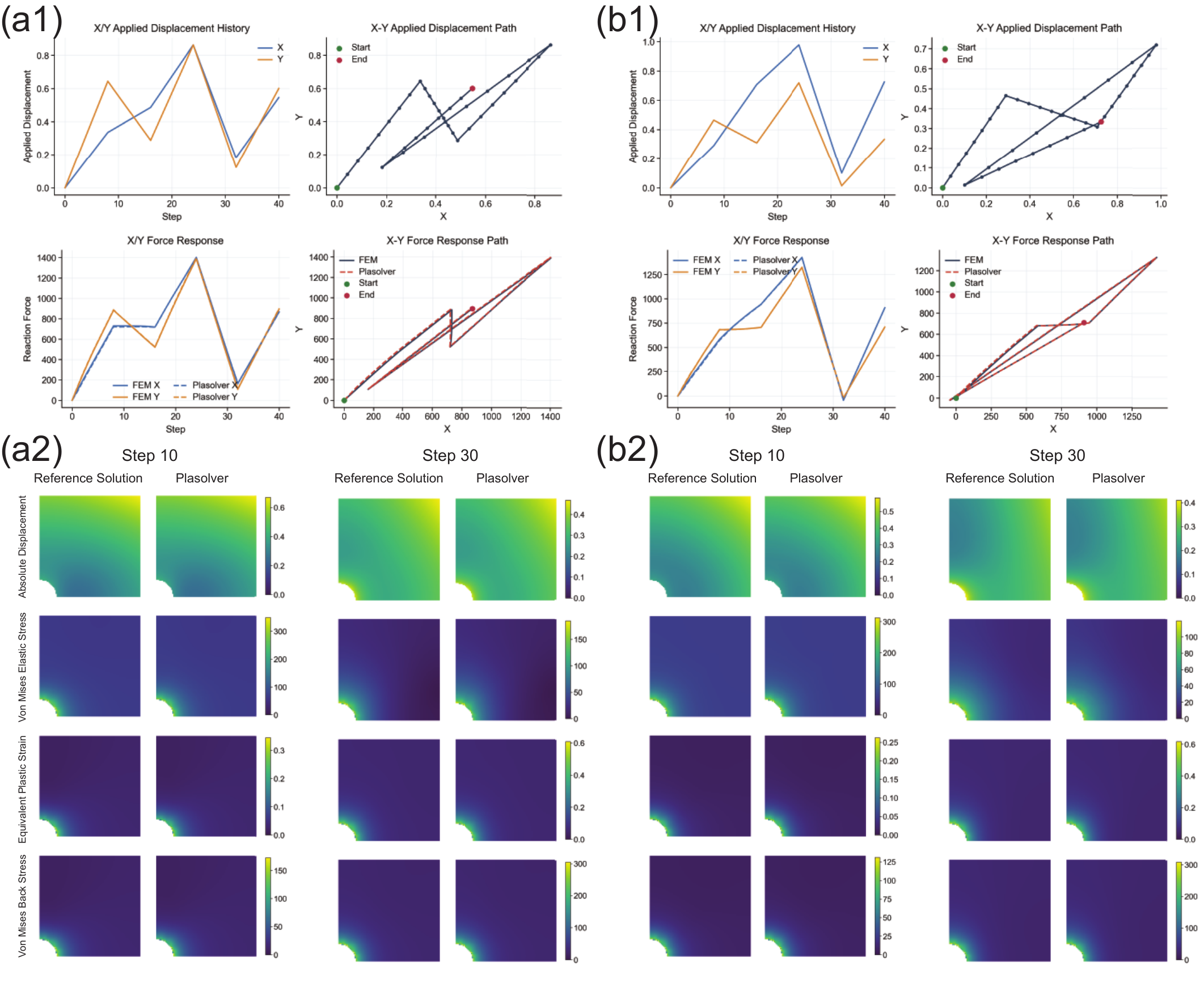}
		\par\end{centering}
	\caption{Performance of the pretrained Plasolver model under loading--unloading paths. In (a1) and (b1), the first row shows the prescribed displacement-loading paths, and the second row compares the reaction-force predictions obtained using Plasolver and the finite element reference solution. Panels (a2) and (b2) compare the contour fields predicted by the pretrained Plasolver model and FEM. From the first to the fourth row, the fields are the displacement magnitude, von Mises elastic stress, equivalent plastic strain, and von Mises equivalent backstress. The loading paths in (a1) and (b1) correspond to the contour fields in (a2) and (b2), respectively.
		\label{fig:Plasovler_loading_unloading_path}}
\end{figure}

The loading path in the first experiment is non-proportional. Loading and unloading are also important in plasticity, and the second experiment therefore considers loading--unloading paths, as illustrated in the ``Loading--Unloading'' column of \Cref{fig:Plasovler_path_intro}b. Six points are randomly specified in the two-dimensional loading space. Further details are provided in the accompanying code and are not repeated here. The loading--unloading dataset contains 1000 trajectories and occupies $32.56$ GB. Among these trajectories, 900 are used for physics-based training and 100 are retained as an independent test set. Although labeled finite element solutions are not required for training, the independent test set is retained to evaluate generalization to unseen loading paths. \Cref{fig:Plasovler_loading_unloading_path}a and b show the performance of the pretrained Plasolver model under loading--unloading paths, demonstrating that Plasolver accurately predicts the responses along unseen loading paths. Additional loading-path predictions are presented in \Cref{fig:Plasovler_loading_unloading_morepath}. Panels (a2) and (b2) of \Cref{fig:Plasovler_loading_unloading_path} compare the displacement magnitude, von Mises elastic stress, equivalent plastic strain, and von Mises equivalent backstress predicted by Plasolver and FEM. The contour fields predicted by the pretrained Plasolver model are almost identical to the finite element results. The corresponding relative errors are listed in \Cref{tab:Plasolver_diff_problem}, as exemplified by the second experiment, “P: UL”. The errors in the backstress and equivalent plastic strain exhibit similar trends because the material remains elastic during most of the unloading process, for which $\dot{\gamma}=0$. When loading resumes, the backstress evolves in an approximately consistent direction. \Cref{fig:Plasovler_error_evolution} shows that the pretrained Plasolver model converges after approximately 20 epochs.

\begin{figure}
	\begin{centering}
		\includegraphics[scale=0.47]{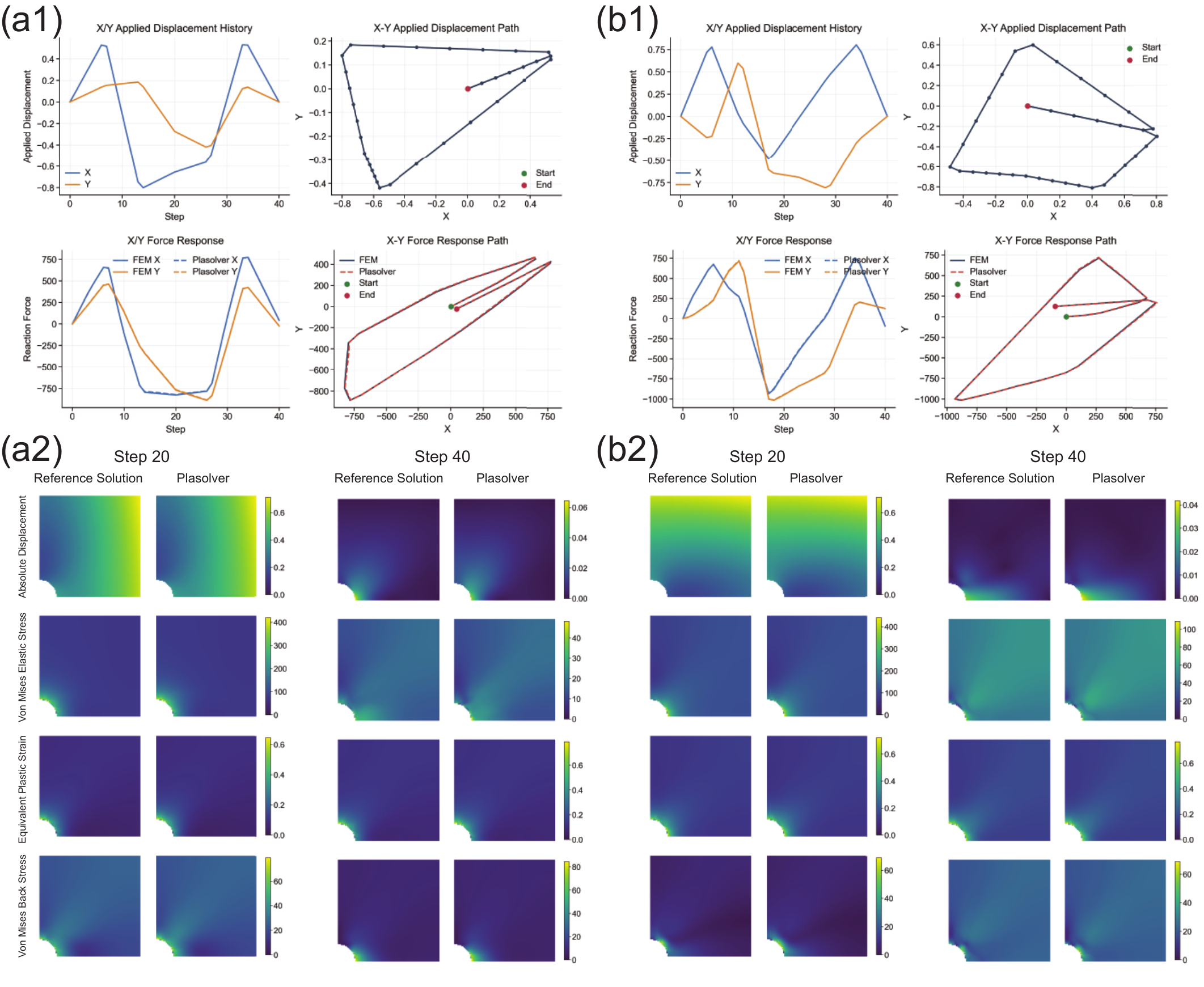}
		\par\end{centering}
	\caption{Performance of the pretrained Plasolver model under cyclic loading. In (a1) and (b1), the first row shows the prescribed displacement-loading paths, and the second row compares the reaction-force predictions obtained using the pretrained Plasolver model and the finite element reference solution. Panels (a2) and (b2) compare the contour fields predicted by the pretrained Plasolver model and FEM. From the first to the fourth row, the fields are the displacement magnitude, von Mises elastic stress, equivalent plastic strain, and von Mises equivalent backstress. The loading paths in (a1) and (b1) correspond to the contour fields in (a2) and (b2), respectively.
		\label{fig:Plasovler_cycling_loading_path}}
\end{figure}

The third loading condition is cyclic loading, as illustrated in the ``Cyclic'' column of \Cref{fig:Plasovler_path_intro}b. Cyclic loading is important for plastic-shakedown problems and is common in engineering applications. A total of 1000 cyclic loading trajectories are randomly generated, producing a dataset of $35.48$ GB. Among these trajectories, 900 are used for training and 100 are retained for testing. \Cref{fig:Plasovler_cycling_loading_path}a and b show the performance of the pretrained Plasolver model under cyclic loading, demonstrating that Plasolver accurately predicts unseen loading-path responses. Additional predictions are presented in \Cref{fig:Plasovler_cycling_loading_morepath}. \Cref{fig:Plasovler_error_evolution} shows that the pretrained Plasolver model converges after approximately 35 epochs. Unlike non-proportional loading and loading--unloading, cyclic loading involves substantial changes in the loading direction. Consequently, the convergence histories of the equivalent plastic strain and backstress errors differ considerably. The prescribed displacements at both the beginning and end of each training trajectory are zero, meaning that unloading is also included. Cyclic loading is therefore more challenging than the preceding non-proportional and loading--unloading cases. Consequently, the pretrained Plasolver model requires more training epochs and exhibits larger relative errors. Nevertheless, the relative displacement error remains below $5\%$, as shown in \Cref{tab:Plasolver_diff_problem}, as exemplified by the second experiment, “P: CY”. Panels (a2) and (b2) of \Cref{fig:Plasovler_cycling_loading_path} compare the displacement magnitude, von Mises elastic stress, equivalent plastic strain, and von Mises equivalent backstress predicted by the pretrained Plasolver model and FEM. The contour fields predicted by Plasolver are almost identical to the finite element results. The contours at the final load step, namely the 40th step, are presented because the structure has undergone substantial unloading by this stage.

The preceding experiments use the same geometry and material properties while varying the loading path. Three common loading conditions are considered: non-proportional loading, loading--unloading, and cyclic loading. The results in \Cref{tab:Plasolver_diff_problem} and \Cref{fig:Plasovler_error_evolution} demonstrate that the pretrained Plasolver model can train a neural operator through unsupervised learning based exclusively on physical equations and can successfully generalize to unseen loading paths.

\begin{figure}
	\begin{centering}
		\includegraphics[scale=0.45]{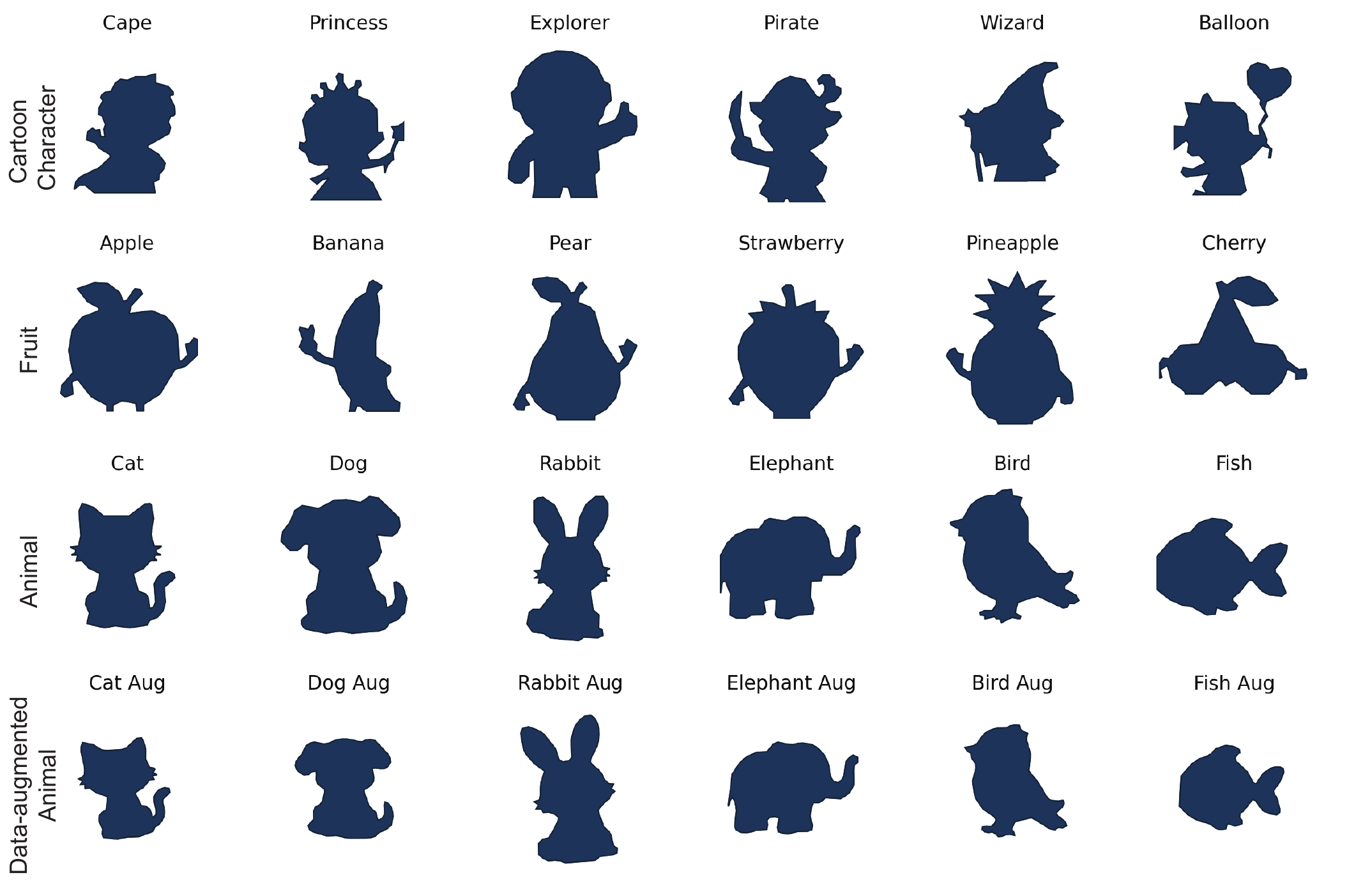}
		\par\end{centering}
	\caption{Examples of complex loading paths. A total of 1,000 distinct loading paths are generated from the outlines of cartoon images. The first, second, and third rows show loading paths based on cartoon characters, fruits, and animals, respectively. The fourth row illustrates data augmentation for the animal category, where the dataset is expanded through rotation and scaling. Similar data-augmentation operations are also applied to the cartoon-character and fruit categories.
		\label{fig:cartoon_path}}
\end{figure}

\begin{figure}
	\begin{centering}
		\includegraphics[scale=0.45]{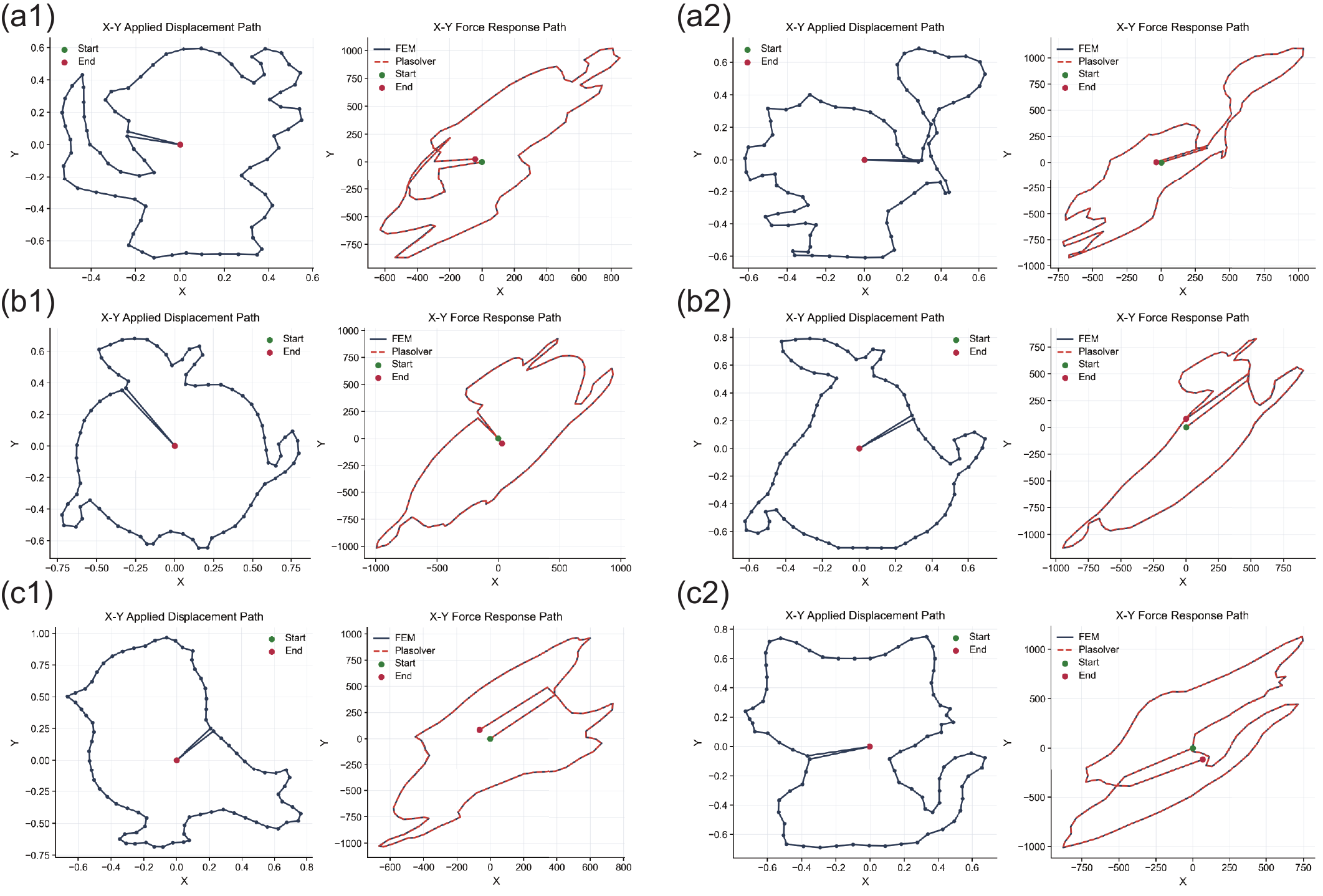}
		\par\end{centering}
	\caption{Performance of the pretrained Plasolver under cartoon-shaped loading paths. In each panel, the left subfigure shows the prescribed displacement loading path, while the right subfigure compares the contour fields predicted by the pretrained Plasolver with the FEM results. Panels (a1) and (a2) correspond to cartoon-character loading paths, namely a pirate captain and a wizard holding a balloon, respectively. Panels (b1) and (b2) correspond to fruit-shaped loading paths, namely an apple and a pear. Panels (c1) and (c2) correspond to animal-shaped loading paths, namely a bird and a cat.
		\label{fig:Plasovler_cartoon_path}}
\end{figure}

Next, we further increase the difficulty by considering highly complex loading paths. Specifically, we use the outlines of cartoon characters, fruits, and animals to define a total of 1,000 distinct loading paths, as illustrated in \Cref{fig:cartoon_path}.
The complex loading dataset contains 1000 trajectories and occupies $72.36$ GB.
Among these, 900 are used for training and the remaining 100 for testing. To adequately represent these complex paths, we increase the number of loading steps from 40 to 80. \Cref{fig:Plasovler_cartoon_path} demonstrates that Plasolver can readily generalize to these highly complex and previously unseen loading paths. The corresponding errors are reported in the row labeled ``P: Cartoon'' in \Cref{tab:Plasolver_diff_problem}.

\begin{figure}
	\begin{centering}
		\includegraphics[scale=0.47]{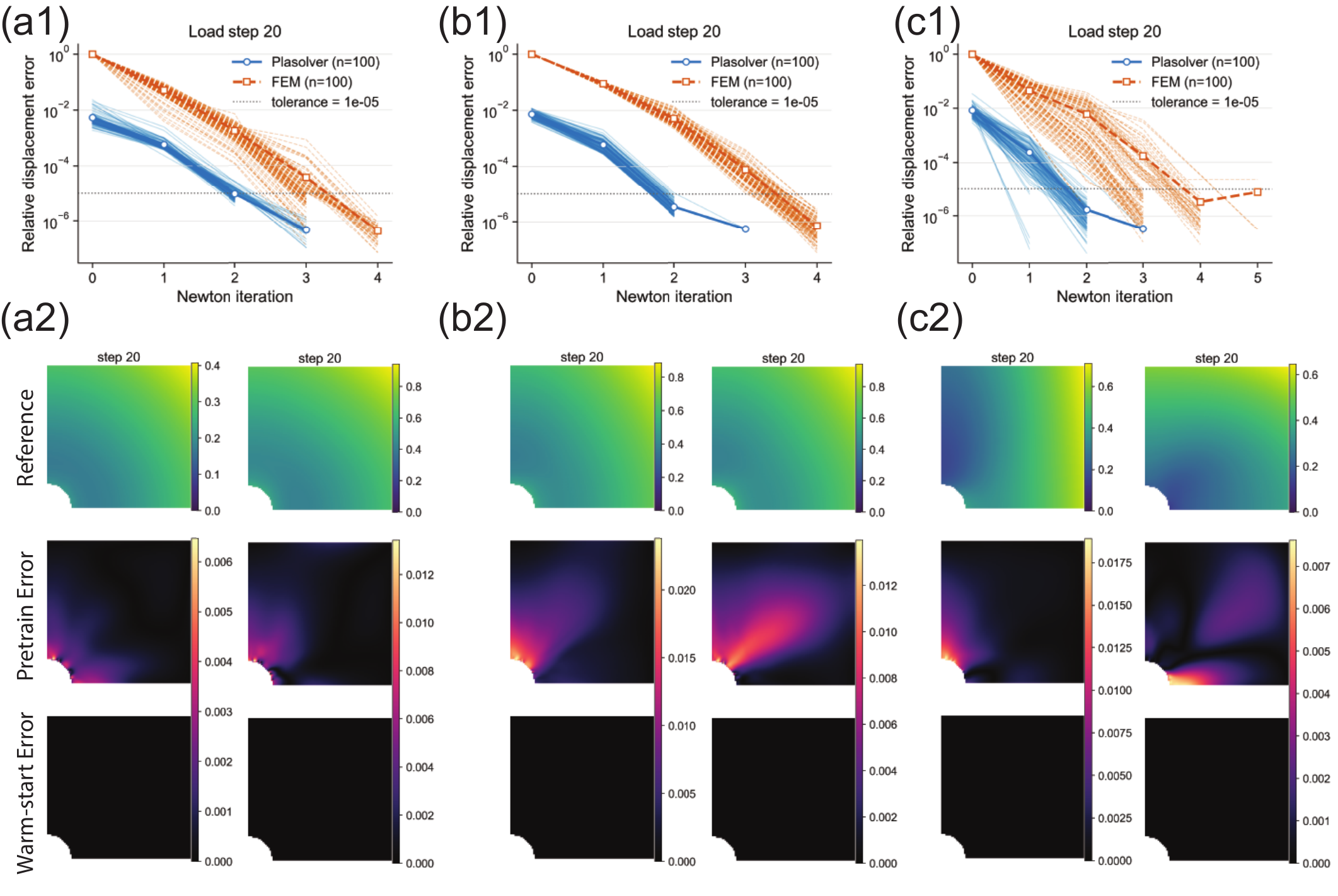}
		\par\end{centering}
	\caption{Performance of the Plasolver warm-start phase under different loading paths. Panels (a1), (b1), and (c1) show the iterative convergence histories at the 20th load step under non-proportional loading, loading--unloading, and cyclic loading, respectively. The vertical axis represents the relative displacement error, and the horizontal axis represents the number of Newton iterations. The statistics are calculated over 100 test trajectories, and the convergence threshold for the warm-start relative displacement error is $0.00001$. Panels (a2), (b2), and (c2) show the reference solutions, the absolute-error fields produced by the pretrained Plasolver model, and the absolute-error fields after warm-start refinement at the 20th step for non-proportional loading, loading--unloading, and cyclic loading, respectively.
		\label{fig:Plasolver_warm_start_path}}
\end{figure}

Although the displacement error of the pretrained Plasolver model is already small, at approximately $1\%$, the accuracy can be further improved using the Plasolver warm-start phase when required. The warm-start-refined solution can approach the finite element reference solution because the prediction from the pretraining phase is used as the initial solution of the iterative solver. \Cref{fig:Plasolver_warm_start_path} demonstrates the performance of the Plasolver warm-start phase. Using the Plasolver prediction as the initial solution substantially reduces the number of nonlinear Newton iterations. The corresponding quantitative iteration speedup ratios are reported in \Cref{tab:Plasolver_diff_problem}.

\begin{figure}
	\begin{centering}
		\includegraphics[scale=0.48]{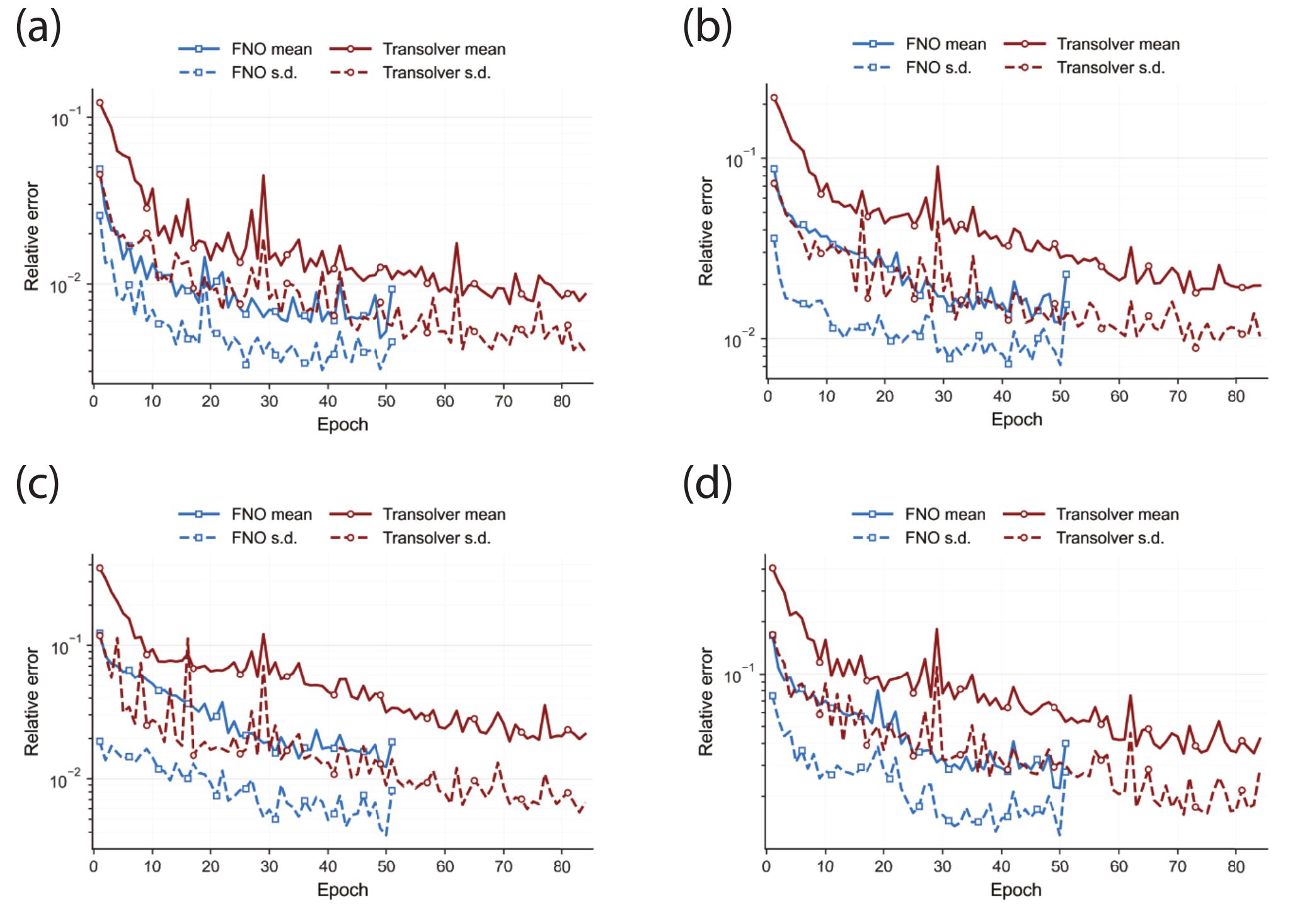}
		\par\end{centering}
	\caption{Comparison of the Transolver and FNO neural operators during the Plasolver pretraining phase. The evolution of the relative errors and their standard deviations over the training epochs is shown for (a) the displacement magnitude, (b) the von Mises elastic stress, (c) the equivalent plastic strain, and (d) the von Mises equivalent backstress.
		\label{fig:Plasolver_fno_transolver}}
\end{figure}

Because the geometry and material properties are fixed in these experiments and only the loading paths are varied, the Transolver neural operator can be replaced by the well-known FNO \citet{li2020fourier}. \Cref{fig:Plasolver_fno_transolver} compares FNO and Transolver during the Plasolver pretraining phase by showing the evolution of their mean relative errors and standard deviations over the training epochs. FNO converges faster than Transolver. After convergence, the two architectures achieve almost identical accuracy, although FNO is slightly more accurate. FNO and Transolver contain 758,606 and 747,314 parameters, respectively, and therefore have similar model sizes. The training times per epoch for FNO and Transolver are 6 and 21 minutes, respectively. Thus, FNO provides a clear advantage for this example, particularly in terms of training efficiency.

However, FNO requires inputs on a regular grid. Therefore, compared with the Transolver input, the FNO input contains an additional mask channel indicating whether each grid point lies within the physical domain. Transolver does not have this limitation and requires only a point cloud sampled from the physical structure, providing substantially greater flexibility for complex geometries. Consequently, only the Transolver architecture is employed as the neural operator in the Plasolver pretraining experiments involving complex geometries in \Cref{subsec:geo_path}.

In summary, this subsection demonstrates that the pretrained Plasolver model can train a neural operator through unsupervised learning based exclusively on physical equations and can successfully generalize to different loading paths. The Plasolver warm-start phase substantially reduces the number of Newton iterations while recovering the accuracy of the finite element solution. Plasolver therefore exhibits remarkably strong performance for complex elastoplastic problems. One reason is that elastoplastic problems reduce to elastic problems under many loading states, as discussed in \Cref{subsec:DEM_optimization}. Consequently, energy optimization for elastoplasticity shares important characteristics with that for linear elasticity.

\subsection{Different geometries and loading paths\label{subsec:geo_path}}

\begin{figure}
	\begin{centering}
		\includegraphics[scale=0.60]{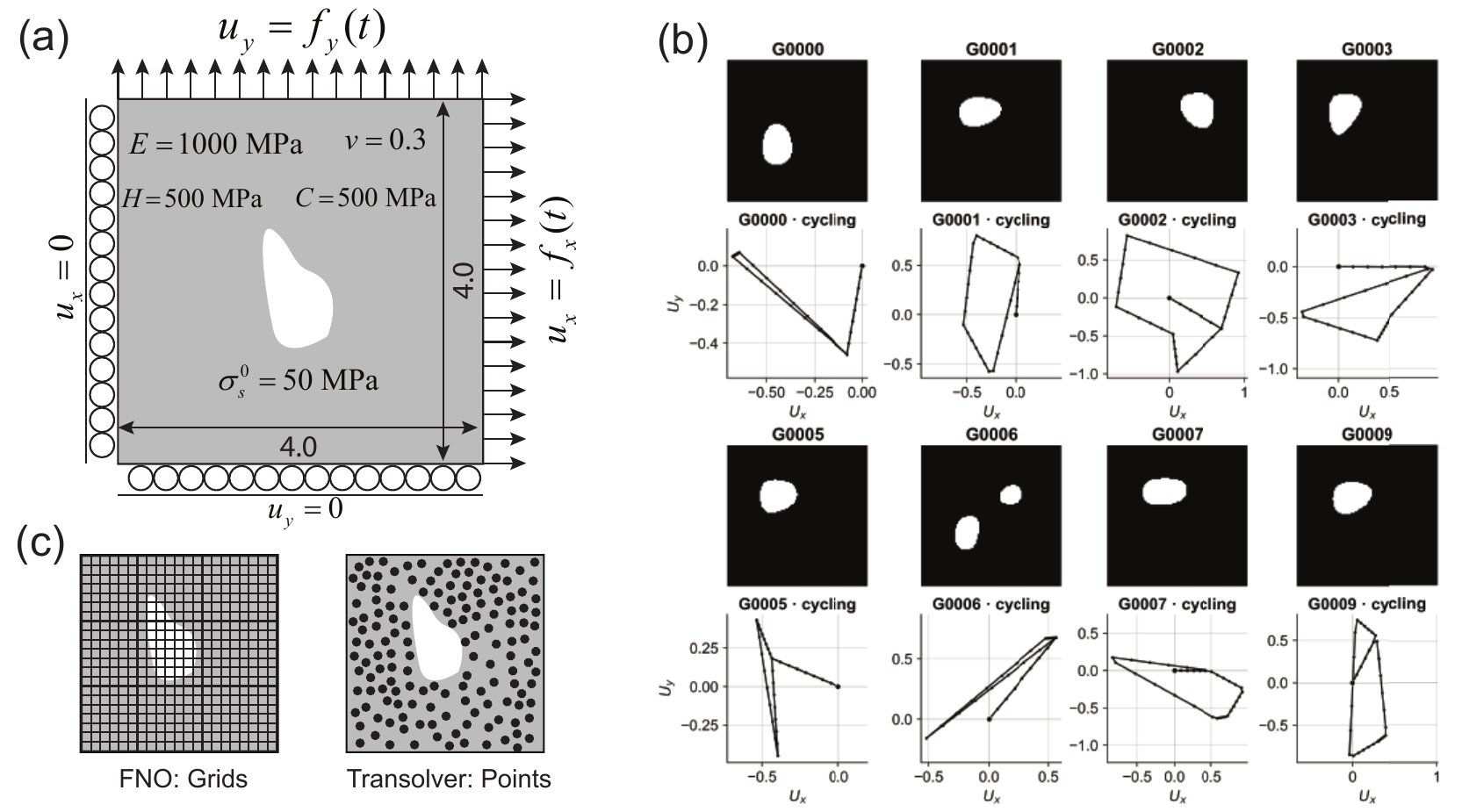}
		\par\end{centering}
	\caption{Elastoplastic problems involving different geometries and loading paths. (a) Geometry, material properties, and boundary conditions of the elastoplastic problem. The geometry is a square plate with a length and width of $4$ mm and a hole of arbitrary shape. The material parameters are the elastic modulus $E=1000$ MPa, Poisson's ratio $\nu=0.3$, plastic modulus $H=500$ MPa, and kinematic hardening modulus $C=500$ MPa. The initial yield stress is $\sigma^{0}_{y}=500$ MPa. The displacement in the $x$ direction is constrained on the left boundary, and the displacement in the $y$ direction is constrained on the bottom boundary. Displacement loading $u_{x}=f_{x}(t)$ is applied on the right boundary, and displacement loading $u_{y}=f_{y}(t)$ is applied on the top boundary. Both $f_{x}(t)$ and $f_{y}(t)$ are spatially uniform and follow cyclic loading trajectories. (b) Examples of different hole shapes and loading paths for the perforated square plates. The loading paths are represented by trajectories in the $u_{x}$--$u_{y}$ loading space. (c) FNO uses inputs defined on a regular grid, whereas Transolver uses point-based inputs.
		\label{fig:geo_intro}}
\end{figure}

The previous experiments in \Cref{subsec:path} considered only different loading paths. In practical simulations, however, the geometry, material properties, and loading paths may all vary. This subsection investigates whether Plasolver can generalize across different geometries and loading paths. \Cref{fig:geo_intro} presents the problems involving different geometries and loading paths. For the geometry, we consider square plates containing holes with arbitrarily varying shapes generated using Gaussian random fields (GRFs). Cyclic loading is used for the loading paths. The dataset contains 1000 instances comprising 50 different geometries and 20 cyclic loading paths for each geometry. The spatial resolution is 100, and the complete dataset occupies $37.22$ GB. In principle, the training and test sets need not be distinguished for physics-based Plasolver training because no labeled data are used and training relies entirely on the governing physics, namely Simo's incremental potential. However, because this subsection compares data-driven and physics-based training, the dataset is divided consistently into 900 training instances and 100 test instances.

\begin{figure}
	\begin{centering}
		\includegraphics[scale=0.47]{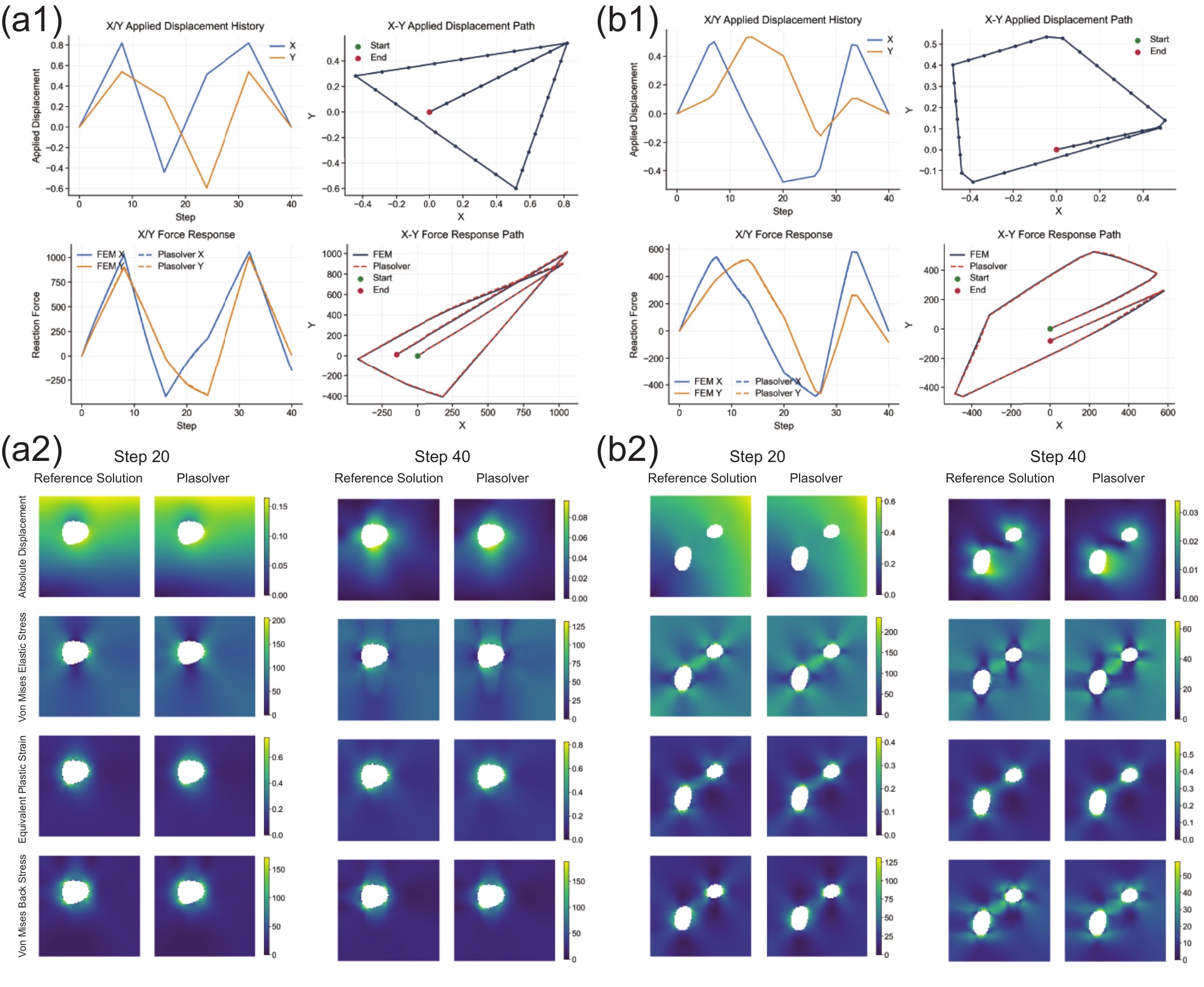}
		\par\end{centering}
	\caption{Performance of the pretrained Plasolver model for different geometries and loading paths. In (a1) and (b1), the first row shows the prescribed displacement-loading paths, and the second row compares the reaction-force predictions obtained using the pretrained Plasolver model and the finite element reference solution. Panels (a2) and (b2) compare the contour fields predicted by Plasolver and FEM. From the first to the fourth row, the fields are the displacement magnitude, von Mises elastic stress, equivalent plastic strain, and von Mises equivalent backstress. The loading paths in (a1) and (b1) correspond to the contour fields in (a2) and (b2), respectively.
		\label{fig:Plasovler_cycling_loading_geo}}
\end{figure}

\Cref{fig:Plasovler_cycling_loading_geo} shows the performance of the pretrained Plasolver model under different geometries and loading paths. The training details are provided in “G+P:CY” of \Cref{tab:Plasolver_diff_problem}. \Cref{fig:Plasovler_geo_cycling_loading_morepath} presents predictions for additional loading paths, and \Cref{fig:Plasolver_geo_more_contourf} presents additional displacement-field predictions obtained using the pretrained Plasolver model. The quantitative results are reported in \Cref{tab:Plasolver_diff_problem}. These results demonstrate that the pretrained Plasolver model generalizes effectively across different geometries and loading paths.

\begin{figure}
	\begin{centering}
		\includegraphics[scale=0.47]{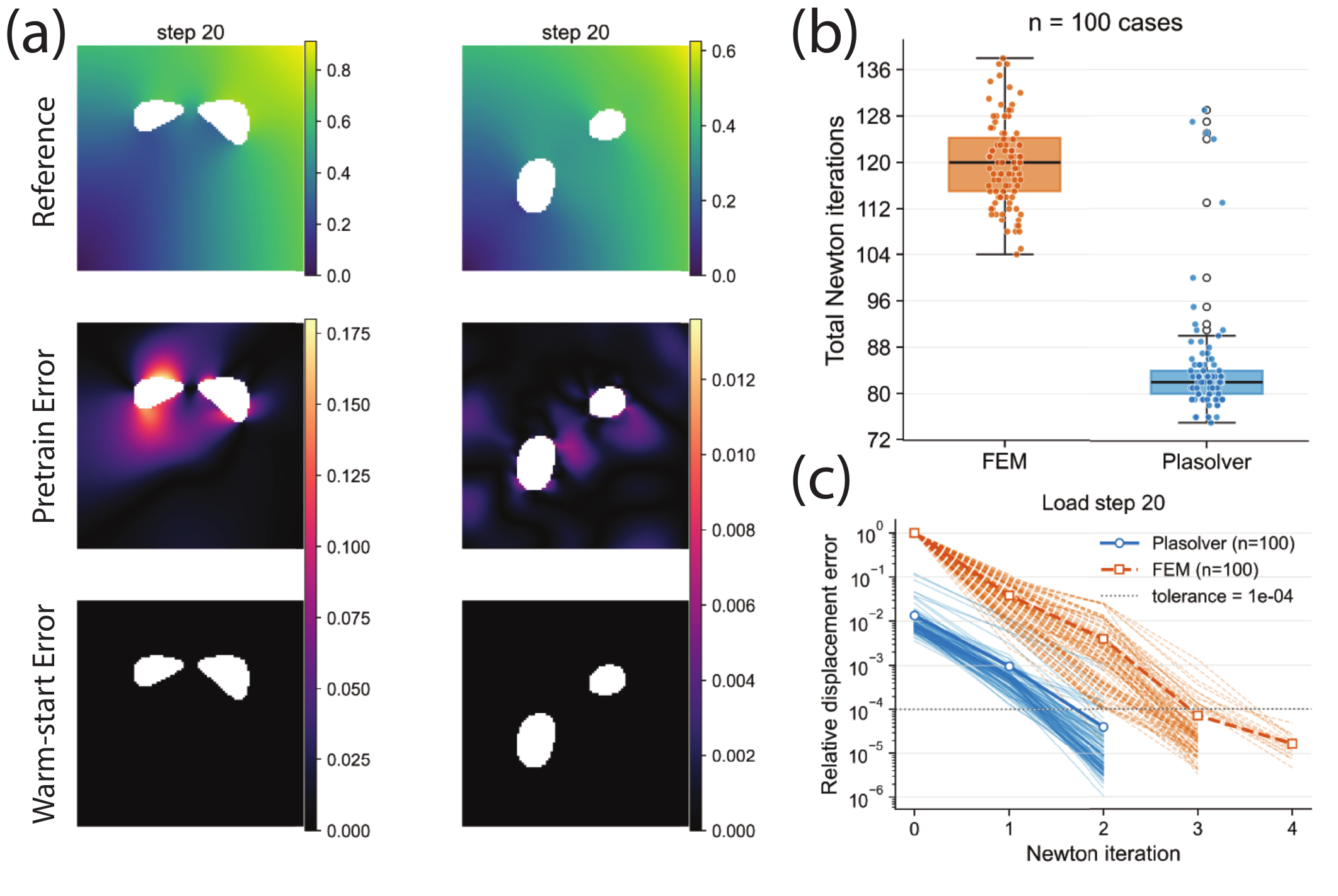}
		\par\end{centering}
	\caption{Performance of the Plasolver warm-start phase for different geometries and loading paths. (a) Finite element reference solution at the 20th load step, absolute-error field produced by the pretrained Plasolver model, and absolute-error field after warm-start refinement. (b) Mean numbers of Newton iterations over all load steps. The convergence threshold for the warm-start relative displacement error is $0.0001$. (c) Iterative convergence curves at the 20th load step. The vertical axis represents the relative displacement error, and the horizontal axis represents the number of Newton iterations. The statistics are calculated over 100 test instances.
		\label{fig:Plasolver_warm_start_geo}}
\end{figure}

Although the error of the pretrained Plasolver model is already small, at approximately $1\%$, the solution accuracy can be further improved using the Plasolver warm-start phase when required. The warm-start-refined solution can approach the finite element reference solution because the prediction from the pretraining phase is used as the initial solution of the iterative solver. \Cref{fig:Plasolver_warm_start_geo} demonstrates the performance of the Plasolver warm-start phase. Using the Plasolver prediction as the initial solution substantially reduces the number of nonlinear Newton iterations. The corresponding quantitative iteration speedup ratio is reported in “G+P:CY” of \Cref{tab:Plasolver_diff_problem}.

\begin{figure}
	\begin{centering}
		\includegraphics[scale=0.47]{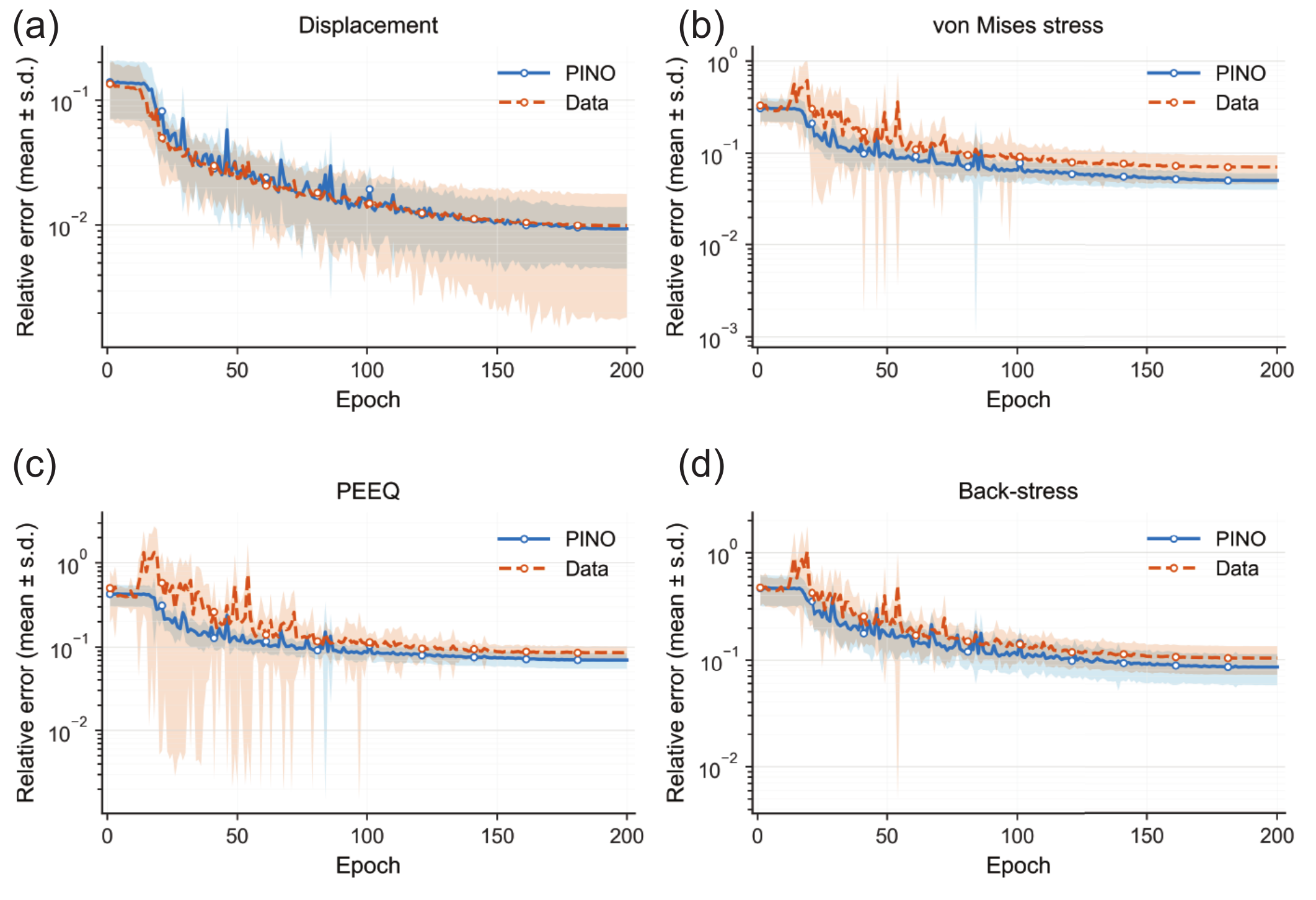}
		\par\end{centering}
	\caption{Comparison between data-driven and physics-based training of the pretrained Plasolver model for different geometries and loading paths. The dark curves represent the mean values, and the shaded regions represent the ranges defined by the standard deviations. The evolution of the relative errors and standard deviations over the training epochs is shown for (a) the displacement field, (b) the von Mises elastic stress, (c) the equivalent plastic strain, and (d) the von Mises equivalent backstress.
		\label{fig:Plasolver_data_physics}}
\end{figure}

The current Plasolver model is pretrained using only the governing physical equations, represented by Simo's incremental potential. Plasolver can also be trained using a data-driven approach. However, data-driven training first requires a conventional solver, such as FEM, to generate a large dataset. We next compare physics-based and data-driven training. To ensure a consistent comparison, the same dataset of 1000 instances is divided into 900 training instances and 100 test instances, and all other training settings are identical. \Cref{fig:Plasolver_data_physics} compares the data-driven and physics-based approaches. The physics-based approach exhibits almost the same convergence trend as the data-driven approach, and the two approaches achieve comparable accuracy for the displacement field. However, the physics-based approach is more accurate for derivative-related physical fields, including the von Mises stress, equivalent plastic strain, and backstress. This is because derivative information is incorporated during physics-based training, whereas it is not explicitly considered during data-driven training. The data-driven loss function measures only the approximation error of the displacement field. Consequently, data-driven training does not directly constrain the corresponding derivative information.

It should be emphasized that each epoch of data-driven training requires 26 minutes, whereas each epoch of physics-based training requires 22 minutes. The high efficiency of physics-based training is achieved by evaluating the required derivatives explicitly using shape functions, as discussed in detail in \Cref{subsec:Shape_function_not_need_distance}. In addition, because the dataset is extremely large, repeated data transfer during data-driven training introduces non-negligible computational overhead. Physics-based training does not require this interaction with finite element data and instead evaluates only the physics-based loss. Generating each finite element instance requires approximately 37 seconds, as shown in “G+P:CY” of \Cref{tab:Plasolver_diff_problem}. Consequently, generating 1000 instances requires more than 10 hours. Considering the additional cost of data generation, the longer training time per epoch, and the nearly identical convergence curves of the two approaches, physics-based training provides an overwhelming advantage for problems whose physical processes are well understood.

\begin{figure}
	\begin{centering}
		\includegraphics[scale=0.47]{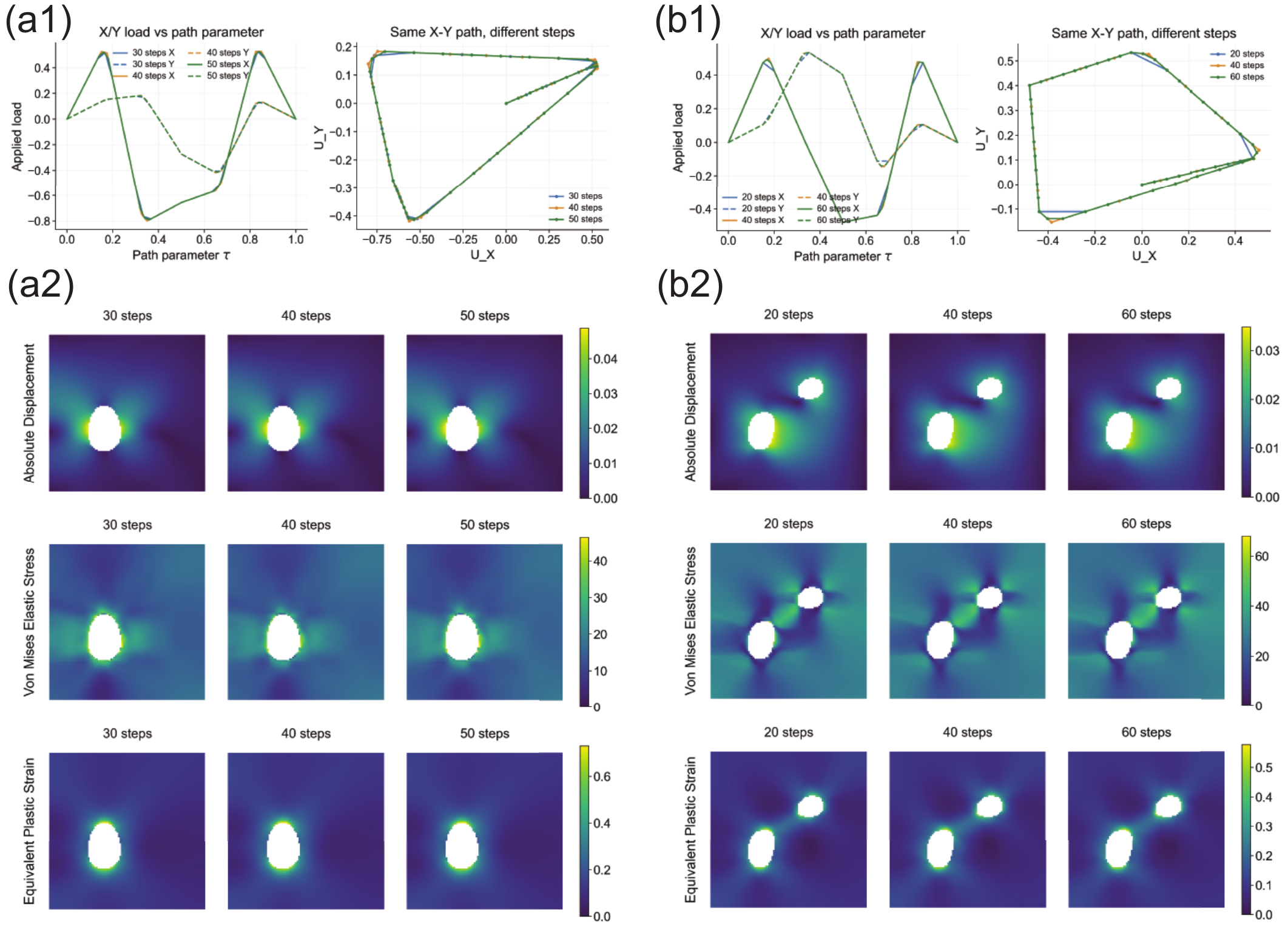}
		\par\end{centering}
	\caption{Path-discretization invariance of the pretrained Plasolver model, namely its robustness for the same loading path represented using different numbers of load steps. (a1) Reaction-force predictions obtained using Plasolver with 30, 40, and 50 load steps. (b1) Reaction-force predictions obtained using Plasolver with 20, 40, and 60 load steps. (a2) Predictions of the displacement magnitude, von Mises elastic stress, and equivalent plastic strain at the final state of the same loading path discretized using 30, 40, and 50 load steps. (b2) Predictions of the displacement magnitude, von Mises elastic stress, and equivalent plastic strain at the final state of the same loading path discretized using 20, 40, and 60 load steps.
		\label{fig:Plasolver_path_discretization_invariance}}
\end{figure}

It must be emphasized that plasticity problems are strongly dependent on their loading paths. In practical numerical simulations, the same loading path may be discretized using different numbers of load steps. For the same loading path with different load-step discretizations, the pretrained Plasolver model is expected to provide robust predictions. We therefore evaluate the performance of the pretrained Plasolver model for the same loading path discretized using different numbers of load steps. \Cref{fig:Plasolver_path_discretization_invariance}(a1) and (a2) show the performance of the pretrained Plasolver model using 30, 40, and 50 load steps. \Cref{fig:Plasolver_path_discretization_invariance}(b1) and (b2) show its performance using 20, 40, and 60 load steps. It should be noted that Plasolver is trained exclusively using trajectories containing 40 load steps. Nevertheless, the results in \Cref{fig:Plasolver_path_discretization_invariance} demonstrate that the pretrained Plasolver model possesses path-discretization invariance. In other words, it provides robust predictions even when the number of load steps is changed. Further discussion of path-discretization invariance is provided in \Cref{subsec:space_path_discretization}.

In summary, this subsection demonstrates that the pretrained Plasolver model can train a neural operator through unsupervised learning based exclusively on physical equations and can successfully generalize across different geometries and loading paths. The Plasolver warm-start phase also substantially reduces the number of Newton iterations while recovering the accuracy of the finite element solution.

\subsection{Different materials, geometries, and loading paths}

\Cref{subsec:geo_path} demonstrated the strong generalization capability of the pretrained Plasolver model across different geometries and loading paths. A natural question is whether Plasolver can generalize over all information defining an elastoplastic problem. In this subsection, we evaluate the comprehensive generalization capability of the pretrained Plasolver model across different materials, geometries, and loading paths.

\begin{figure}
	\begin{centering}
		\includegraphics[scale=0.60]{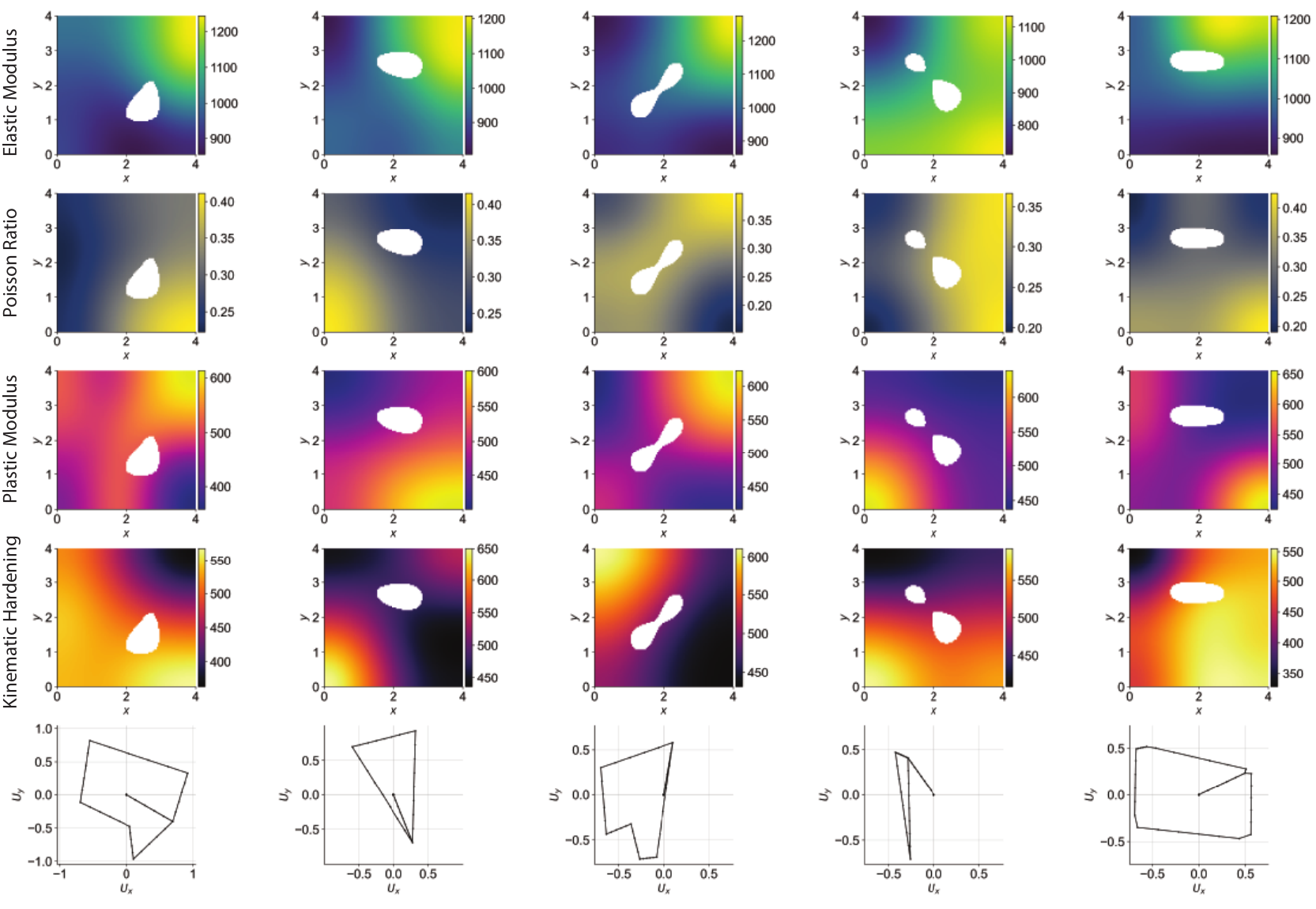}
		\par\end{centering}
	\caption{Elastoplastic problems involving different materials, geometries, and loading paths. The geometry is a square plate with a length and width of $4$ mm and a hole of arbitrary shape. From the first to the fourth row, the fields represent the elastic modulus, Poisson's ratio, plastic modulus, and kinematic hardening modulus, respectively. The initial yield stress is $\sigma^{0}_{y}=500$ MPa. The fifth row shows the loading paths. The displacement in the $x$ direction is constrained on the left boundary, and the displacement in the $y$ direction is constrained on the bottom boundary. Displacement loading $u_{x}=f_{x}(t)$ is applied on the right boundary, and displacement loading $u_{y}=f_{y}(t)$ is applied on the top boundary. Both $f_{x}(t)$ and $f_{y}(t)$ are spatially uniform and follow cyclic loading trajectories.
		\label{fig:material_intro}}
\end{figure}

\Cref{fig:material_intro} presents the problems involving different materials, geometries, and loading paths. The spatial distributions of the elastic modulus, Poisson's ratio, plastic modulus, and kinematic hardening modulus are generated using Gaussian random fields (GRFs). The loading paths adopt cyclic loading, which is the most complex loading condition considered in this work. A total of 1000 problem instances are generated, of which 900 are used for training and 100 are reserved for testing. The complete dataset occupies $37.49$ GB.

\begin{figure}
	\begin{centering}
		\includegraphics[scale=0.47]{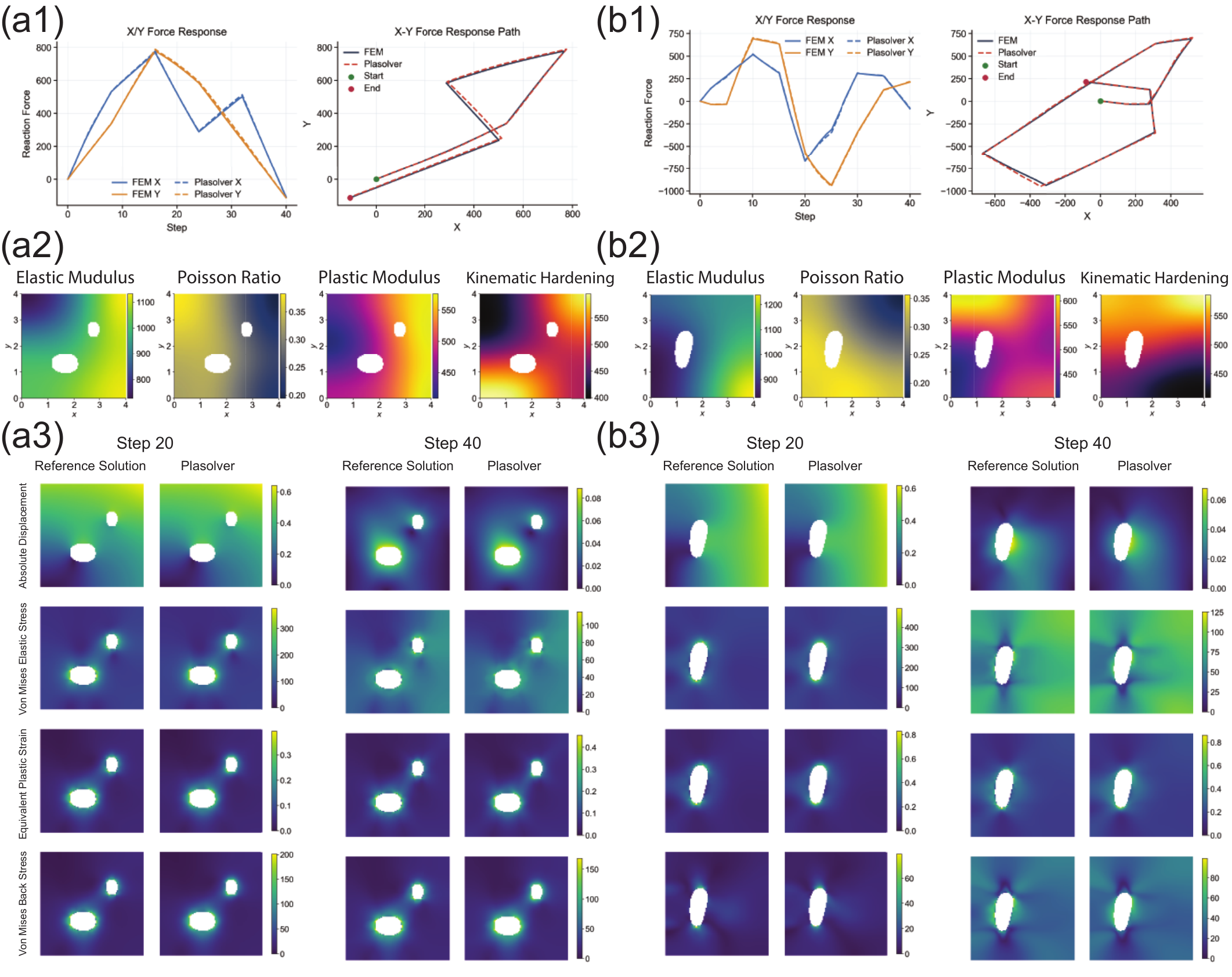}
		\par\end{centering}
	\caption{Performance of the pretrained Plasolver model for elastoplastic problems involving different materials, geometries, and loading paths. Panels (a1) and (b1) compare the reaction-force predictions obtained using Plasolver and the finite element reference solution. Panels (a2) and (b2) show the spatial distributions of the elastic modulus, Poisson's ratio, plastic modulus, and kinematic hardening modulus. Panels (a3) and (b3) compare the contour fields predicted by Plasolver and FEM. From the first to the fourth row, the fields are the displacement magnitude, von Mises elastic stress, equivalent plastic strain, and von Mises equivalent backstress. The loading path in (a1) corresponds to the material distributions in (a2) and the contour predictions in (a3). The loading path in (b1) corresponds to the material distributions in (b2) and the contour predictions in (b3).
		\label{fig:Plasovler_cycling_loading_material}}
\end{figure}

\Cref{fig:Plasovler_cycling_loading_material} demonstrates the generalization capability of the pretrained Plasolver model, and the corresponding quantitative results are reported in “M+G+P: L” of \Cref{tab:Plasolver_diff_problem}. \Cref{fig:Plasovler_material_cycling_loading_morepath} and \ref{fig:Plasolver_material_more_contourf} present additional results obtained using the pretrained Plasolver model. These results demonstrate that the pretrained Plasolver model can generalize effectively and simultaneously across different materials, geometries, and loading paths.

\begin{figure}
	\begin{centering}
		\includegraphics[scale=0.47]{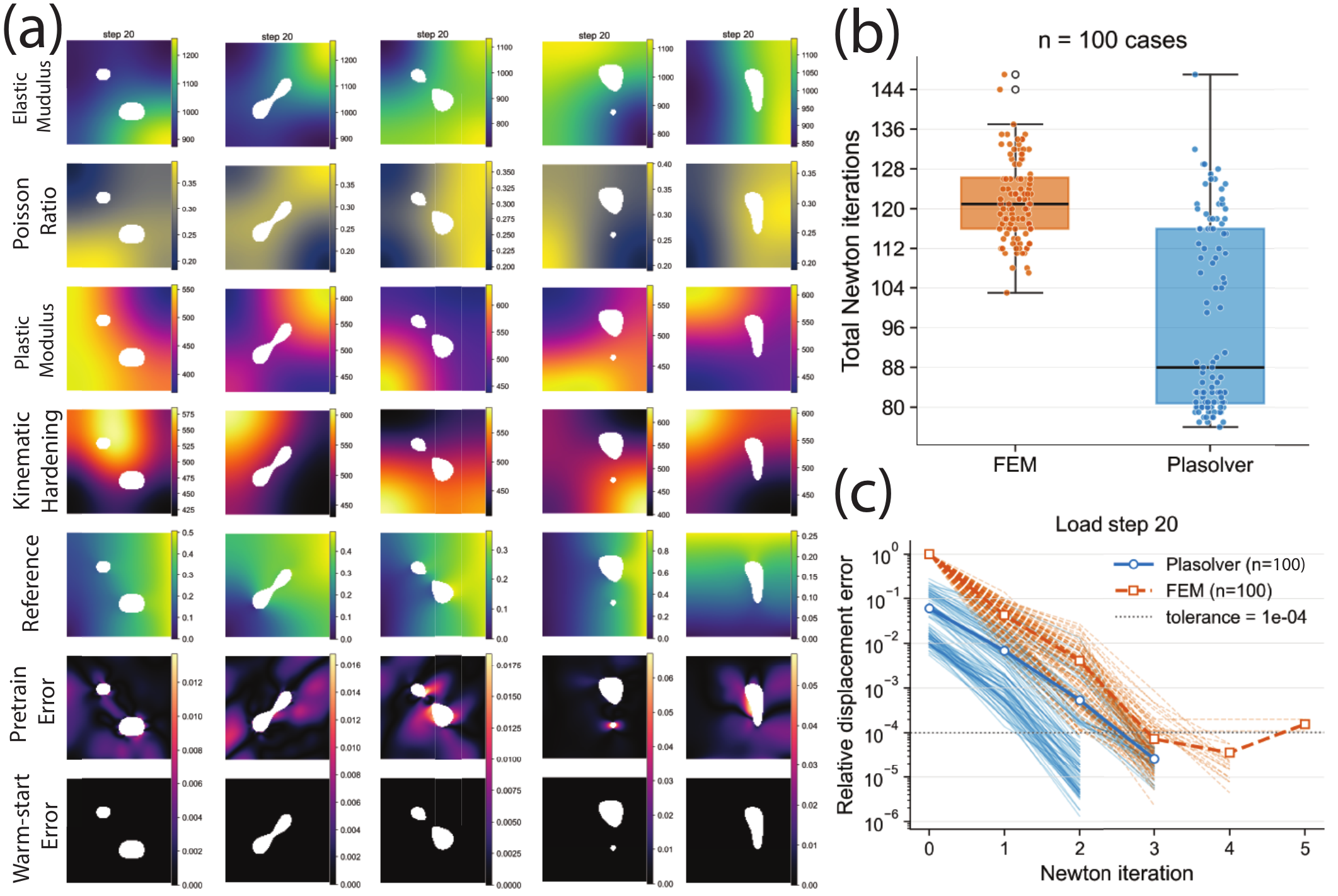}
		\par\end{centering}
	\caption{Performance of the Plasolver warm-start phase for different materials, geometries, and loading paths. (a) Elastic-modulus field, Poisson's-ratio field, plastic-modulus field, kinematic-hardening-modulus field, finite element reference solution, absolute-error field produced by the pretrained Plasolver model, and absolute-error field after warm-start refinement at the 20th load step. (b) Mean numbers of Newton iterations over all load steps. The convergence threshold for the warm-start relative displacement error is $0.0001$. (c) Iterative convergence curves at the 20th load step. The vertical axis represents the relative displacement error, and the horizontal axis represents the number of Newton iterations. The statistics are calculated over 100 test instances.
		\label{fig:Plasolver_warm_start_material}}
\end{figure}

Although the error of the pretrained Plasolver model is already small, at approximately $1\%$, the solution accuracy can be further improved using the Plasolver warm-start phase when required. The warm-start-refined solution can approach the finite element reference solution because the prediction from the pretraining phase is used as the initial solution of the iterative solver. \Cref{fig:Plasolver_warm_start_material} demonstrates the performance of the Plasolver warm-start phase. Using the Plasolver prediction as the initial solution substantially reduces the number of nonlinear Newton iterations. The corresponding quantitative iteration speedup ratio is reported in “M+G+P: L” of \Cref{tab:Plasolver_diff_problem}.

\begin{figure}
	\begin{centering}
		\includegraphics[scale=0.47]{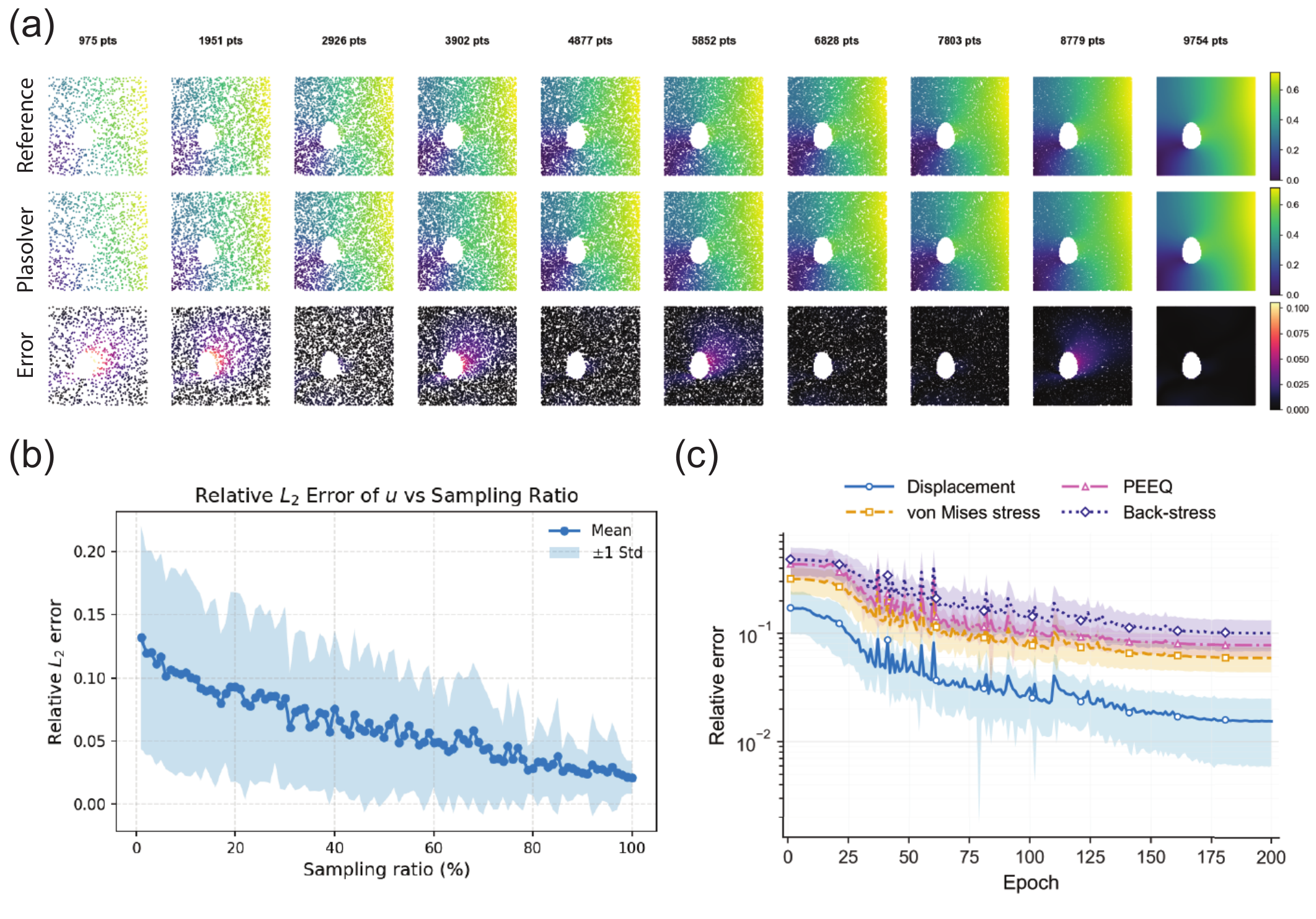}
		\par\end{centering}
	\caption{Spatial-discretization invariance of the pretrained Plasolver model. (a) Predicted displacement-magnitude fields obtained using different proportions of test points, ranging from $10\%$ to $100\%$ in increments of $10\%$. (b) Displacement-magnitude error distributions obtained using different proportions of test points, ranging from $1\%$ to $100\%$ in increments of $1\%$. (c) Evolution of the relative errors during Plasolver training.
		\label{fig:Plasolver_space_discretization_invariance}}
\end{figure}

\Cref{fig:Plasolver_path_discretization_invariance} demonstrated that the pretrained Plasolver model possesses path-discretization invariance. We next evaluate the spatial-discretization invariance of the pretrained Plasolver model. \Cref{fig:Plasolver_space_discretization_invariance}a and b demonstrate that the pretrained Plasolver model possesses strong spatial-discretization invariance. Even when the number of input points is reduced to $10\%$ of the original number, the model maintains good accuracy and robustness. \Cref{fig:Plasolver_space_discretization_invariance}c shows the convergence history during Plasolver pretraining. The displacement error approaches $1\%$ after approximately 175 epochs. Further discussion of spatial-discretization invariance is provided in \Cref{subsec:space_path_discretization}.

The above experiments all employed the radial return algorithm, as they considered linear isotropic hardening. We next evaluate the performance of Plasolver on nonlinear isotropic hardening. Unlike the linear isotropic hardening case, nonlinear isotropic hardening cannot be integrated using the radial return algorithm and instead requires the return mapping algorithm. The constitutive model consists of nonlinear isotropic hardening and linear kinematic hardening:
\begin{equation}
	\begin{aligned}
		\sigma_{y}(\bar{\varepsilon}^{p}) & =\sigma^{0}_{y}+H*[1-\exp(-\bar{\varepsilon}^{p})],\\
		\dot{q}_{ij} & =\frac{2}{3}C\dot{\varepsilon}^{p}_{ij}=\dot{\gamma}\sqrt{\frac{2}{3}}C\frac{\boldsymbol{\eta}}{\sqrt{\boldsymbol{\eta}:\boldsymbol{\eta}}}.
	\end{aligned}
\end{equation}

\begin{figure}
	\begin{centering}
		\includegraphics[scale=0.47]{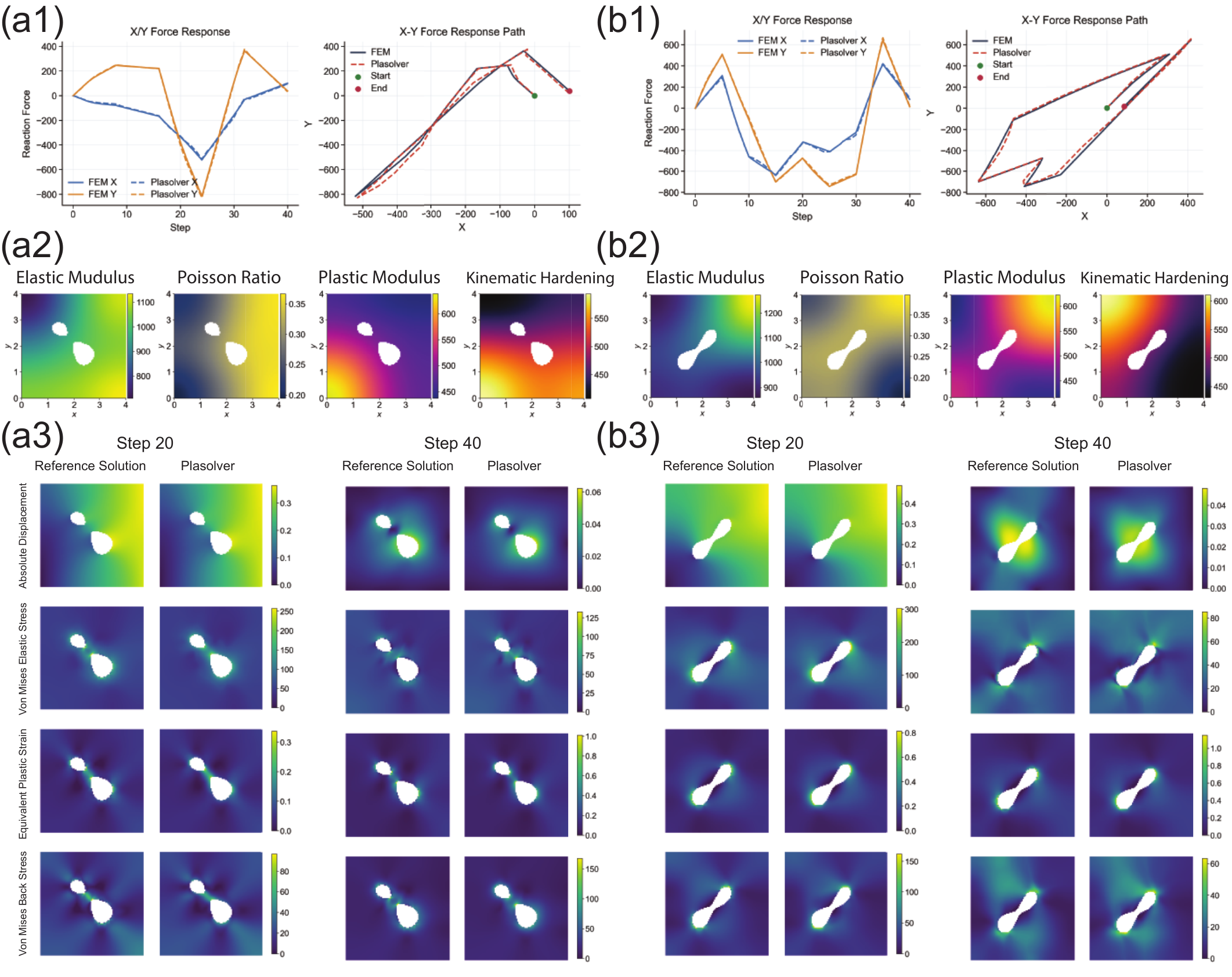}
		\par\end{centering}
	\caption{Performance of Plasolver during the pretraining stage for elastoplastic problems with different material properties, geometries, and loading paths under nonlinear isotropic hardening. (a1,b1) Predicted reaction forces by Plasolver and the FEM reference solution. (a2,b2) Distributions of Young's modulus, Poisson's ratio, isotropic hardening modulus, and kinematic hardening modulus. (a3,b3) Contour predictions of Plasolver and FEM. From the first to the fourth row, the fields correspond to the displacement magnitude, von Mises stress, equivalent plastic strain, and von Mises back stress, respectively. (a1) corresponds to the loading path associated with the material distribution in (a2) and the contour predictions in (a3), while (b1) corresponds to the loading path associated with (b2) and (b3).\label{fig:Plasovler_cycling_loading_material_nonlinear}}
\end{figure}

\begin{figure}
	\begin{centering}
		\includegraphics[scale=0.47]{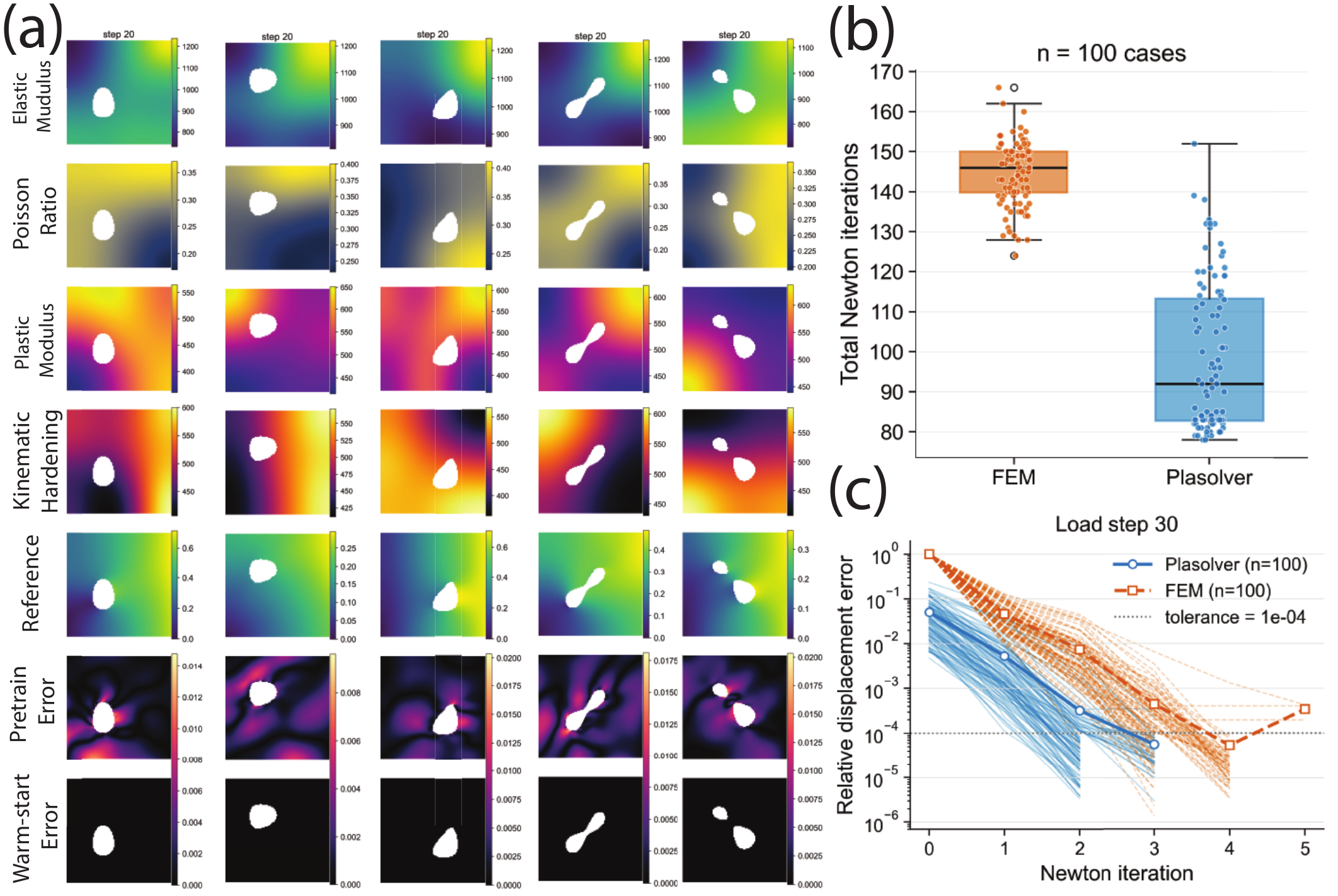}
		\par\end{centering}
	\caption{Performance of Plasolver during the warm-start stage for nonlinear isotropic hardening with different material properties, geometries, and loading paths. (a) Young's modulus field, Poisson's ratio field, isotropic hardening modulus field, kinematic hardening modulus field, FEM reference solution, absolute error of the pretrained Plasolver, and absolute error after the warm-start stage at the twentieth loading increment. (b) Average number of Newton iterations over all loading increments. The convergence threshold for the relative displacement error during the warm-start stage is set to $0.0001$. (c) Newton convergence histories at the thirtieth loading increment, where the vertical axis denotes the relative displacement error and the horizontal axis denotes the number of Newton iterations. The statistics are collected over all 100 test cases.\label{fig:Plasolver_warm_start_material_nonlinear}}
\end{figure}

We adopt the most challenging cyclic loading paths in this experiment. A total of 1,000 cases are generated, with 900 used for training and the remaining 100 reserved for testing, resulting in a dataset of 37.71 GB. The performance of Plasolver during the pretraining and warm-start stages is presented in \ref{fig:Plasovler_cycling_loading_material_nonlinear} and \ref{fig:Plasolver_warm_start_material_nonlinear}, respectively. As can be seen, Plasolver maintains excellent performance even for nonlinear hardening. Notably, the accuracy achieved for nonlinear hardening is comparable to that of the linear hardening model, as summarized in the \textquotedblleft M+G+P: N\textquotedblright~column of \ref{tab:Plasolver_diff_problem}. This is because the constitutive integration procedure is independent of Plasolver itself. Consequently, as long as the corresponding incremental energy functional is available, namely within the framework of the associated $J_{2}$ plasticity flow theory, Plasolver is expected to exhibit strong generalization capability across a wide range of elastoplastic constitutive models.

In summary, \Cref{sec:Result} demonstrates the strong generalization capability of the pretrained Plasolver model. Plasolver can generalize simultaneously across different materials, geometries, and loading paths. The error of the pretrained model is approximately $1\%$, while its simulation speed is approximately 100 times higher than that of the conventional finite element method. In addition, the Plasolver warm-start phase substantially reduces the number of Newton iterations while recovering the accuracy of the finite element solution. It should be noted that the warm-start phase is optional. When higher accuracy is required, warm-start refinement can be performed after the pretraining phase.

\section{Discussion\label{sec:Discussion}}

\subsection{From PINNs to PINO\label{subsec:PINNs-to-PINOs}}

\begin{figure}
	\begin{centering}
		\includegraphics[scale=0.80]{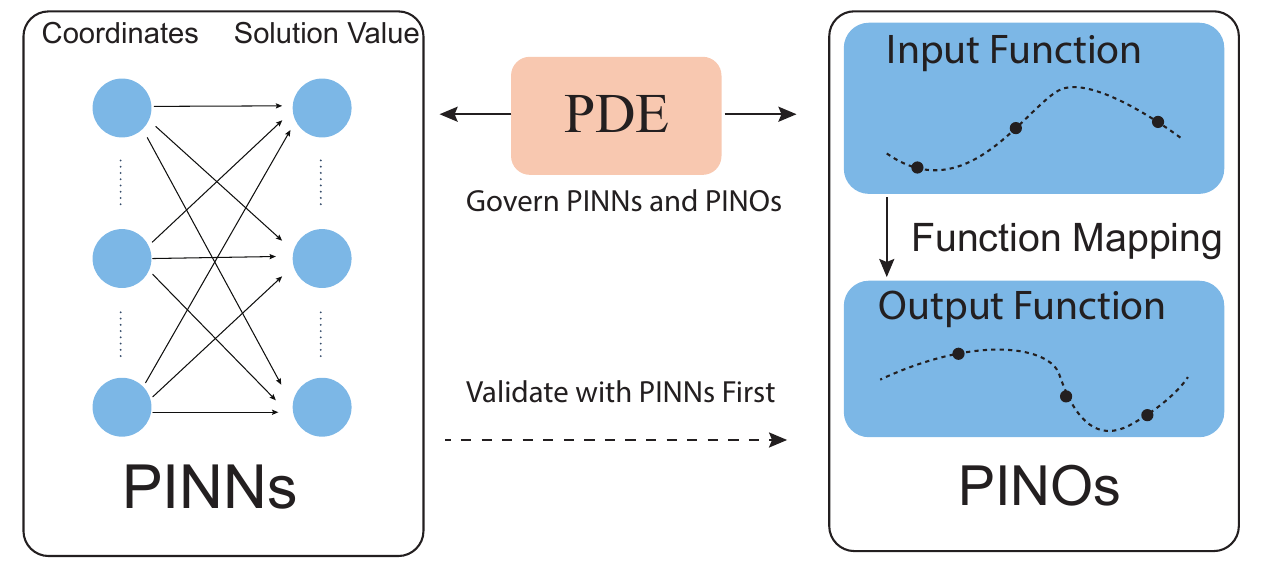}
		\par\end{centering}
	\caption{PINNs serve as a prerequisite and testbed for PINO. Before training a PINO, it is recommended to test the proposed formulation using a PINN to reduce the time required for methodological iterations.
		\label{fig:PINNs2PINOs}}
\end{figure}

PINO trains neural operators entirely using physical equations without relying on labeled data. However, training a PINO is generally computationally expensive, making the iterative development of new ideas time-consuming. We therefore recommend first conducting experiments using PINNs. If the PINN experiment is successful, the approximation function used in the PINN, such as an MLP, can be replaced by a neural operator while keeping the loss function unchanged, as illustrated in \Cref{fig:PINNs2PINOs}.

It should also be noted that we recommend initially testing the PINO using only one problem instance. After this test succeeds, the method can be extended to the complete dataset. Further details are provided in Section 5.1, ``Patch tests for Pretrain Finite Element Method'', of \citet{wang2026pfem}.

\subsection{Shape-function differentiation eliminates the need for distance functions\label{subsec:Shape_function_not_need_distance}}

The displacement boundary conditions in PINO can be imposed strongly, as in the finite element method. If a point lies on the displacement boundary, the model output at that point can be replaced directly by the prescribed displacement. Therefore, the commonly used distance-function construction is not strictly required \citet{boundary_conditions_distance_functions}.

Energy-based DEM and PINO formulations generally employ distance functions to satisfy Dirichlet boundary conditions. However, when Dirichlet boundary conditions are imposed strongly, differentiation using automatic differentiation (AD) \citet{automatic_differential} often produces incorrect results. It should be noted that both DEM and PINO constructed using the DEM loss are based on energy formulations. The difference is that DEM solves an individual problem, whereas PINO solves a family of problems. As discussed in \Cref{subsec:PINNs-to-PINOs}, DEM, which is the energy-based form of PINNs, can be used as a preliminary test for PINO. We therefore focus on DEM and illustrate the issue using the following simple problem:
\begin{equation}
	\begin{aligned}
		-\Delta T
		&=
		0,
		&&
		\boldsymbol{x}\in[0,1]^{2},
		\\
		T
		&=
		0,
		&&
		x=0,
		\\
		\frac{\partial T}{\partial\boldsymbol{n}}
		&=
		0,
		&&
		y=0,\;y=1,
		\\
		\frac{\partial T}{\partial\boldsymbol{n}}
		&=
		1,
		&&
		x=1.
	\end{aligned}
\end{equation}
Here, $T$ is the field variable to be solved, and $\boldsymbol{n}$ is the outward unit normal to the boundary.

\begin{figure}
	\begin{centering}
		\includegraphics[scale=0.54]{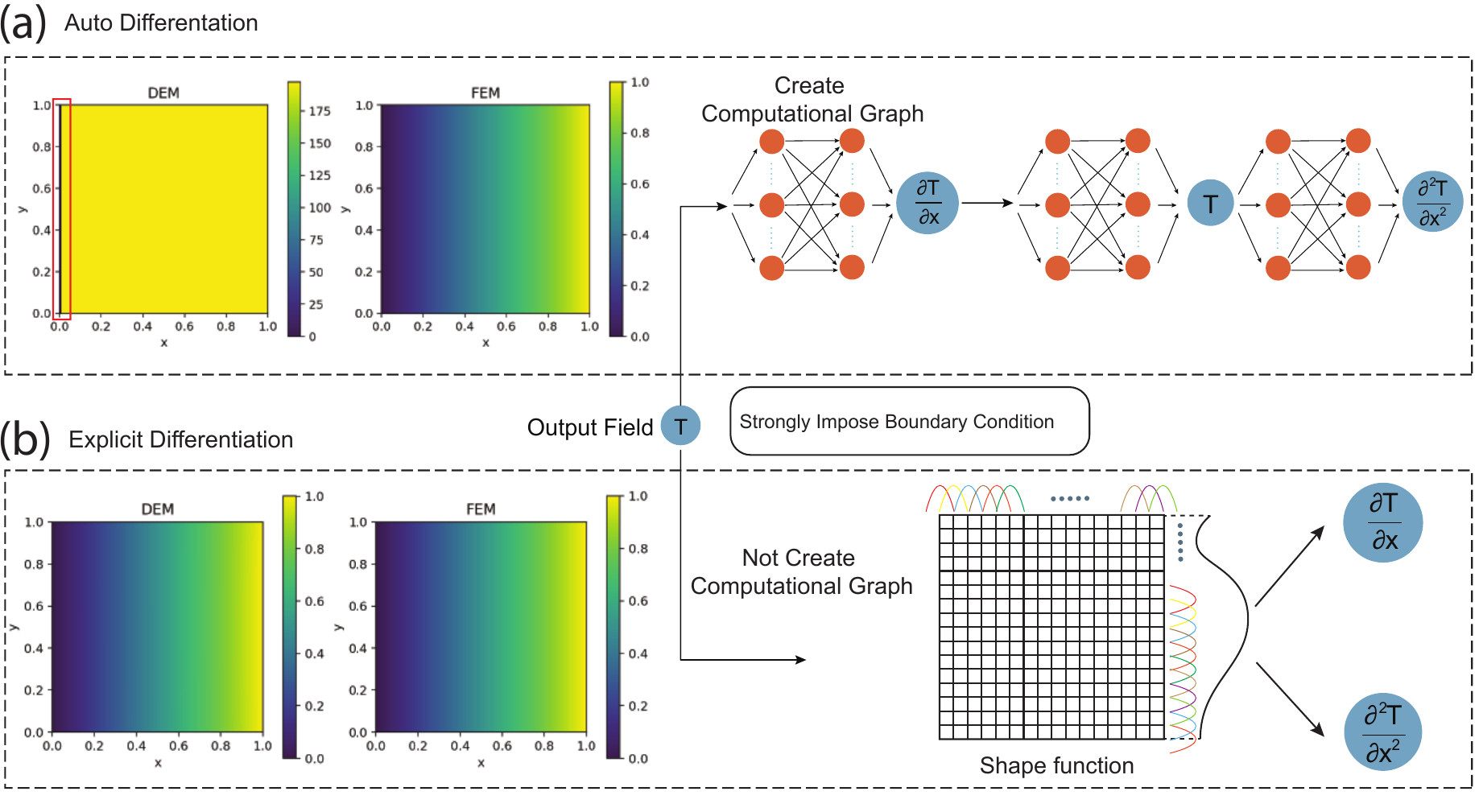}
		\par\end{centering}
	\caption{Considerations when strongly imposing boundary conditions in the deep energy method. (a) When derivatives are obtained using automatic differentiation, a nonphysical discontinuity frequently appears on the Dirichlet boundary, as indicated by the red box. In addition, the computational graph constructed by AD grows rapidly as the derivative order increases. (b) When derivatives are obtained through explicit differentiation, typically using finite element shape functions, no computational graph is constructed.
		\label{fig:Strongly-impose-boundary}}
\end{figure}

\Cref{fig:Strongly-impose-boundary}a shows the DEM prediction obtained by strongly imposing the boundary conditions and calculating derivatives using AD. A nonphysical numerical discontinuity can be clearly observed near $x=0$, as indicated by the red box. When the boundary conditions are imposed strongly, DEM tends to reduce the internal energy toward zero to decrease the loss function because the internal energy cannot be negative. For the present problem, the internal energy is
\begin{equation}
	W_{e}
	=
	\int_{\Omega}
	\frac{1}{2}
	(\nabla T)\cdot(\nabla T)
	\,d\Omega.
\end{equation}
The potential of the external force is
\begin{equation}
	W_{\mathrm{ext}}
	=
	\int_{x=1}
	T
	\frac{\partial T}{\partial\boldsymbol{n}}
	\,d\Gamma
	=
	\int_{x=1}
	T
	\,d\Gamma.
\end{equation}
The energy minimized by DEM is
\begin{equation}
	\begin{aligned}
		\Pi
		&=
		W_{e}
		-
		W_{\mathrm{ext}},
		\\
		\textrm{s.t.}\quad
		T
		&=
		0,
		\qquad
		x=0.
	\end{aligned}
\end{equation}
Because the essential boundary condition is imposed strongly, DEM reduces the energy by driving the internal energy toward zero while simultaneously increasing $W_{\mathrm{ext}}$. As the internal energy approaches zero, $T$ approaches a constant within the domain. To increase $W_{\mathrm{ext}}$, this constant continues to increase during DEM optimization. Consequently, the energy becomes unbounded from below. Numerically, this behavior produces a nonphysical discontinuity at the essential boundary, whereas the solution in the remaining domain becomes much larger than the analytical solution, as shown in \Cref{fig:Strongly-impose-boundary}a. Therefore, when the boundary conditions are imposed strongly, a DEM formulation using AD-based differentiation cannot theoretically recover the correct solution.

Fortunately, accurate results can be obtained when explicit differentiation is used. In this work, derivatives are calculated using finite element shape functions. \Cref{fig:Strongly-impose-boundary}b shows that DEM produces an accurate result when the boundary conditions are imposed strongly and the derivatives are evaluated using shape functions. Because shape-function differentiation requires only the neural-network predictions at the corresponding nodes, the procedure is essentially similar to FEM and avoids the problems associated with AD-based differentiation.

\begin{figure}
	\begin{centering}
		\includegraphics[scale=0.55]{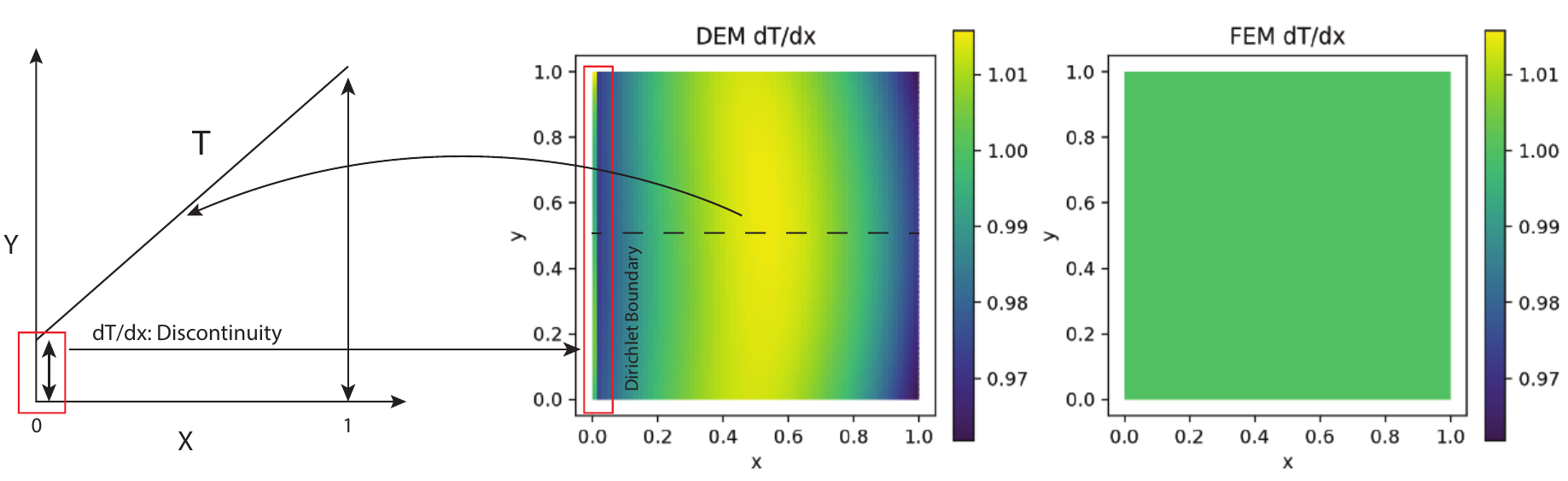}
		\par\end{centering}
	\caption{When boundary conditions are imposed strongly, the derivatives obtained by the deep energy method may be discontinuous at the Dirichlet boundary, as indicated by the red box.
		\label{fig:Strongly-impose-boundary_discontinuity}}
\end{figure}

Although explicit differentiation performs well for the original field when the boundary conditions are imposed strongly, its derivative predictions are less accurate. Strongly enforcing the Dirichlet boundary conditions can produce discontinuities in the derivatives, as shown in \Cref{fig:Strongly-impose-boundary_discontinuity}. Such derivative discontinuities can produce discontinuities in the stress field of a plasticity problem, thereby affecting subsequent updates of the internal variables. For example, errors may arise in the equivalent plastic strain. Although the error introduced in a single step may be small, errors accumulate over the loading history in plasticity problems. Therefore, distance functions are still used in the present PINO formulation. This choice is motivated not by the accuracy of the original displacement field, but by the improved accuracy of its derivatives.

Future work may investigate more stable approaches for training Plasolver when boundary conditions are imposed strongly. Distance functions can be difficult to construct for problems involving complex boundaries. If the boundary conditions can be imposed directly and strongly, model training can be simplified substantially.

\subsection{Analysis of the optimization difficulty of elastoplastic problems\label{subsec:DEM_optimization}}

\Cref{sec:Result} shows that the elastoplastic problems converge rapidly, reaching a relative error of approximately $1\%$ within ten epochs. In general, non-quasistatic problems are considerably more difficult than quasistatic problems. However, PINO performs almost as well for the considered elastoplastic problems as it does for quasistatic problems, which appears counterintuitive. Plasticity problems also involve error accumulation, suggesting that such strong performance should not necessarily be expected.

This behavior can be explained by the fact that the elastoplastic energy can readily reduce to the elastic energy. If none of the material points enter the plastic regime, \Cref{eq:simo} reduces to
\begin{equation}
	\begin{aligned}
		\Pi^{(n+1)}
		&=
		\int_{\Omega}
		\left[
		W^{(n+1)}_{e}
		+
		\frac{1}{2}
		\boldsymbol{v}^{(n+1)}
		\cdot
		\boldsymbol{D}^{-1}
		\cdot
		\boldsymbol{v}^{(n+1)}
		\right]
		\,dV
		-
		W_{\mathrm{ext}}.
	\end{aligned}
	\label{eq:simo-elas}
\end{equation}
Because the material remains elastic,
$\boldsymbol{\varepsilon}^{p(n+1)}
=
\boldsymbol{\varepsilon}^{p(n)}$
and
$\boldsymbol{v}^{(n+1)}
=
\boldsymbol{v}^{(n)}$.
The term
$\boldsymbol{v}^{(n+1)}
\cdot
\boldsymbol{D}^{-1}
\cdot
\boldsymbol{v}^{(n+1)}/2$
in \Cref{eq:simo-elas} is therefore constant. Consequently, optimizing $\Pi^{(n+1)}$ is essentially equivalent to optimizing the energy of a linear elastic problem. Under these conditions, elastoplastic energy optimization has the same fundamental form as that of a quasistatic elastic problem.

When some material points enter the plastic regime, the plastic multiplier is determined during constitutive integration using the consistency condition in \Cref{eq:consistent_condition}. This calculation is essentially equivalent to the constitutive integration performed by a UMAT in a conventional finite element analysis. Furthermore, the proof in \ref{sec:plasticity_variational} shows that the incremental potential of the elastoplastic problem is a single-extremum problem. Such a problem is considerably easier to optimize than a problem with multiple extrema. When the deep energy method is applied to a multi-extremum problem, the optimization process must often overcome energy barriers. These problems are therefore substantially more difficult than single-extremum problems, as demonstrated by deep-energy formulations of phase-field fracture \citet{goswami2020transfer}.

Fortunately, although both elastoplasticity and phase-field fracture are path-dependent, the elastoplastic energy considered here is a single-extremum problem. It is therefore considerably easier to optimize than the phase-field fracture energy. Moreover, elastoplastic problems contain multiple load steps, meaning that each loading trajectory effectively contributes a number of optimization instances equal to its number of load steps. Consequently, most examples in \Cref{sec:Result} converge within ten epochs and may even converge faster than many quasistatic problems.

\subsection{Spatial- and path-discretization invariance\label{subsec:space_path_discretization}}

The Transolver neural operator is used in most of the Plasolver pretraining experiments. \Cref{fig:Plasolver_space_discretization_invariance} demonstrates the spatial-discretization invariance of Plasolver. This property originates from the neural-operator architecture itself, particularly the permutation equivariance of the attention mechanism.

\begin{figure}
	\begin{centering}
		\includegraphics[scale=0.47]{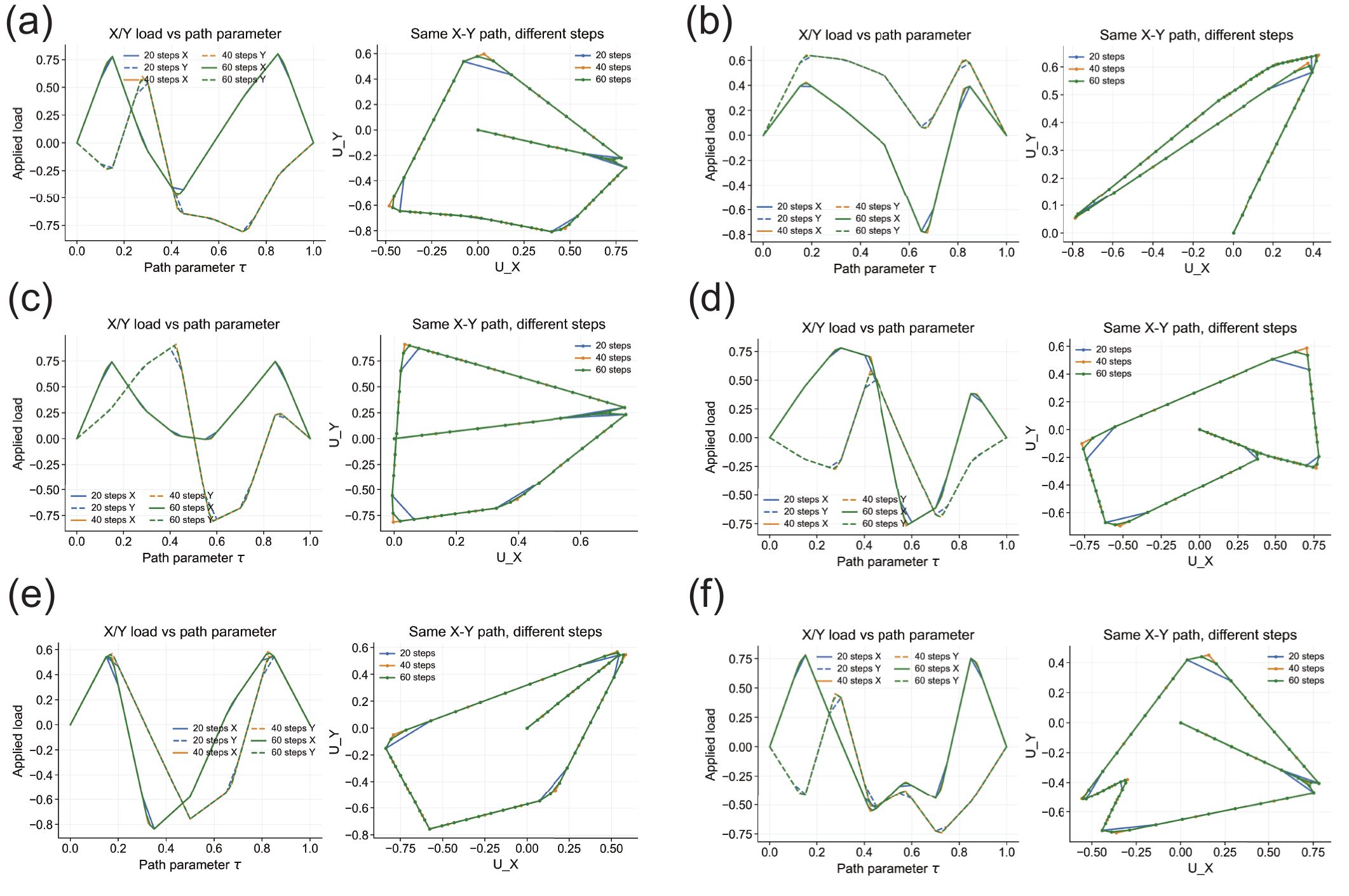}
		\par\end{centering}
	\caption{Predictions obtained using the pretrained Plasolver model for the same loading path discretized using different numbers of load steps.
		\label{fig:Plasolver_path_discretization_more}}
\end{figure}

Interestingly, Plasolver possesses not only spatial-discretization invariance but also path-discretization invariance, as shown in \Cref{fig:Plasolver_path_discretization_invariance}. Path-discretization invariance arises from the inputs and outputs of the neural operator. The number of load steps is not explicitly included as an input. Instead, the information associated with the current load increment is provided directly to the model. Through training on numerous problem instances while enforcing the governing laws of plasticity, the neural operator learns the local mapping associated with incremental plastic loading. This is the central reason why Plasolver possesses path-discretization invariance. \Cref{fig:Plasolver_path_discretization_more} presents additional results demonstrating the robustness of Plasolver for the same loading trajectory discretized using different numbers of load steps.

\subsection{Effect of batch size on PINO\label{subsec:diff_batch}}

\begin{figure}
	\begin{centering}
		\includegraphics[scale=0.47]{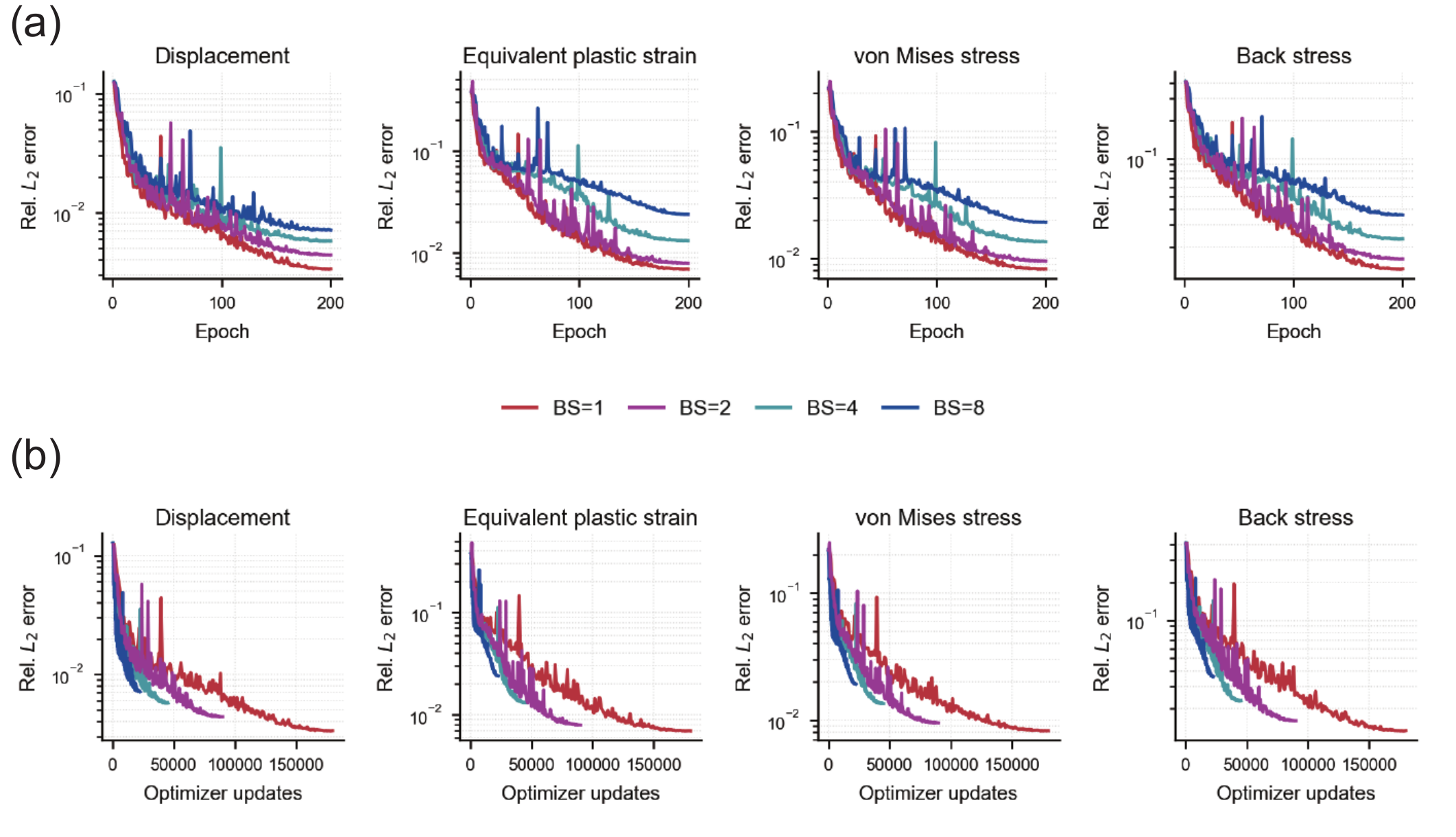}
		\par\end{centering}
	\caption{Effects of different batch sizes during the Plasolver pretraining phase. (a) Evolution of the relative error over the training epochs. (b) Evolution of the relative error with respect to the number of optimization steps.
		\label{fig:Plasolver_batch_size}}
\end{figure}

All Plasolver pretraining experiments in \Cref{sec:Result} use a batch size of one. However, the available GPU memory can support larger batch sizes, which can substantially improve training efficiency. We therefore evaluate the effect of batch size on Plasolver. The cyclic loading problem in \Cref{subsec:path} is used to test batch sizes of 1, 2, 4, and 8. \Cref{fig:Plasolver_batch_size}a shows the evolution of the relative error over the training epochs. Smaller batch sizes produce lower relative errors and better performance when the models are compared at the same epoch. However, with a batch size of one, the model parameters are updated 1000 times during each epoch because each epoch contains 1000 training instances. By contrast, with a batch size of eight, only 125 parameter updates are required. Therefore, increasing the batch size substantially reduces the time required for each epoch.

A more objective comparison should therefore be based on the number of optimization steps rather than the number of epochs. \Cref{fig:Plasolver_batch_size}b shows the evolution of the relative error with respect to the number of optimization steps. For the same number of optimization steps, larger batch sizes achieve better performance. These results indicate that future pretrained physics foundation models may achieve faster convergence by using larger batch sizes.

\subsection{Comparison of the Deep Energy Method, the Finite Element Method, and Plasolver\label{subsec:DEM_FEM_Plasolver}}

\begin{figure}
	\begin{centering}
		\includegraphics[scale=0.45]{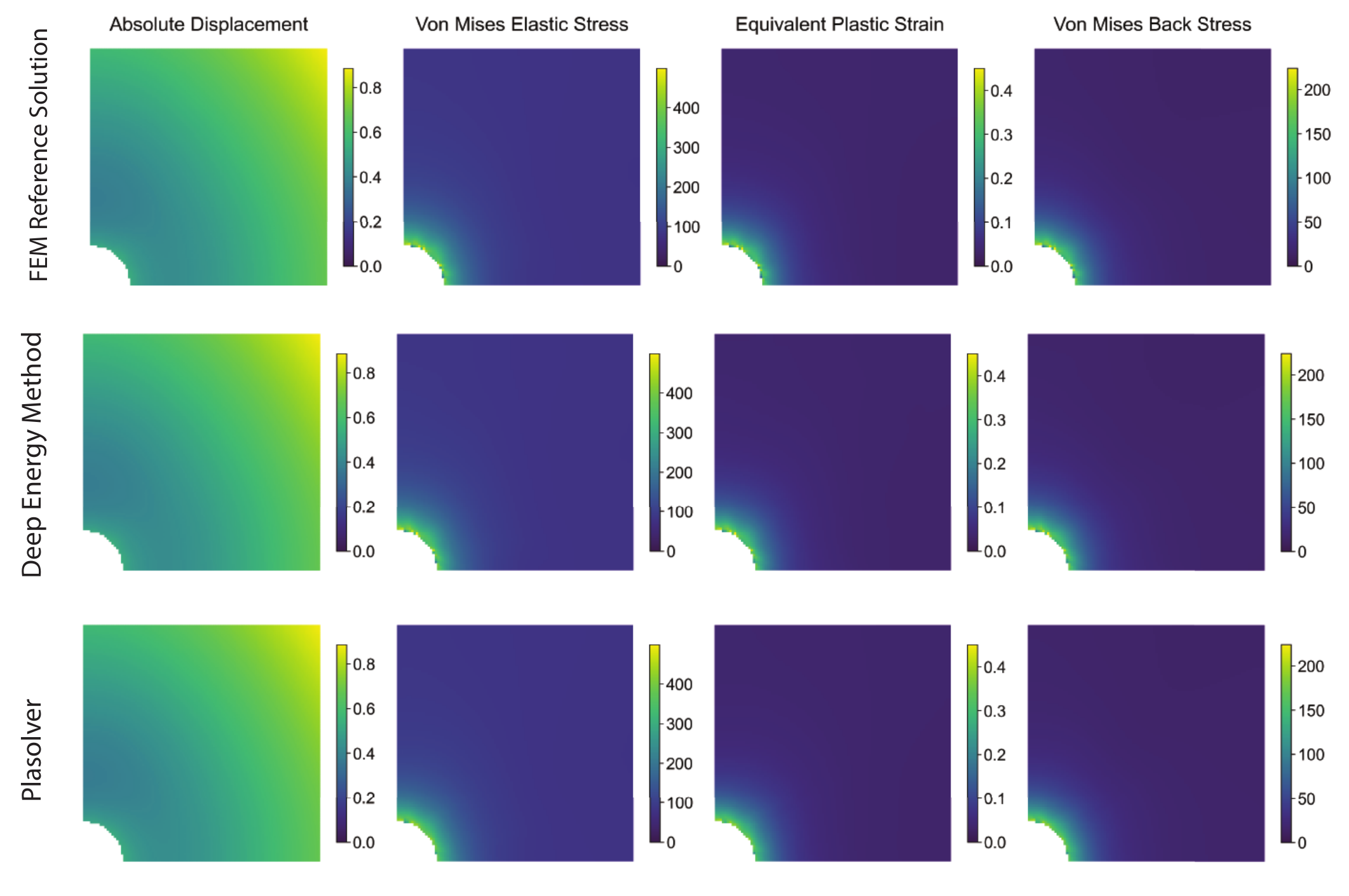}
		\par\end{centering}
	\caption{Predictions obtained by the finite element method (FEM) as the reference solution, the Deep Energy Method (DEM), and Plasolver.\label{fig:FEM_DEM_Plasolver}}
\end{figure}

The Deep Energy Method (DEM) is the energy-based formulation of physics-informed neural networks (PINNs). Similar to the finite element method (FEM), DEM does not possess generalization capability. Therefore, when the geometry, material properties, or boundary/loading conditions change, both DEM and FEM need to solve the new problem from scratch. In contrast, Plasolver is a neural operator with generalization capability across different geometries, material properties, and loading conditions, as demonstrated in \Cref{sec:Result}. For a specific problem with the loading--unloading path described in \Cref{subsec:path}, \Cref{fig:FEM_DEM_Plasolver} shows that DEM, FEM, and Plasolver produce nearly identical predictions. This demonstrates that all three methods can accurately solve elastoplastic problems.

\begin{figure}
	\begin{centering}
		\includegraphics[scale=0.70]{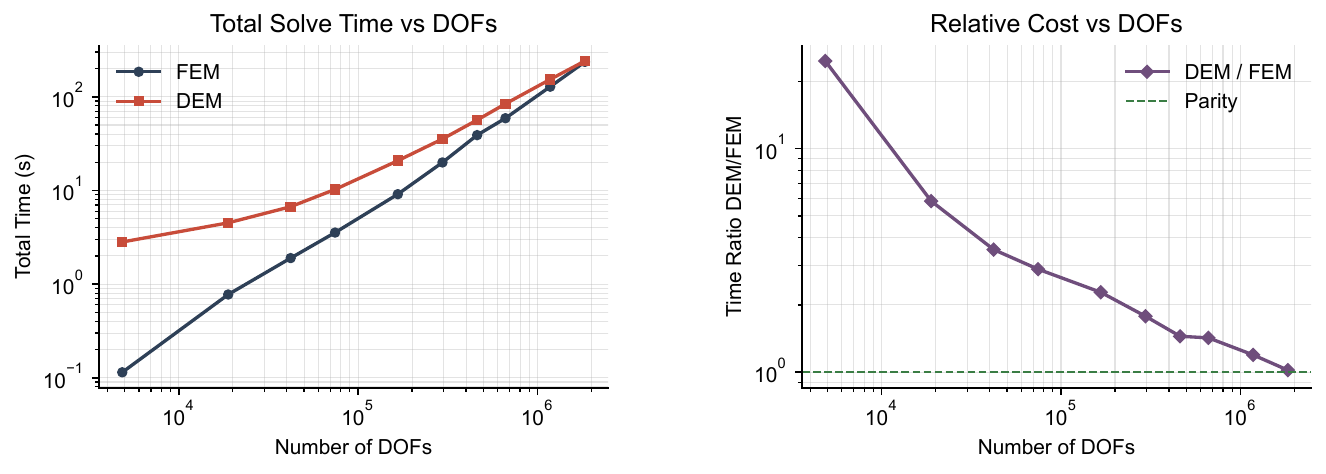}
		\par\end{centering}
	\caption{Computational efficiency of the finite element method (FEM) and the Deep Energy Method (DEM) for different numbers of degrees of freedom (DOFs): (a) computational time as a function of the number of DOFs; (b) ratio of the computational time of DEM to that of FEM as a function of the number of DOFs.\label{fig:FEM_DEM_dof}}
\end{figure}

As shown in \Cref{tab:DEM_plasticity}, FEM is more computationally efficient than DEM for problems with relatively few degrees of freedom (DOFs). However, this advantage only holds for low-DOF problems. For problems involving extremely large numbers of DOFs, DEM becomes more computationally efficient than FEM, as shown in \Cref{tab:Time_comparision_dem_fem_plasolver}. This is because the computational complexity of FEM is strongly dependent on the number of nodal DOFs, whereas that of DEM is primarily associated with the number of neural network parameters. Specifically, DEM optimizes the neural network parameters rather than directly solving for the nodal displacements as in FEM. Therefore, for problems with extremely large numbers of DOFs, DEM can achieve an efficiency advantage using a relatively small number of neural network parameters. \Cref{fig:FEM_DEM_dof} compares the computational efficiency of FEM and DEM for different numbers of DOFs. It can be observed that the efficiency advantage of DEM becomes increasingly pronounced as the number of DOFs increases. It is worth noting that many complex geometries require extremely large numbers of DOFs to achieve an accurate geometric representation.

Neither FEM nor DEM possesses generalization capability; therefore, a new computation is required for each unseen problem. This becomes particularly computationally expensive for problems with extremely high spatial resolutions. We therefore further investigate the computational efficiency of DEM, FEM, and Plasolver at a spatial resolution of 1000. The results are summarized in \Cref{tab:Time_comparision_dem_fem_plasolver}. It can be seen that the pretrained Plasolver achieves an approximately 300-fold speedup over conventional FEM during inference. Although training Plasolver requires a computational cost equivalent to approximately three FEM simulations at this resolution, once trained, Plasolver can provide highly efficient predictions for unseen problems at extremely high spatial resolutions. It is also worth noting that at a spatial resolution of 100, Plasolver achieves an approximately 100-fold speedup over FEM, as shown in \Cref{tab:Plasolver_diff_problem}. These results suggest that the computational benefit of the neural operator becomes increasingly pronounced as the spatial resolution increases. This behavior can be attributed to the more favorable computational scaling of neural-operator inference compared with conventional FEM at extremely high resolutions. A rigorous quantitative characterization of the computational complexity and scaling behavior of these methods will be investigated in future work.

\begin{table}
	\caption{Comparison of the computational efficiency of the Deep Energy Method (DEM), the finite element method (FEM), and Plasolver at a spatial resolution of 1000. Each training epoch of Plasolver requires approximately 22 minutes, and the reported model is obtained after 24 training epochs. Since DEM and FEM do not possess generalization capability and must solve each new problem independently, the concept of inference time is not applicable to these two methods. The inference time reported for Plasolver corresponds only to its pretrained inference stage.\label{tab:Time_comparision_dem_fem_plasolver}}
	
	\centering{}%
	\begin{tabular}{ccc}
		\toprule 
		Method & Training Time (s) & Inference Time (s)\tabularnewline
		\midrule
		DEM & 9652.84 & N/A\tabularnewline
		FEM & 9444.08 & N/A\tabularnewline
		Plasolver & 31680 & 32.8\tabularnewline
		\bottomrule
	\end{tabular}
\end{table}

\section{Conclusion\label{sec:Conclusion}}

We proposed Plasolver, a framework specifically designed to accelerate the solution of elastoplastic problems. Plasolver consists of two stages: pretraining and warm start. The pretraining phase exploits the computational efficiency of operator learning, whereas the warm-start phase incorporates the accuracy of the conventional finite element method. The pretrained Plasolver model is trained using Simo's incremental potential. This training process is entirely physics-based and requires no labeled data, thereby eliminating the substantial time required to generate datasets for conventional data-driven approaches. In the warm-start phase, the prediction produced by the pretrained model is used as the initial solution of the conventional finite element iterative solver, substantially reducing the number of Newton iterations. We systematically evaluated Plasolver across different loading paths, geometries, and materials. The results demonstrate that the pretrained Plasolver model achieves approximately $99\%$ accuracy and an approximately 100-fold speedup. The warm-start phase recovers the accuracy of FEM while providing an approximately $50\%$ speedup. Plasolver also possesses both spatial- and path-discretization invariance, enabling it to operate across different spatial and loading-path resolutions. Overall, we systematically demonstrated the generalization capability of the pretrained Plasolver model across loading paths, geometries, and materials, as well as the ability of the warm-start phase to retain finite-element-level accuracy.

The current pretrained Plasolver model primarily employs Transolver as its operator-learning architecture. Recently, AB-UPT \citet{alkin2025ab} has demonstrated state-of-the-art performance on the DrivAerNet++ automotive-flow dataset \citet{elrefaie2024drivaernet++}. Therefore, Transolver could be replaced by more advanced operator-learning architectures in future implementations. Although operator-learning architectures are developing rapidly, we expect future operator-learning methods to remain centered on Transformer-based architectures. The central contribution of this work is a computational framework for applying PINO to elastoplasticity, thereby accelerating the complete simulation of plastic deformation. However, applying PINO to plasticity requires a sufficiently clear understanding of the underlying physical mechanisms so that the corresponding PDEs and evolution equations can be formulated. For plastic processes whose physical mechanisms remain unclear, such as those without known internal variables or evolution equations, physics-based frameworks such as PINO cannot be used to train the neural operator. The main reason is that the applicability of Simo's incremental potential in \Cref{eq:simo} is limited. For example, it is applicable to associative $J_{2}$ plasticity but does not generally apply to non-associated plastic flow. This limitation arises from the underlying theory rather than from the proposed algorithm. Such problems must instead be addressed using purely data-driven methods. For example, \citet{guo2025history} employed data-driven operator learning to construct plastic constitutive models. However, \citet{guo2025history} did not consider operator learning for the equilibrium equations. Future work could therefore combine Plasolver with the approach of \citet{guo2025history}. In addition, Plasolver evaluates derivatives explicitly using finite element shape functions, providing highly efficient derivative calculations. This capability is important for training future pretrained physics foundation models. \Cref{subsec:Shape_function_not_need_distance} also demonstrates that shape-function differentiation does not require a distance function when boundary conditions are imposed strongly. However, numerical discontinuities may appear in the derivative fields near the boundary. Future work may investigate more stable approaches for training Plasolver with strongly imposed boundary conditions. Distance functions can be difficult to construct for problems involving complex boundaries, whereas directly and strongly imposing the boundary conditions would substantially simplify model training.

\section*{Declaration of competing interest}
The authors declare that they have no known competing financial interests or personal relationships that could
have appeared to influence the work reported in this paper.

\section*{Acknowledgement}
The study was supported by the Key Project of the National Natural Science Foundation of China (12332005) and scholarship from Bauhaus University in Weimar.  We would like to thank Zerui Chen and Yuzhou Lin for the insightful discussions.

\section*{CRediT authorship contribution statement}

\textbf{Yizheng Wang}: Conceptualization, Methodology, Formal analysis, Investigation, Data curation, Validation, Visualization, Writing – original draft, Writing – review \& editing.

\appendix

\section{Basic derivations in plasticity\label{sec:derivative_in_plasticity}}

In this section, we derive the forms of the von Mises equivalent stress
$\bar{\sigma}
=
\sqrt{(3/2)\boldsymbol{\eta}:\boldsymbol{\eta}}$,
the equivalent plastic strain increment
$d\bar{\varepsilon}^{p}
=
\sqrt{(2/3)d\boldsymbol{\varepsilon}^{p}:d\boldsymbol{\varepsilon}^{p}}$,
and the backstress evolution equation
$\dot{q}_{ij}
=
(2/3)C\dot{\varepsilon}^{p}_{ij}$
in $J_{2}$ plasticity. These coefficients arise from the requirement that the three-dimensional tensor formulation be consistent with uniaxial plastic deformation.

We consider uniaxial tension in which only the normal stress $\sigma$ acts along the $x$ axis:
\begin{equation}
	\boldsymbol{\sigma}
	=
	\begin{bmatrix}
		\sigma & 0 & 0
		\\
		0 & 0 & 0
		\\
		0 & 0 & 0
	\end{bmatrix}.
	\label{eq:sigma_one_d}
\end{equation}
In $J_{2}$ plasticity, only the deviatoric stress causes plastic deformation. The deviatoric part $\boldsymbol{s}$ of the stress tensor in \Cref{eq:sigma_one_d} is therefore
\begin{equation}
	\boldsymbol{s}
	=
	\boldsymbol{\sigma}'
	=
	\begin{bmatrix}
		\dfrac{2}{3}\sigma & 0 & 0
		\\
		0 & -\dfrac{1}{3}\sigma & 0
		\\
		0 & 0 & -\dfrac{1}{3}\sigma
	\end{bmatrix}.
\end{equation}

\subsection{Equivalent plastic strain}

Because uniaxial tension is considered, the plastic strain increment along the $x$ axis is denoted by $d\varepsilon^{p}_{x}$. For an isotropic material, the plastic strain increments in the $y$ and $z$ directions are equal because these two directions are equivalent. In addition, the plastic strain tensor is deviatoric because the plastic strain increment is determined by the deviatoric stress, as shown in \Cref{eq:plasticity_strain_increment}. The plastic strain increment tensor is therefore
\begin{equation}
	d\boldsymbol{\varepsilon}^{p}
	=
	\begin{bmatrix}
		d\varepsilon^{p}_{x} & 0 & 0
		\\
		0 & -\dfrac{1}{2}d\varepsilon^{p}_{x} & 0
		\\
		0 & 0 & -\dfrac{1}{2}d\varepsilon^{p}_{x}
	\end{bmatrix}.
\end{equation}

The equivalent plastic strain increment is
\begin{equation}
	d\bar{\varepsilon}^{p}
	=
	\sqrt{
		\frac{2}{3}
		d\boldsymbol{\varepsilon}^{p}
		:
		d\boldsymbol{\varepsilon}^{p}
	}
	=
	d\varepsilon^{p}_{x}.
	\label{eq:equal_plastic_strain}
\end{equation}
The coefficient $2/3$ in \Cref{eq:equal_plastic_strain} therefore ensures consistency between the three-dimensional plastic strain tensor and the uniaxial plastic strain increment $d\varepsilon^{p}_{x}$. It should be noted that the equivalent plastic strain increases monotonically and cannot decrease. By contrast, changes in the individual components of the plastic strain depend on the loading direction. Therefore, an individual plastic strain component may decrease or even return to zero during loading and unloading.

\subsection{Von Mises equivalent stress}

We continue to consider uniaxial tension. According to the backstress evolution equation
$\dot{q}_{ij}
=
(2/3)C\dot{\varepsilon}^{p}_{ij}$,
the backstress increment is
\begin{equation}
	d\boldsymbol{q}
	=
	\begin{bmatrix}
		\dfrac{2}{3}C\,d\varepsilon^{p}_{x} & 0 & 0
		\\
		0 & -\dfrac{1}{3}C\,d\varepsilon^{p}_{x} & 0
		\\
		0 & 0 & -\dfrac{1}{3}C\,d\varepsilon^{p}_{x}
	\end{bmatrix}.
\end{equation}
The shifted deviatoric stress $\boldsymbol{\eta}$ can then be written as
\begin{equation}
	\boldsymbol{\eta}
	=
	\boldsymbol{s}
	-
	\boldsymbol{q}
	=
	\begin{bmatrix}
		\dfrac{2}{3}\sigma
		-
		\dfrac{2}{3}C\varepsilon^{p}_{x}
		& 0 & 0
		\\
		0
		&
		-\dfrac{1}{3}\sigma
		+
		\dfrac{1}{3}C\varepsilon^{p}_{x}
		& 0
		\\
		0 & 0
		&
		-\dfrac{1}{3}\sigma
		+
		\dfrac{1}{3}C\varepsilon^{p}_{x}
	\end{bmatrix}.
	\label{eq:eta_stress}
\end{equation}

The von Mises equivalent stress is therefore
\begin{equation}
	\bar{\sigma}
	=
	\sqrt{
		\frac{3}{2}
		\boldsymbol{\eta}
		:
		\boldsymbol{\eta}
	}
	=
	\left|
	\sigma
	-
	C\varepsilon^{p}_{x}
	\right|.
	\label{eq:mises_stress}
\end{equation}
The coefficient $3/2$ in \Cref{eq:mises_stress} ensures consistency between the three-dimensional tensor $\boldsymbol{\eta}$ and the corresponding uniaxial stress measure.

\subsection{Backstress evolution equation}

The coefficient $2/3$ in the backstress evolution equation
$\dot{q}_{ij}
=
(2/3)C\dot{\varepsilon}^{p}_{ij}$
may initially appear unusual. However, \Cref{eq:mises_stress} shows that the kinematic-hardening correction to the uniaxial stress is directly given by
$C\varepsilon^{p}_{x}$.
The coefficient multiplying $C$ in this uniaxial correction is exactly one.

Therefore, the coefficient $2/3$ in the three-dimensional backstress evolution equation ensures that the uniaxial von Mises equivalent stress is corrected by
$C\varepsilon^{p}_{x}$.
This consistency requirement explains the appearance of the coefficient $2/3$ in
$\dot{q}_{ij}
=
(2/3)C\dot{\varepsilon}^{p}_{ij}$.

\section{Derivation of the analytical plastic multiplier for $J_{2}$ plasticity\label{sec:proof_J2_plasticity_multify}}

In this section, we derive the analytical expression for the plastic multiplier in \Cref{eq:plasticity_factor}. This expression is valid only for associative $J_{2}$ plasticity with linear isotropic and kinematic hardening.

Substituting the yield function in \Cref{eq:yield_function_j2} into the consistency condition in \Cref{eq:consistent_condition} gives
\begin{equation}
	f^{(n+1)}
	=
	\sqrt{
		\frac{3}{2}
		\boldsymbol{\eta}^{(n+1)}
		:
		\boldsymbol{\eta}^{(n+1)}
	}
	-
	\left(
	\sigma^{0}_{y}
	+
	H\bar{\varepsilon}^{p(n+1)}
	\right)
	=
	0.
	\label{eq:consistent_condition_derivative}
\end{equation}

Using the plastic evolution equations in \Cref{eq:plasticity_strain_increment}, we obtain
\begin{equation}
	\begin{aligned}
		\boldsymbol{\eta}^{(n+1)}
		&=
		\boldsymbol{\eta}^{\mathrm{trial}(n+1)}
		-
		\boldsymbol{C}
		:
		\left(
		\Delta\boldsymbol{\varepsilon}^{p(n+1)}
		\right)'
		-
		\Delta\boldsymbol{q}^{(n+1)}
		\\
		&=
		\boldsymbol{\eta}^{\mathrm{trial}(n+1)}
		-
		\Delta\gamma
		\sqrt{\frac{3}{2}}
		\,2G
		\boldsymbol{z}^{\mathrm{trial}(n+1)}
		-
		\Delta\gamma
		\sqrt{\frac{2}{3}}
		\,C
		\boldsymbol{z}^{\mathrm{trial}(n+1)},
		\\
		\bar{\varepsilon}^{p(n+1)}
		&=
		\bar{\varepsilon}^{p(n)}
		+
		\Delta\gamma.
	\end{aligned}
	\label{eq:next_state}
\end{equation}

For an isotropic material,
\begin{equation}
	\begin{aligned}
		\boldsymbol{C}:\boldsymbol{\eta}
		&=
		\left[
		\lambda\delta_{ij}\delta_{kl}
		+
		G
		\left(
		\delta_{ik}\delta_{jl}
		+
		\delta_{il}\delta_{jk}
		\right)
		\right]
		\eta_{kl}
		=
		2G\boldsymbol{\eta},
		\\
		\frac{\boldsymbol{\eta}}
		{\sqrt{\boldsymbol{\eta}:\boldsymbol{\eta}}}
		:
		\boldsymbol{C}
		:
		\frac{\boldsymbol{\eta}}
		{\sqrt{\boldsymbol{\eta}:\boldsymbol{\eta}}}
		&=
		z_{ij}
		\left[
		\lambda\delta_{ij}\delta_{kl}
		+
		G
		\left(
		\delta_{ik}\delta_{jl}
		+
		\delta_{il}\delta_{jk}
		\right)
		\right]
		z_{kl}
		=
		2G.
	\end{aligned}
	\label{eq:plastic_derivative_material}
\end{equation}
Here,
\begin{equation}
	\boldsymbol{z}
	=
	\frac{\boldsymbol{\eta}}
	{\sqrt{\boldsymbol{\eta}:\boldsymbol{\eta}}}.
\end{equation}

Substituting \Cref{eq:next_state} into \Cref{eq:consistent_condition_derivative} gives
\begin{equation}
	\begin{aligned}
		f^{(n+1)}
		&=
		\left[
		\sqrt{
			\frac{3}{2}
			\boldsymbol{\eta}^{\mathrm{trial}(n+1)}
			:
			\boldsymbol{\eta}^{\mathrm{trial}(n+1)}
		}
		-
		3
		\left(
		G+\frac{1}{3}C
		\right)
		\Delta\gamma
		\right]
		\\
		&\quad
		-
		\left[
		\sigma^{0}_{y}
		+
		H
		\left(
		\bar{\varepsilon}^{p(n)}
		+
		\Delta\gamma
		\right)
		\right]
		=
		0.
	\end{aligned}
\end{equation}

Therefore, for a plastic step,
\begin{align}
	\Delta\gamma
	&=
	\frac{f_{\mathrm{trial}}}
	{3G+C+H},
	\nonumber
	\\
	f_{\mathrm{trial}}
	&=
	\sqrt{
		\frac{3}{2}
		\boldsymbol{\eta}^{\mathrm{trial}(n+1)}
		:
		\boldsymbol{\eta}^{\mathrm{trial}(n+1)}
	}
	-
	\left[
	\sigma^{0}_{y}
	+
	H\bar{\varepsilon}^{p(n)}
	\right].
\end{align}
Including both the elastic and plastic cases, the plastic multiplier can be written as
\begin{equation}
	\Delta\gamma
	=
	\left\langle
	\frac{f_{\mathrm{trial}}}
	{3G+C+H}
	\right\rangle_{+}.
\end{equation}

This completes the derivation.

\section{Variational principle of the incremental plasticity formulation\label{sec:plasticity_variational}}

In this section, we take the variation of \Cref{eq:simo} to analyze the strong-form PDE corresponding to the incremental energy and determine whether the formulation constitutes an extremum problem. In general, an energy-form physics loss is easier to optimize than a strong-form loss when training a PINO \citet{wang2026pfem}. There are two primary reasons. First, the energy formulation requires lower-order derivatives than the strong formulation. Second, it involves considerably fewer hyperparameters \citet{the_comparision_of_strong_and_energy_form}.

However, the energy formulation also has two unavoidable limitations compared with the strong formulation. First, it is less general because not every PDE possesses a corresponding energy formulation. A detailed discussion is provided in Appendix A of \citet{wang2025physics}. Second, the energy formulation should define a minimization problem rather than merely a stationarity problem. A stationary energy formulation can be extremely difficult to optimize because the neural network is itself nonconvex, and the stationarity problem may introduce many additional stationary points. Consequently, identifying the stationary point corresponding to the physical solution can be difficult during optimization \citet{wang2024artificial}. Therefore, successful energy-based optimization generally requires the energy functional to define a minimization problem.

\Cref{eq:simo} is a Lagrange-multiplier functional and can be rewritten as
\begin{equation}
	\begin{aligned}
		\Pi^{(n+1)}
		&=
		\mathcal{H}(\boldsymbol{\chi})
		-
		\Delta\gamma f^{(n+1)},
		\\
		\mathcal{H}(\boldsymbol{\chi})
		&=
		\int_{\Omega}
		\left\{
		W^{(n+1)}
		+
		\frac{1}{2}
		\boldsymbol{v}^{(n+1)}
		\cdot
		\boldsymbol{D}^{-1}
		\cdot
		\boldsymbol{v}^{(n+1)}
		\right.
		\\
		&\qquad
		+
		\left(
		\boldsymbol{\varepsilon}^{p(n+1)}
		-
		\boldsymbol{\varepsilon}^{p(n)}
		\right)
		:
		\boldsymbol{\sigma}^{(n+1)}
		\\
		&\qquad\left.
		-
		\boldsymbol{v}^{(n+1)}
		\cdot
		\boldsymbol{D}^{-1}
		\cdot
		\left(
		\boldsymbol{v}^{(n+1)}
		-
		\boldsymbol{v}^{(n)}
		\right)
		\right\}
		\,dV
		-
		W_{\mathrm{ext}}.
	\end{aligned}
\end{equation}
Here, $\boldsymbol{\chi}$ denotes all optimization variables except $\Delta\gamma$. The Hessian with respect to $\boldsymbol{\chi}$ and $\Delta\gamma$ is
\begin{equation}
	\operatorname{Hess}
	=
	\begin{bmatrix}
		\dfrac{\partial^{2}\mathcal{H}(\boldsymbol{\chi})}
		{\partial\boldsymbol{\chi}^{2}}
		&
		-\dfrac{\partial f}{\partial\boldsymbol{\chi}}
		\\[8pt]
		-\left(
		\dfrac{\partial f}{\partial\boldsymbol{\chi}}
		\right)^{T}
		&
		0
	\end{bmatrix}.
\end{equation}
This Hessian is indefinite and is therefore not positive definite. Consequently, when $\Delta\gamma$ is treated as an unknown optimization variable, \Cref{eq:simo} defines a stationarity problem rather than an extremum problem.

To eliminate the Lagrange-multiplier term
$\Delta\gamma f^{(n+1)}$,
the return-mapping algorithm is applied at every integration point, ensuring that $\delta f=0$ on the active yield surface. After applying return mapping at every integration point, the first variation of $\Pi^{(n+1)}$ is
\begin{equation}
	\begin{aligned}
		\delta\Pi^{(n+1)}
		&=
		\int_{\Omega}
		\left\{
		\boldsymbol{\sigma}^{(n+1)}
		:
		\delta\boldsymbol{\varepsilon}^{e(n+1)}
		+
		H\bar{\varepsilon}^{p(n+1)}
		\delta\bar{\varepsilon}^{p(n+1)}
		+
		\frac{3}{2C}
		\boldsymbol{q}^{(n+1)}
		\cdot
		\delta\boldsymbol{q}^{(n+1)}
		\right.
		\\
		&\qquad
		+
		\delta
		\left(
		\Delta\boldsymbol{\varepsilon}^{p(n+1)}
		\right)
		:
		\boldsymbol{\sigma}^{(n+1)}
		+
		\Delta\boldsymbol{\varepsilon}^{p(n+1)}
		:
		\delta\boldsymbol{\sigma}^{(n+1)}
		\\
		&\qquad
		-
		H
		\delta\bar{\varepsilon}^{p(n+1)}
		\Delta\bar{\varepsilon}^{p(n+1)}
		-
		H
		\bar{\varepsilon}^{p(n+1)}
		\delta
		\left(
		\Delta\bar{\varepsilon}^{p(n+1)}
		\right)
		\\
		&\qquad
		-
		\frac{3}{2C}
		\delta\boldsymbol{q}^{(n+1)}
		\cdot
		\Delta\boldsymbol{q}^{(n+1)}
		\\
		&\qquad\left.
		-
		\frac{3}{2C}
		\boldsymbol{q}^{(n+1)}
		\cdot
		\delta
		\left(
		\Delta\boldsymbol{q}^{(n+1)}
		\right)
		\right\}
		\,dV
		-
		\delta W_{\mathrm{ext}}
		\\
		&=
		\int_{\Omega}
		\left\{
		\boldsymbol{\sigma}^{(n+1)}
		:
		\delta
		\left(
		\boldsymbol{\varepsilon}^{(n+1)}
		-
		\boldsymbol{\varepsilon}^{p(n+1)}
		\right)
		+
		H\bar{\varepsilon}^{p(n+1)}
		\delta\bar{\varepsilon}^{p(n+1)}
		\right.
		\\
		&\qquad
		+
		\frac{3}{2C}
		\boldsymbol{q}^{(n+1)}
		\cdot
		\delta\boldsymbol{q}^{(n+1)}
		+
		\delta\boldsymbol{\varepsilon}^{p(n+1)}
		:
		\boldsymbol{\sigma}^{(n+1)}
		\\
		&\qquad
		+
		\Delta\boldsymbol{\varepsilon}^{p(n+1)}
		:
		\delta\boldsymbol{\sigma}^{(n+1)}
		-
		H
		\delta\bar{\varepsilon}^{p(n+1)}
		\Delta\bar{\varepsilon}^{p(n+1)}
		\\
		&\qquad
		-
		H
		\bar{\varepsilon}^{p(n+1)}
		\delta
		\left(
		\Delta\bar{\varepsilon}^{p(n+1)}
		\right)
		-
		\frac{3}{2C}
		\delta\boldsymbol{q}^{(n+1)}
		\cdot
		\Delta\boldsymbol{q}^{(n+1)}
		\\
		&\qquad\left.
		-
		\frac{3}{2C}
		\boldsymbol{q}^{(n+1)}
		\cdot
		\delta
		\left(
		\Delta\boldsymbol{q}^{(n+1)}
		\right)
		\right\}
		\,dV
		-
		\delta W_{\mathrm{ext}}
		\\
		&=
		\int_{\Omega}
		\left\{
		\boldsymbol{\sigma}^{(n+1)}
		:
		\delta\boldsymbol{\varepsilon}^{(n+1)}
		+
		H\bar{\varepsilon}^{p(n+1)}
		\delta\bar{\varepsilon}^{p(n+1)}
		\right.
		\\
		&\qquad
		+
		\frac{3}{2C}
		\boldsymbol{q}^{(n+1)}
		\cdot
		\delta\boldsymbol{q}^{(n+1)}
		+
		\Delta\boldsymbol{\varepsilon}^{p(n+1)}
		:
		\delta\boldsymbol{\sigma}^{(n+1)}
		\\
		&\qquad
		-
		H
		\delta\bar{\varepsilon}^{p(n+1)}
		\Delta\bar{\varepsilon}^{p(n+1)}
		-
		H
		\bar{\varepsilon}^{p(n+1)}
		\delta\bar{\varepsilon}^{p(n+1)}
		\\
		&\qquad
		-
		\frac{3}{2C}
		\delta\boldsymbol{q}^{(n+1)}
		\cdot
		\Delta\boldsymbol{q}^{(n+1)}
		\\
		&\qquad\left.
		-
		\frac{3}{2C}
		\boldsymbol{q}^{(n+1)}
		\cdot
		\delta\boldsymbol{q}^{(n+1)}
		\right\}
		\,dV
		-
		\delta W_{\mathrm{ext}}
		\\
		&=
		\int_{\Omega}
		\left\{
		\boldsymbol{\sigma}^{(n+1)}
		:
		\delta\boldsymbol{\varepsilon}^{(n+1)}
		-
		H
		\delta\bar{\varepsilon}^{p(n+1)}
		\Delta\bar{\varepsilon}^{p(n+1)}
		\right.
		\\
		&\qquad\left.
		+
		\Delta\boldsymbol{\varepsilon}^{p(n+1)}
		:
		\delta\boldsymbol{\sigma}^{(n+1)}
		-
		\frac{3}{2C}
		\delta\boldsymbol{q}^{(n+1)}
		\cdot
		\Delta\boldsymbol{q}^{(n+1)}
		\right\}
		\,dV
		-
		\delta W_{\mathrm{ext}}.
	\end{aligned}
	\label{eq:first_dirivative}
\end{equation}

In the preceding derivation, we used
\begin{equation}
\delta
\left(
\Delta\boldsymbol{\varepsilon}^{p(n+1)}
\right)
=
\delta
\left(
\boldsymbol{\varepsilon}^{p(n+1)}
-
\boldsymbol{\varepsilon}^{p(n)}
\right)
=
\delta\boldsymbol{\varepsilon}^{p(n+1)},
\end{equation}
\begin{equation}
\delta
\left(
\Delta\bar{\varepsilon}^{p(n+1)}
\right)
=
\delta
\left(
\bar{\varepsilon}^{p(n+1)}
-
\bar{\varepsilon}^{p(n)}
\right)
=
\delta\bar{\varepsilon}^{p(n+1)},
\end{equation}
and
\begin{equation}
\delta
\left(
\Delta\boldsymbol{q}^{(n+1)}
\right)
=
\delta
\left(
\boldsymbol{q}^{(n+1)}
-
\boldsymbol{q}^{(n)}
\right)
=
\delta\boldsymbol{q}^{(n+1)}.
\end{equation}
The plastic strain $\boldsymbol{\varepsilon}^{p(n)}$ and backstress $\boldsymbol{q}^{(n)}$ from the previous step are known. Therefore,
$\delta\boldsymbol{\varepsilon}^{p(n)}=0$
and
$\delta\boldsymbol{q}^{(n)}=0$.
For associative plasticity, \Cref{eq:first_dirivative} becomes
\begin{equation}
	\begin{aligned}
		\delta\Pi^{(n+1)}
		&=
		\int_{\Omega}
		\left\{
		\boldsymbol{\sigma}^{(n+1)}
		:
		\delta\boldsymbol{\varepsilon}^{(n+1)}
		+
		\frac{\partial f}
		{\partial\bar{\varepsilon}^{p}}
		\delta\bar{\varepsilon}^{p(n+1)}
		\Delta\gamma
		\right.
		\\
		&\qquad
		+
		\Delta\gamma
		\frac{\partial f}
		{\partial\boldsymbol{\sigma}}
		:
		\delta\boldsymbol{\sigma}^{(n+1)}
		\\
		&\qquad\left.
		+
		\Delta\gamma
		\frac{\partial f}
		{\partial\boldsymbol{q}}
		:
		\delta\boldsymbol{q}^{(n+1)}
		\right\}
		\,dV
		-
		\delta W_{\mathrm{ext}}
		\\
		&=
		\int_{\Omega}
		\left\{
		\boldsymbol{\sigma}^{(n+1)}
		:
		\delta\boldsymbol{\varepsilon}^{(n+1)}
		+
		\Delta\gamma\,\delta f
		\right\}
		\,dV
		-
		\delta W_{\mathrm{ext}}.
	\end{aligned}
	\label{eq:first_dirivative-1}
\end{equation}

Because the radial return algorithm is employed, the yield condition $f=0$ is automatically satisfied during plastic loading, and hence $\delta f=0$. \Cref{eq:first_dirivative} therefore reduces to
\begin{equation}
	\delta\Pi^{(n+1)}
	=
	\int_{\Omega}
	\boldsymbol{\sigma}^{(n+1)}
	:
	\delta\boldsymbol{\varepsilon}^{(n+1)}
	\,dV
	-
	\delta W_{\mathrm{ext}}.
	\label{eq:weak_form}
\end{equation}

It can be readily shown that
$\delta\Pi^{(n+1)}=0$
is equivalent to the equilibrium equation. We next analyze the extremum property of the incremental plastic potential:
\begin{equation}
	\begin{aligned}
		\delta^{2}\Pi^{(n+1)}
		&=
		\int_{\Omega}
		\delta\boldsymbol{\sigma}^{(n+1)}
		:
		\delta\boldsymbol{\varepsilon}^{(n+1)}
		\,dV
		\\
		&=
		\int_{\Omega}
		\delta\boldsymbol{\varepsilon}^{e(n+1)}
		:
		\boldsymbol{C}
		:
		\delta\boldsymbol{\varepsilon}^{(n+1)}
		\,dV
		\\
		&=
		\int_{\Omega}
		\delta
		\left(
		\boldsymbol{\varepsilon}^{(n+1)}
		-
		\boldsymbol{\varepsilon}^{p(n+1)}
		\right)
		:
		\boldsymbol{C}
		:
		\delta\boldsymbol{\varepsilon}^{(n+1)}
		\,dV
		\\
		&=
		\int_{\Omega}
		\delta
		\left(
		\boldsymbol{\varepsilon}^{(n+1)}
		-
		\Delta\boldsymbol{\varepsilon}^{p(n+1)}
		\right)
		:
		\boldsymbol{C}
		:
		\delta\boldsymbol{\varepsilon}^{(n+1)}
		\,dV
		\\
		&=
		\int_{\Omega}
		\delta
		\left(
		\boldsymbol{\varepsilon}^{(n+1)}
		-
		\Delta\gamma\boldsymbol{N}
		\right)
		:
		\boldsymbol{C}
		:
		\delta\boldsymbol{\varepsilon}^{(n+1)}
		\,dV
		\\
		&=
		\int_{\Omega}
		\left[
		\delta\boldsymbol{\varepsilon}^{(n+1)}
		-
		\delta(\Delta\gamma)\boldsymbol{N}
		-
		\Delta\gamma\delta\boldsymbol{N}
		\right]
		:
		\boldsymbol{C}
		:
		\delta\boldsymbol{\varepsilon}^{(n+1)}
		\,dV
		\\
		&=
		\int_{\Omega}
		\delta\boldsymbol{\varepsilon}^{(n+1)}
		:
		\boldsymbol{C}
		:
		\delta\boldsymbol{\varepsilon}^{(n+1)}
		\,dV
		\\
		&\quad
		-
		\int_{\Omega}
		\delta(\Delta\gamma)
		\boldsymbol{N}
		:
		\boldsymbol{C}
		:
		\delta\boldsymbol{\varepsilon}^{(n+1)}
		\,dV
		\\
		&\quad
		-
		\int_{\Omega}
		\Delta\gamma
		\delta\boldsymbol{N}
		:
		\boldsymbol{C}
		:
		\delta\boldsymbol{\varepsilon}^{(n+1)}
		\,dV,
		\\
		\boldsymbol{N}
		&=
		\frac{\partial f}
		{\partial\boldsymbol{\sigma}}
		=
		\sqrt{\frac{3}{2}}
		\frac{\boldsymbol{\sigma}'}
		{\sqrt{\boldsymbol{\sigma}':\boldsymbol{\sigma}'}}.
	\end{aligned}
	\label{eq:second_derivative}
\end{equation}

Because the direction of the radial return remains radial,
$\partial f/\partial\boldsymbol{\sigma}$
is assumed to remain unchanged, and hence
$\delta(\partial f/\partial\boldsymbol{\sigma})=0$.
\Cref{eq:second_derivative} therefore reduces to
\begin{equation}
	\begin{aligned}
		\delta^{2}\Pi^{(n+1)}
		&=
		\int_{\Omega}
		\delta\boldsymbol{\varepsilon}^{(n+1)}
		:
		\boldsymbol{C}
		:
		\delta\boldsymbol{\varepsilon}^{(n+1)}
		\,dV
		\\
		&\quad
		-
		\int_{\Omega}
		\delta(\Delta\gamma)
		\boldsymbol{N}
		:
		\boldsymbol{C}
		:
		\delta\boldsymbol{\varepsilon}^{(n+1)}
		\,dV.
	\end{aligned}
	\label{eq:simple_second}
\end{equation}

We next consider the consistency condition:
\begin{equation}
	\delta f
	=
	\boldsymbol{N}:\delta\boldsymbol{\sigma}
	-
	H\delta\bar{\varepsilon}^{p}
	=
	\boldsymbol{N}:\delta\boldsymbol{\sigma}
	-
	H\delta
	\left(
	\Delta\bar{\varepsilon}^{p}
	\right)
	=
	\boldsymbol{N}:\delta\boldsymbol{\sigma}
	-
	H\delta(\Delta\gamma)
	=
	0.
	\label{eq:consistency_condition}
\end{equation}

Substituting
\begin{equation}
\delta\boldsymbol{\sigma}
=
\boldsymbol{C}
:
\delta\boldsymbol{\varepsilon}
-
\delta(\Delta\gamma)
\boldsymbol{C}
:
\boldsymbol{N}
\end{equation}
into \Cref{eq:consistency_condition} gives
\begin{align}
	\boldsymbol{N}
	:
	\left[
	\boldsymbol{C}
	:
	\delta\boldsymbol{\varepsilon}
	-
	\delta(\Delta\gamma)
	\boldsymbol{C}
	:
	\boldsymbol{N}
	\right]
	-
	H\delta(\Delta\gamma)
	&=
	0,
	\nonumber
	\\
	\delta(\Delta\gamma)
	&=
	\frac{
		\boldsymbol{N}
		:
		\boldsymbol{C}
		:
		\delta\boldsymbol{\varepsilon}
	}{
		\boldsymbol{N}
		:
		\boldsymbol{C}
		:
		\boldsymbol{N}
		+
		H
	}.
	\label{eq:delta_gama}
\end{align}

Substituting \Cref{eq:delta_gama} into \Cref{eq:simple_second} gives
\begin{equation}
	\begin{aligned}
		\delta^{2}\Pi^{(n+1)}
		&=
		\int_{\Omega}
		\delta\boldsymbol{\varepsilon}^{(n+1)}
		:
		\boldsymbol{C}
		:
		\delta\boldsymbol{\varepsilon}^{(n+1)}
		\,dV
		\\
		&\quad
		-
		\int_{\Omega}
		\frac{
			\boldsymbol{N}
			:
			\boldsymbol{C}
			:
			\delta\boldsymbol{\varepsilon}^{(n+1)}
		}{
			\boldsymbol{N}
			:
			\boldsymbol{C}
			:
			\boldsymbol{N}
			+
			H
		}
		\boldsymbol{N}
		:
		\boldsymbol{C}
		:
		\delta\boldsymbol{\varepsilon}^{(n+1)}
		\,dV
		\\
		&=
		\int_{\Omega}
		\delta\boldsymbol{\varepsilon}^{(n+1)}
		:
		\boldsymbol{C}_{\mathrm{ep}}
		:
		\delta\boldsymbol{\varepsilon}^{(n+1)}
		\,dV,
		\\
		\boldsymbol{C}_{\mathrm{ep}}
		&=
		\boldsymbol{C}
		-
		\frac{
			\boldsymbol{C}
			:
			\boldsymbol{N}
			\otimes
			\boldsymbol{N}
			:
			\boldsymbol{C}
		}{
			\boldsymbol{N}
			:
			\boldsymbol{C}
			:
			\boldsymbol{N}
			+
			H
		}.
	\end{aligned}
\end{equation}

Because
$\boldsymbol{C}:\boldsymbol{N}=2G\boldsymbol{N}$
and
$\boldsymbol{N}:\boldsymbol{N}=3/2$,
the elastoplastic tangent tensor can be written as
\begin{equation}
	\boldsymbol{C}_{\mathrm{ep}}
	=
	\boldsymbol{C}
	-
	\frac{
		4G^{2}
		\boldsymbol{N}
		\otimes
		\boldsymbol{N}
	}{
		3G+H
	}.
\end{equation}

In general, $H>0$. We next analyze the positive definiteness of
$\boldsymbol{C}_{\mathrm{ep}}$.
Consider an arbitrary direction
\[
\boldsymbol{Z}
=
\boldsymbol{Z}^{\perp}
+
\boldsymbol{Z}^{\parallel},
\]
where $\boldsymbol{Z}^{\perp}$ is the projection of $\boldsymbol{Z}$ onto the plane perpendicular to $\boldsymbol{N}$, and
$\boldsymbol{Z}^{\parallel}=a\boldsymbol{N}$
is the component parallel to $\boldsymbol{N}$. We then have
\begin{equation}
	\begin{aligned}
		\boldsymbol{Z}
		:
		\boldsymbol{C}_{\mathrm{ep}}
		:
		\boldsymbol{Z}
		&=
		\boldsymbol{Z}
		:
		\boldsymbol{C}
		:
		\boldsymbol{Z}
		-
		\frac{4G^{2}}
		{3G+H}
		\left(
		\boldsymbol{N}
		:
		\boldsymbol{Z}
		\right)^{2}
		\\
		&=
		\boldsymbol{Z}^{\perp}
		:
		\boldsymbol{C}
		:
		\boldsymbol{Z}^{\perp}
		+
		\boldsymbol{Z}^{\parallel}
		:
		\boldsymbol{C}
		:
		\boldsymbol{Z}^{\parallel}
		\\
		&\quad
		+
		2
		\boldsymbol{Z}^{\perp}
		:
		\boldsymbol{C}
		:
		\boldsymbol{Z}^{\parallel}
		-
		\frac{4G^{2}}
		{3G+H}
		\left(
		\boldsymbol{N}
		:
		\boldsymbol{Z}^{\parallel}
		\right)^{2}
		\\
		&=
		\boldsymbol{Z}^{\perp}
		:
		\boldsymbol{C}
		:
		\boldsymbol{Z}^{\perp}
		+
		3a^{2}G
		-
		\frac{4G^{2}}
		{3G+H}
		\left(
		\frac{3}{2}a
		\right)^{2}
		\\
		&=
		\boldsymbol{Z}^{\perp}
		:
		\boldsymbol{C}
		:
		\boldsymbol{Z}^{\perp}
		+
		a^{2}
		\left(
		3G
		-
		\frac{9G^{2}}
		{3G+H}
		\right)
		>
		0.
	\end{aligned}
\end{equation}

Therefore, for associative plastic flow with a positive plastic modulus $H>0$, the elastoplastic tangent tensor
$\boldsymbol{C}_{\mathrm{ep}}$
is positive definite. Consequently, the incremental plastic potential
$\Pi^{(n+1)}$
is a single-extremum problem.

\section{Deep energy method examples for elastoplasticity\label{sec:DEM-for-elastoplasticity}}

In this section, we present two examples in which the deep energy method is used to solve elastoplastic problems with mixed hardening. Unlike \citet{he2023deep}, who considered isotropic and kinematic hardening separately, we consider models containing both types of hardening. In addition, \citet{he2023deep} considered only linear isotropic hardening and therefore employed the radial return algorithm without considering the return-mapping procedure required for nonlinear isotropic hardening. We consider more complex material models. The first example combines linear isotropic hardening with kinematic hardening, whereas the second combines nonlinear isotropic hardening with kinematic hardening. Because the first example employs linear isotropic hardening, its constitutive integration is performed using the radial return algorithm, whose derivation is provided in \ref{sec:proof_J2_plasticity_multify}. In the second example, nonlinear isotropic hardening is considered, and the plastic multiplier must therefore be determined iteratively using a return-mapping algorithm.

\begin{figure}
	\begin{centering}
		\includegraphics[scale=0.47]{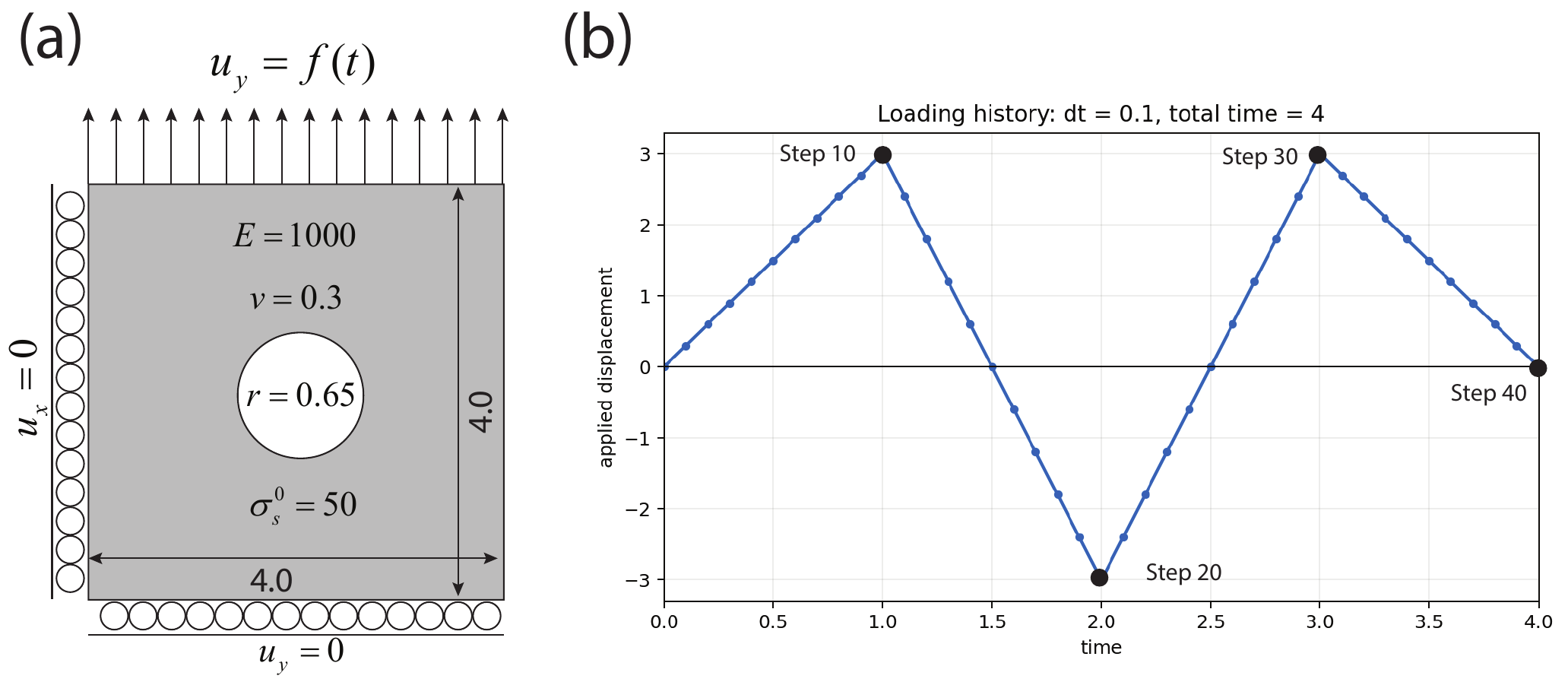}
		\par\end{centering}
	\caption{Definition of the elastoplastic problem. (a) A perforated square plate with a length and width of $4$ mm. The elastic modulus is $E=1000$ MPa, Poisson's ratio is $\nu=0.3$, the radius of the circular hole is $r=0.65$ mm, and the initial yield stress is $\sigma^{0}_{y}=50$ MPa. The displacement in the $y$ direction is constrained on the bottom boundary, $u_{y}=0$, and the displacement in the $x$ direction is constrained on the left boundary, $u_{x}=0$. The loading $t_{y}=f(t)$ is applied on the top boundary. (b) Evolution of the loading history $f(t)$, which is discretized into 40 load steps.
		\label{fig:intro_dem_plasticity}}
\end{figure}

The geometry, material properties, and boundary conditions are illustrated in \Cref{fig:intro_dem_plasticity}a, and the loading history $f(t)$ is shown in \Cref{fig:intro_dem_plasticity}b. In the first example, the plastic modulus is $H=500$, and the kinematic hardening modulus is $C=500$. In the second example, the coefficient controlling isotropic hardening is 500, and the nonlinear hardening law is
\begin{equation}
	\sigma_{y}
	\left(
	\bar{\varepsilon}^{p}
	\right)
	=
	\sigma^{0}_{y}
	+
	500
	\left[
	1
	-
	\exp
	\left(
	-\bar{\varepsilon}^{p}
	\right)
	\right].
\end{equation}

Following the distance-function approach of \citet{wang2022cenn}, we construct a kinematically admissible displacement field that satisfies the essential boundary conditions in advance:
\begin{equation}
	\begin{aligned}u_{x}(x,y) & =\frac{x}{4}*\mathrm{NN}_{x}(x,y;\boldsymbol{\theta})\\
		u_{y}(x,y) & =\frac{y}{4}*(1-\frac{y}{4})*\mathrm{NN}_{y}(x,y;\boldsymbol{\theta})+\frac{y}{4}*f(t)
	\end{aligned}
\end{equation}
Here, $\boldsymbol{\theta}$ denotes the trainable parameters of the neural network.

\begin{figure}
	\begin{centering}
		\includegraphics[scale=0.47]{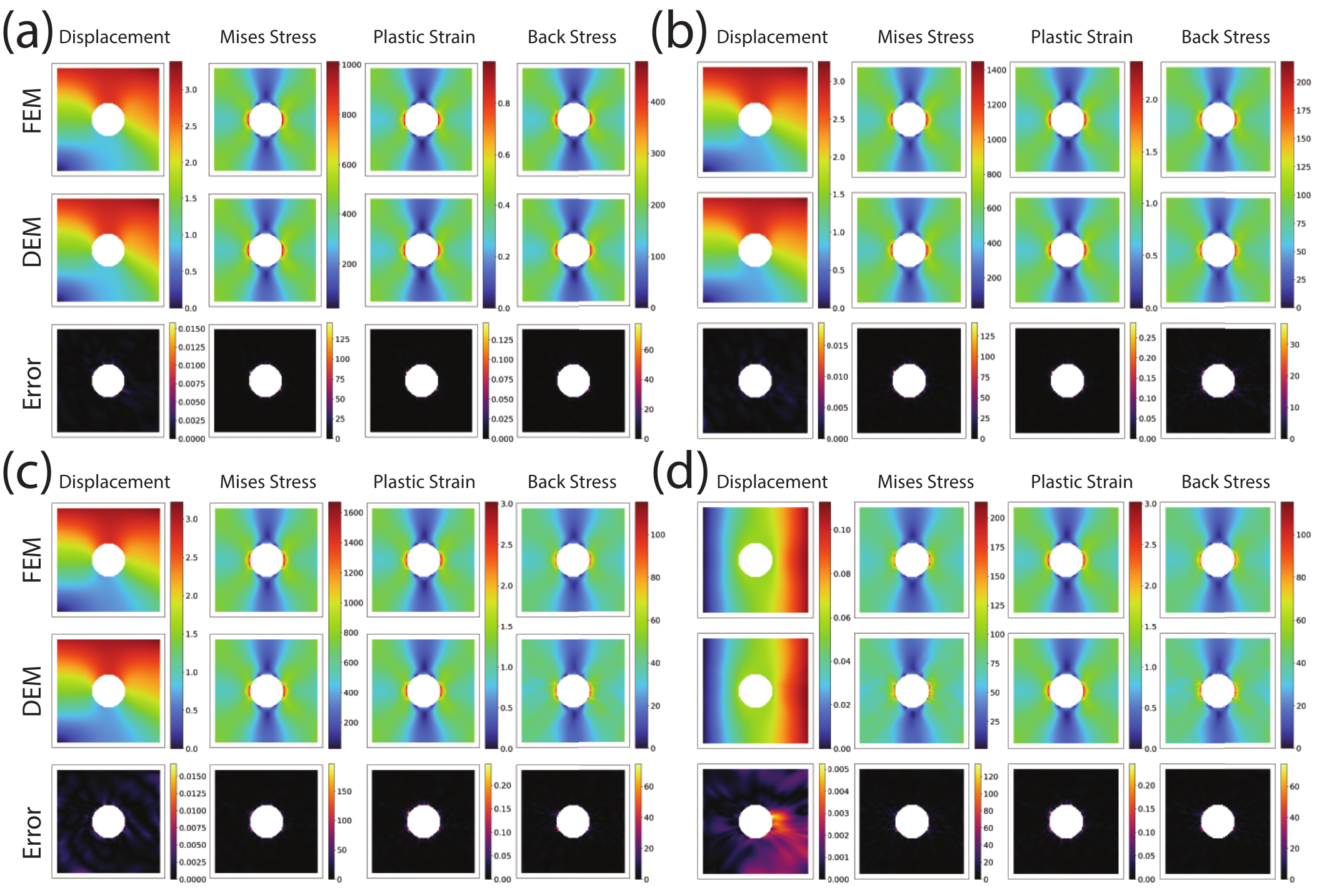}
		\par\end{centering}
	\caption{Results obtained using the deep energy method for the model combining linear isotropic and kinematic hardening. Panels (a)--(d) show the results at load steps 10, 20, 30, and 40, respectively. ``Displacement'' denotes the displacement magnitude, ``Mises Stress'' denotes the von Mises elastic stress, ``Plastic Strain'' denotes the equivalent plastic strain, and ``Back Stress'' denotes the von Mises equivalent backstress.
		\label{fig:DEM_linear}}
\end{figure}

\begin{figure}
	\begin{centering}
		\includegraphics[scale=0.47]{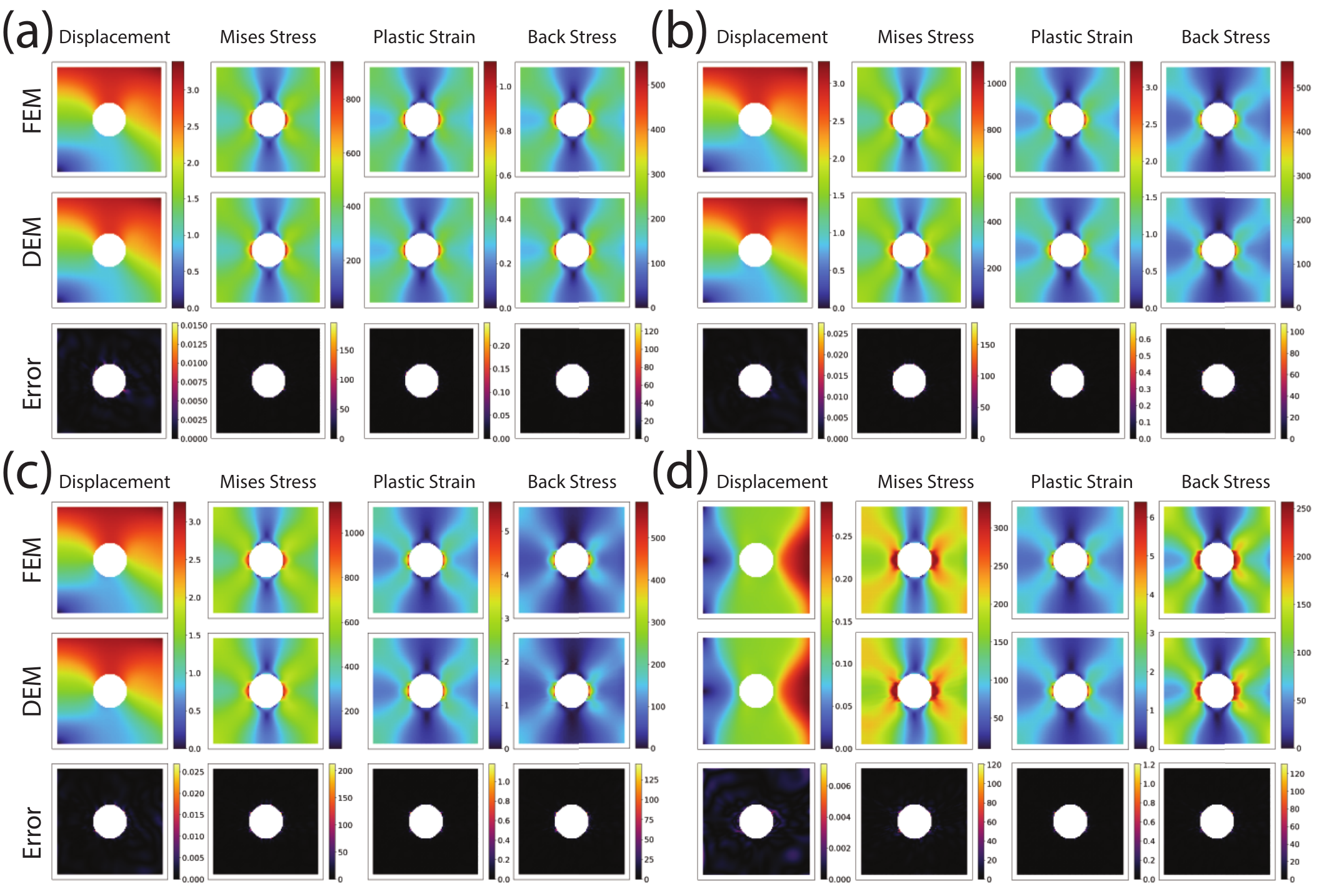}
		\par\end{centering}
	\caption{Results obtained using the deep energy method for the model combining nonlinear isotropic and kinematic hardening. Panels (a)--(d) show the results at load steps 10, 20, 30, and 40, respectively. ``Displacement'' denotes the displacement magnitude, ``Mises Stress'' denotes the von Mises elastic stress, ``Plastic Strain'' denotes the equivalent plastic strain, and ``Back Stress'' denotes the von Mises equivalent backstress.
		\label{fig:DEM_nonlinear}}
\end{figure}

\Cref{fig:DEM_linear} and \ref{fig:DEM_nonlinear} show the displacement magnitude, von Mises elastic stress, equivalent plastic strain, and von Mises equivalent backstress obtained using DEM for the linear and nonlinear isotropic-hardening models, respectively. It should be emphasized that both examples employ mixed-hardening models. The finite element method is used to generate the reference solutions for both examples. The finite element mesh contains 9443 linear elements with $2\times2$ Gauss integration, and the small-strain assumption is adopted \citet{he2023deep}. DEM and FEM use the same spatial discretization. The neural network in DEM is an MLP with the architecture
$[2,40,80,160,80,40,2]$. The number of trainable parameters is 129122.
The inputs are the $x$ and $y$ coordinates, and the outputs are the displacement components $u_{x}$ and $u_{y}$.

\begin{figure}
	\begin{centering}
		\includegraphics[scale=0.45]{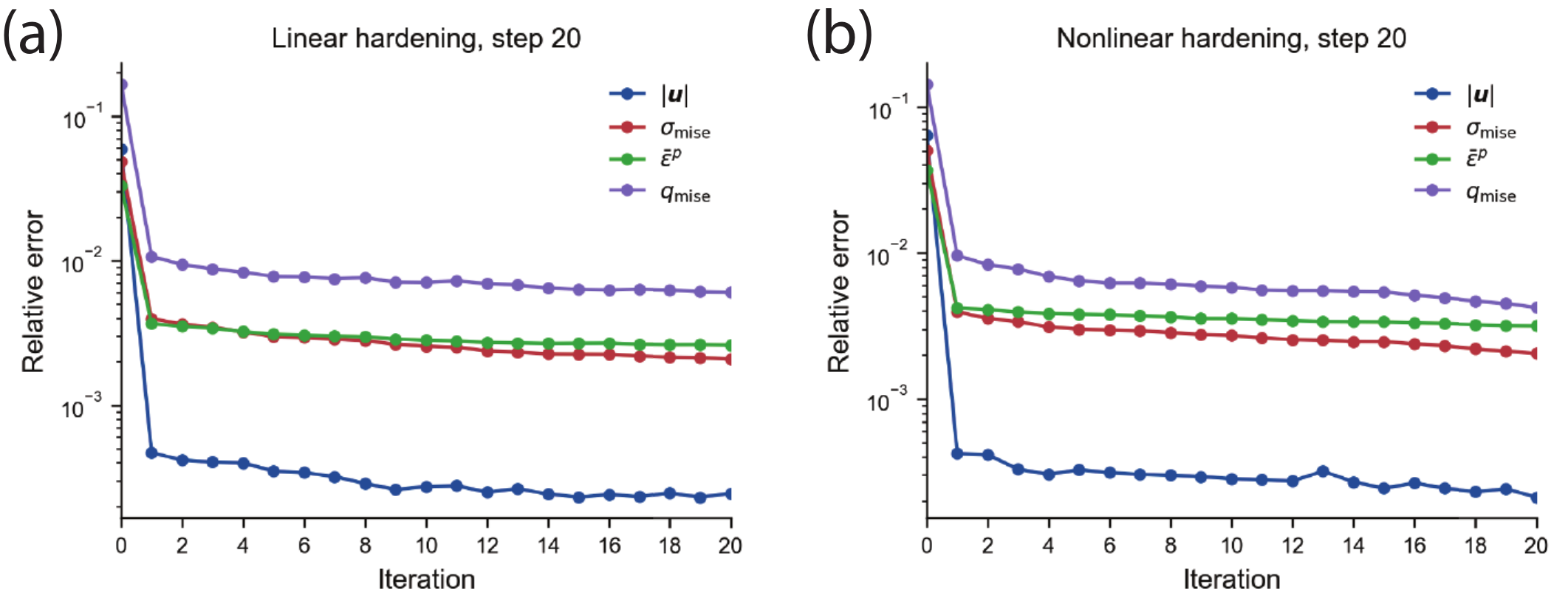}
		\par\end{centering}
	\caption{Evolution of the relative error of the Deep Energy Method (DEM) at the twentieth loading increment: (a) linear isotropic hardening model; (b) nonlinear isotropic hardening model.\label{fig:DEM_error}}
\end{figure}

The DEM results shown in \Cref{fig:DEM_nonlinear} agree closely with the finite element solutions. 
The convergence history of the relative error is presented in \ref{fig:DEM_error}, demonstrating that DEM converges very rapidly. We employ the L-BFGS optimizer, where each optimization iteration performs up to 200 line-search evaluations. Consequently, the model typically converges within only a few optimization iterations.
To further quantify the accuracy and computational efficiency, \Cref{tab:DEM_plasticity} summarizes the performance of DEM. It should be noted that constitutive integration accounts for a larger proportion of the FEM computational time. This is because FEM additionally requires the calculation of the consistent tangent stiffness, denoted by \texttt{DDSDDE} in a UMAT. DEM does not require \texttt{DDSDDE}; therefore, constitutive integration accounts for a smaller proportion of its total computational time.

The constitutive-integration time of the nonlinear isotropic-hardening model is only slightly greater than that of the linear isotropic-hardening model. The nonlinear model requires a Newton iteration, whereas no local Newton iteration is required for the linear model. Theoretically, the nonlinear isotropic-hardening model should therefore require more computational time. However, \Cref{tab:DEM_plasticity} shows that their constitutive-integration times are similar. This is because the Newton iteration is performed only for the scalar plastic multiplier. The inverse of the scalar Jacobian is simply its reciprocal and can be evaluated directly. Consequently, the scalar Newton iteration for the plastic multiplier is highly efficient, resulting in similar constitutive-integration times for the nonlinear and linear isotropic-hardening models.

It is worth noting that although \Cref{tab:DEM_plasticity} shows that FEM is more computationally efficient than DEM, this conclusion only holds for problems with relatively few degrees of freedom. For problems with extremely large numbers of DOFs, DEM becomes more computationally efficient than FEM, as demonstrated in \Cref{fig:FEM_DEM_dof}.

\begin{table}
	\caption{Performance of DEM for elastoplastic problems. The reported relative errors are relative $L_{1}$ errors. The errors in the displacement magnitude, von Mises elastic stress, equivalent plastic strain, and von Mises equivalent backstress are evaluated at load steps 10, 20, and 30. The finite element solution is used as the reference. For both FEM and DEM, the computational times required for constitutive integration and solving the equilibrium equations are reported separately. CI and Eq. denote constitutive integration and equilibrium solution, respectively. All results were obtained using the LBFGS optimizer with a learning rate of $0.5$.
		\label{tab:DEM_plasticity}}
	
	\centering
	\begin{adjustbox}{max width=\textwidth}
	\begin{tabular}{ccccccc}
		\toprule
		Problem
		& Step
		& Relative error: $|\boldsymbol{u}|$, $\sigma_{mise}$, $\bar{\varepsilon}^{p}$, $q_{mise}$
		& FEM time (s): CI / Eq.
		& DEM time (s): CI / Eq.
		\tabularnewline
		\midrule
		\multirow{3}{*}{Linear isotropic + kinematic}
		& 10
		& $0.000267,0.003522,0.004084,0.004062$
		& $0.07395,0.5134$
		& $1.577,68.90$

		\tabularnewline
		& 20
		& $0.000354,0.003604,0.003565,0.010044$
		& $0.07712,0.5096$
		& $0.8096,35.93$

		\tabularnewline
		& 30
		& $0.000435,0.003606,0.003659,0.021834$
		& $0.07254,0.5041$
		& $1.249,53.08$

		\tabularnewline
		\midrule
		\multirow{3}{*}{Nonlinear isotropic + kinematic}
		& 10
		& $0.000209,0.003342,0.004177,0.004169$
		& $0.03213,0.5636$
		& $1.148,30.45$

		\tabularnewline
		& 20
		& $0.000335,0.003559,0.004255,0.008117$
		& $0.1298,0.8365$
		& $1.188,30.48$

		\tabularnewline
		& 30
		& $0.000519,0.004181,0.005091,0.014021$
		& $0.1276,0.8365$
		& $0.6615,14.75$

		\tabularnewline
		\bottomrule
	\end{tabular}
	\end{adjustbox}
\end{table}

\section{FEM Convergence Analysis\label{sec:FEM_convergence_proof}}

In this section, we perform convergence studies for the finite element method (FEM). Three representative loading paths, namely non-proportional loading, loading--unloading, and cyclic loading, are considered, as described in \Cref{subsec:path}.

\begin{figure}
	\begin{centering}
		\includegraphics[scale=0.48]{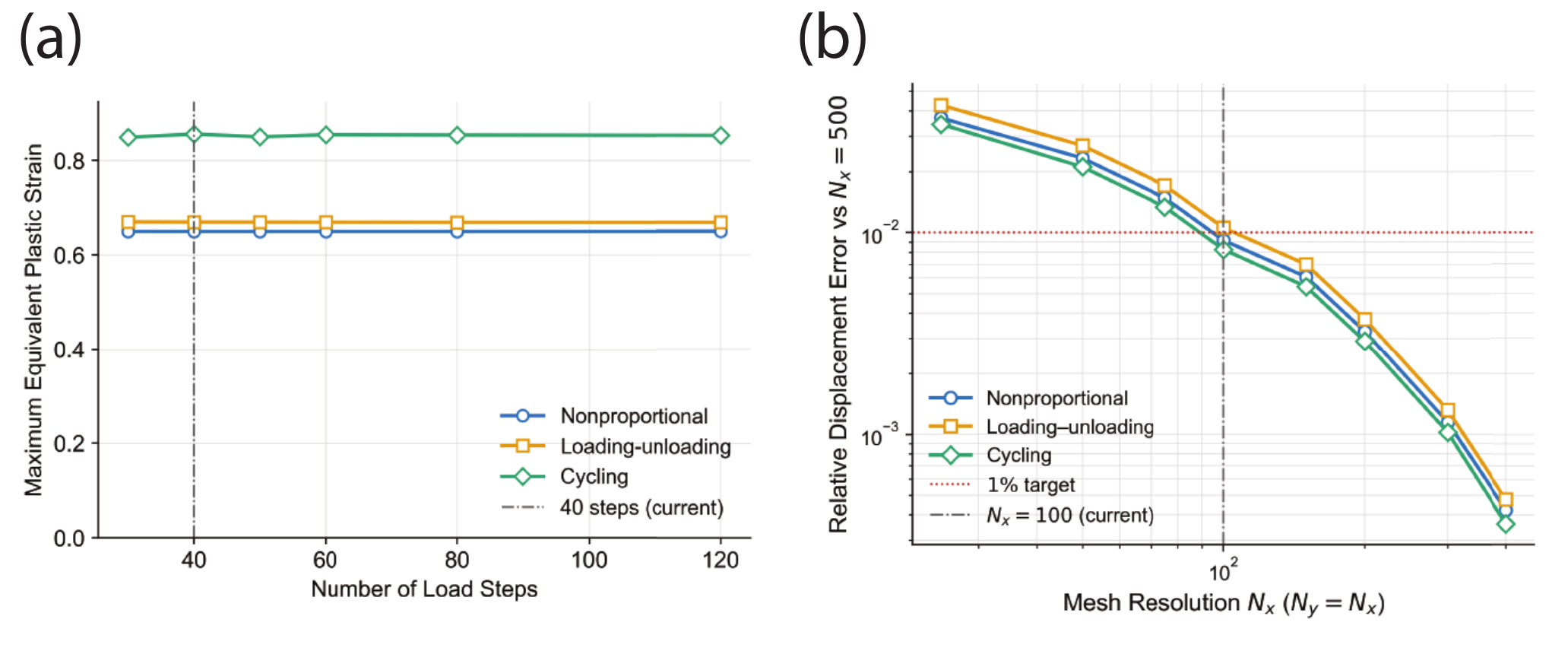}
		\par\end{centering}
	\caption{FEM convergence study: (a) evolution of the maximum equivalent plastic strain at the final loading increment with respect to the number of loading increments. Each curve corresponds to the same loading path discretized using a different number of loading increments. (b) Evolution of the relative error with respect to the spatial resolution. Each curve corresponds to the same benchmark problem solved using different mesh resolutions.\label{fig:FEM_convergence}}
\end{figure}

The convergence results are presented in \Cref{fig:FEM_convergence}, where both the temporal discretization (loading increments) and the spatial discretization (mesh resolution) are investigated. \Cref{fig:FEM_convergence}a shows the maximum equivalent plastic strain at the final loading increment for the same physical loading path discretized using different numbers of loading increments. It can be seen that the maximum equivalent plastic strain at the final loading state remains essentially unchanged, indicating that 40 loading increments are sufficient to accurately capture the loading process. \Cref{fig:FEM_convergence}b presents the influence of the spatial resolution on the numerical solution for the same benchmark problem. The solution obtained with $N_{x}=500$ is taken as the reference solution. It can be observed that when the mesh resolution reaches $N_{x}=100$, the relative error is already approximately $1\%$. Therefore, a mesh resolution of $N_{x}=100$ is adopted throughout this work to generate the training dataset.

\section{Supplementary Figures\label{sec:Supplementary-Figures}}

\begin{figure}
	\begin{centering}
		\includegraphics[scale=0.47]{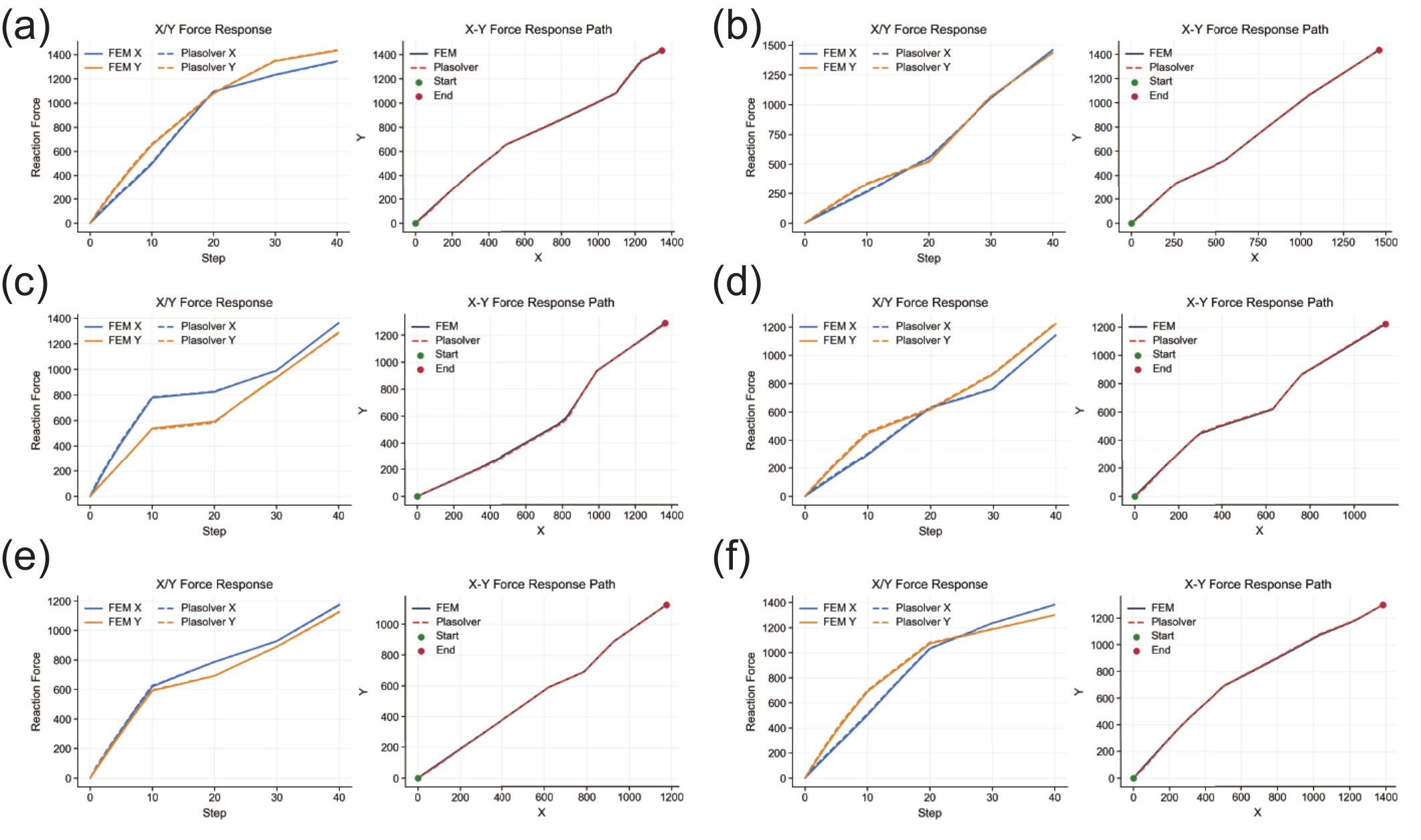}
		\par\end{centering}
	\caption{Comparison of the reaction-force predictions obtained using the pretrained Plasolver model and FEM under non-proportional loading. Panels (a)--(f) show the predictions for different test cases.
		\label{fig:Plasovler_nonproportional_loading_morepath}}
\end{figure}

\begin{figure}
	\begin{centering}
		\includegraphics[scale=0.47]{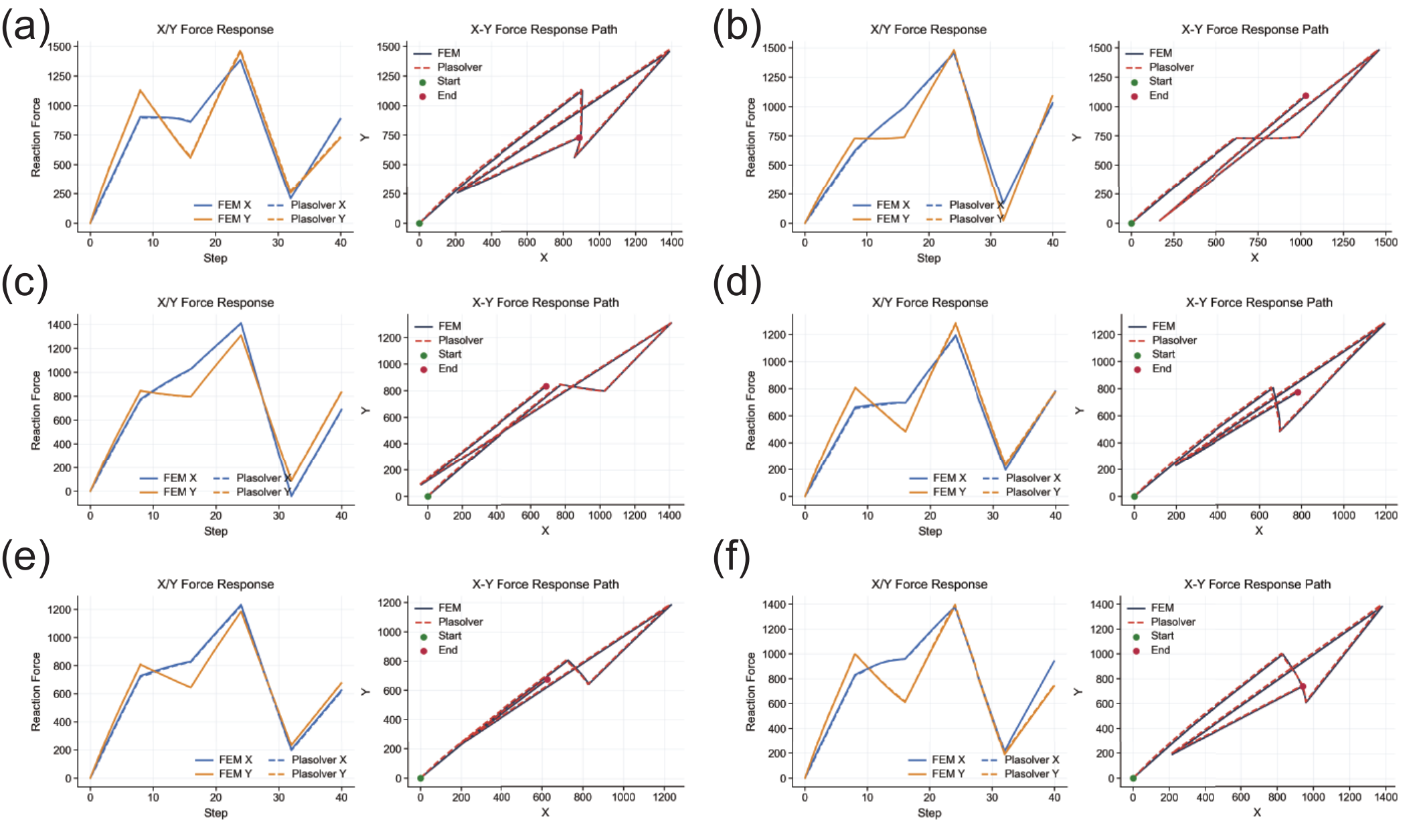}
		\par\end{centering}
	\caption{Comparison of the reaction-force predictions obtained using the pretrained Plasolver model and FEM under loading--unloading paths. Panels (a)--(f) show the predictions for different test cases.
		\label{fig:Plasovler_loading_unloading_morepath}}
\end{figure}

\begin{figure}
	\begin{centering}
		\includegraphics[scale=0.47]{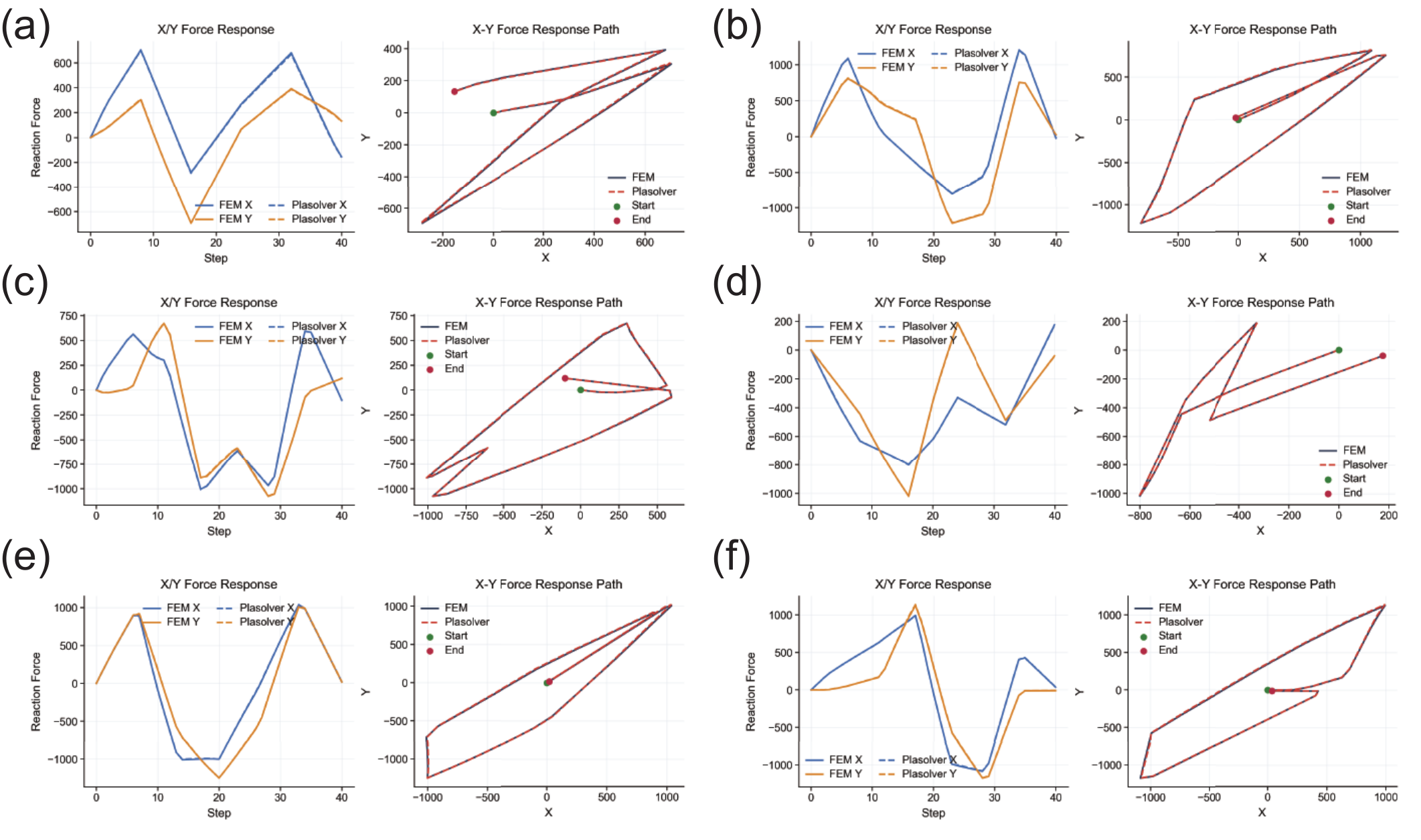}
		\par\end{centering}
	\caption{Comparison of the reaction-force predictions obtained using the pretrained Plasolver model and FEM under cyclic loading. Panels (a)--(f) show the predictions for different test cases.
		\label{fig:Plasovler_cycling_loading_morepath}}
\end{figure}

\begin{figure}
	\begin{centering}
		\includegraphics[scale=0.47]{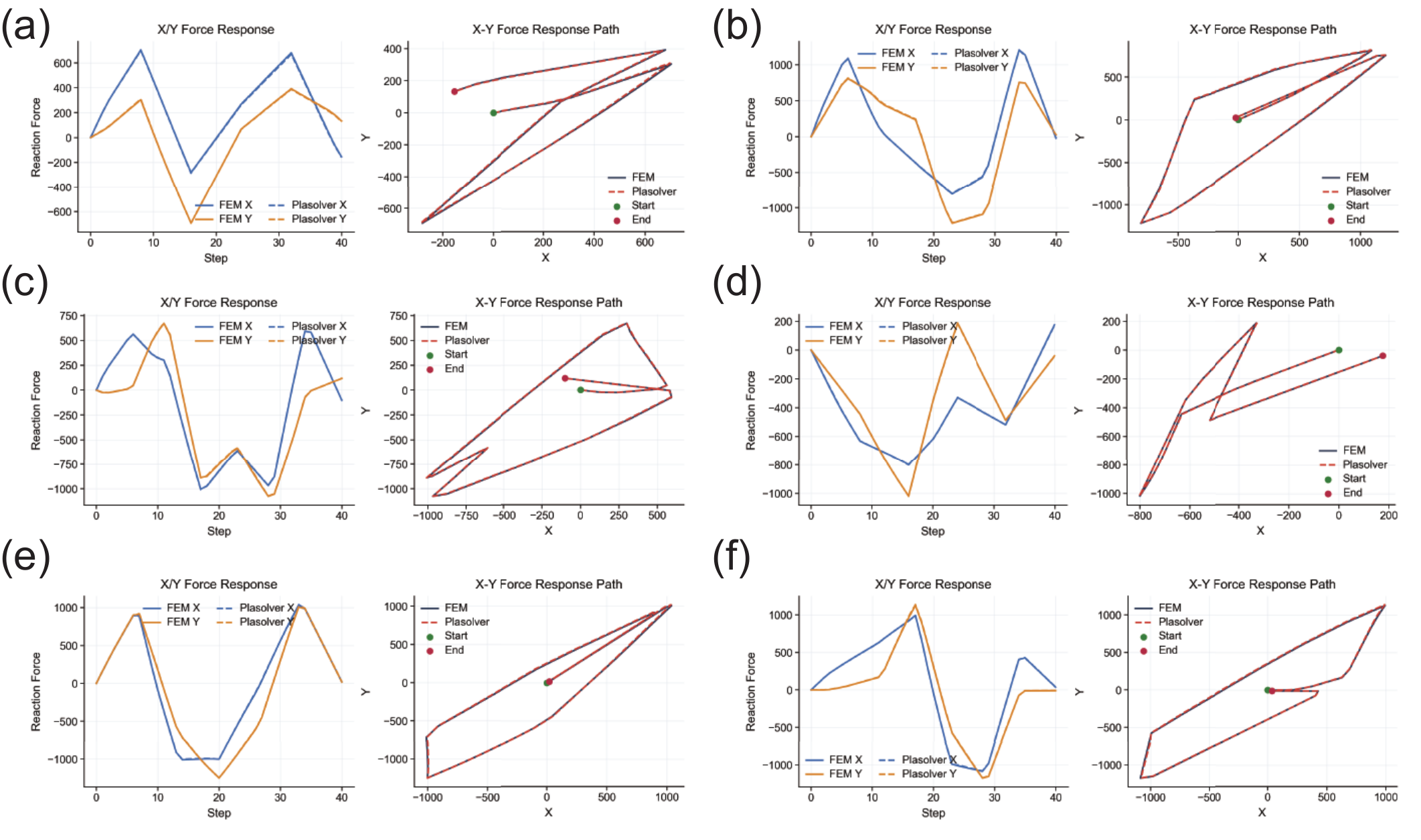}
		\par\end{centering}
	\caption{Comparison of the reaction-force predictions obtained using the pretrained Plasolver model and FEM under cyclic loading with different geometries and loading paths. Panels (a)--(f) show the predictions for different test cases.
		\label{fig:Plasovler_geo_cycling_loading_morepath}}
\end{figure}

\begin{figure}
	\begin{centering}
		\includegraphics[scale=0.47]{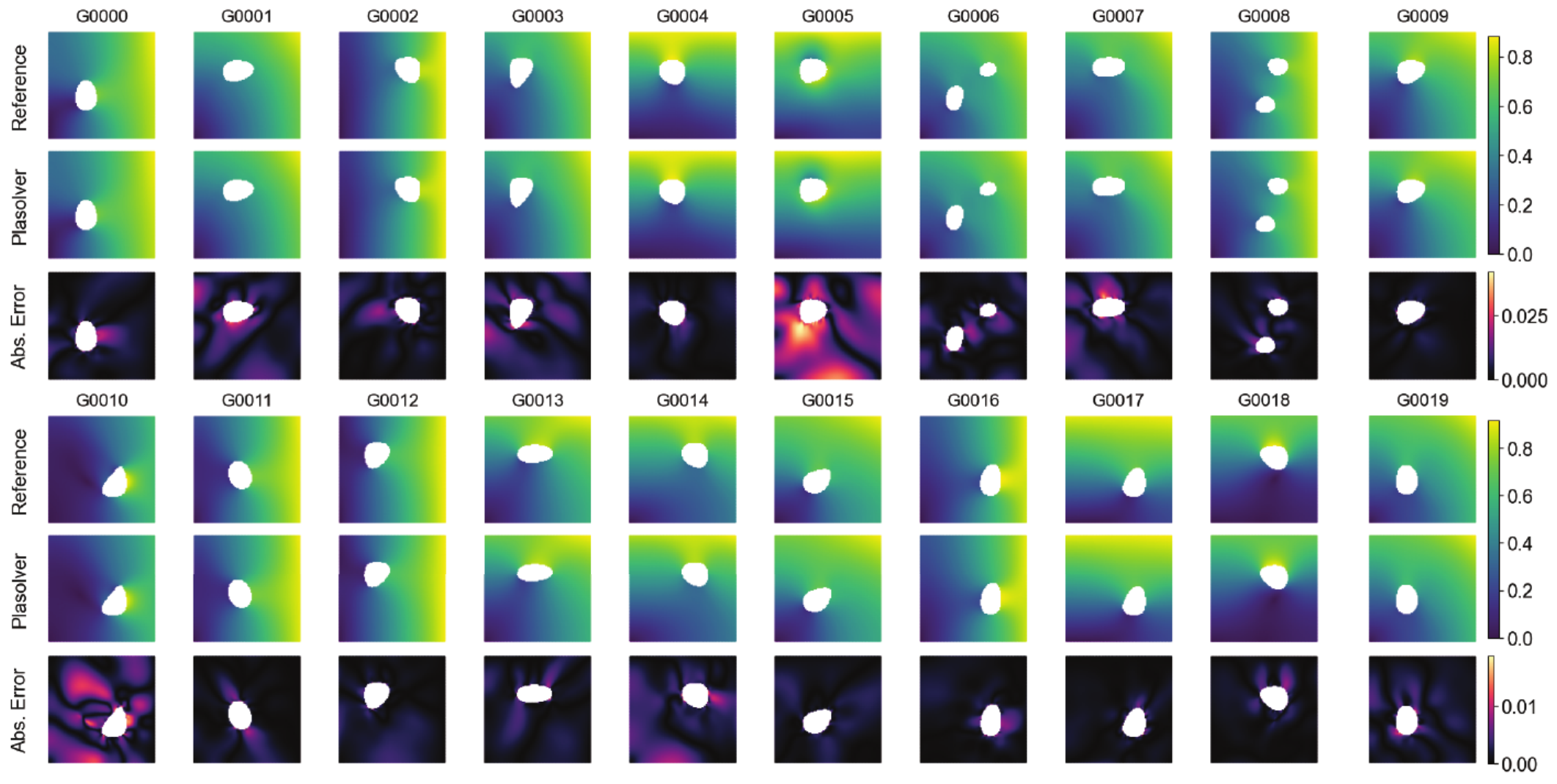}
		\par\end{centering}
	\caption{Comparison of the displacement-magnitude fields predicted by the pretrained Plasolver model and the finite element reference solutions for different geometries and loading paths at the intermediate load step, namely the 20th step.
		\label{fig:Plasolver_geo_more_contourf}}
\end{figure}

\begin{figure}
	\begin{centering}
		\includegraphics[scale=0.47]{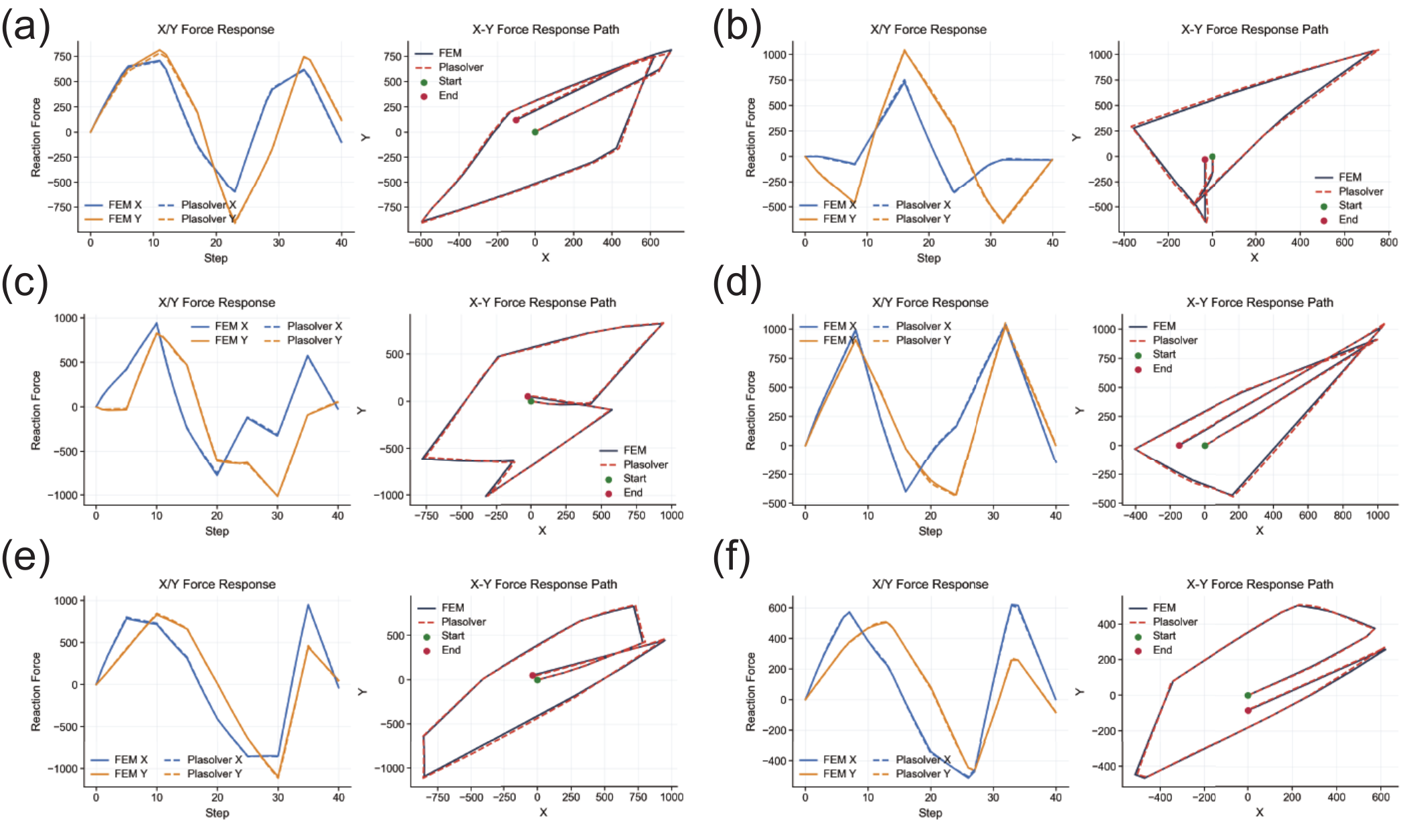}
		\par\end{centering}
	\caption{Comparison of the reaction-force predictions obtained using the pretrained Plasolver model and FEM under cyclic loading with different materials, geometries, and loading paths. Panels (a)--(f) show the predictions for different test cases.
		\label{fig:Plasovler_material_cycling_loading_morepath}}
\end{figure}

\begin{figure}
	\begin{centering}
		\includegraphics[scale=0.47]{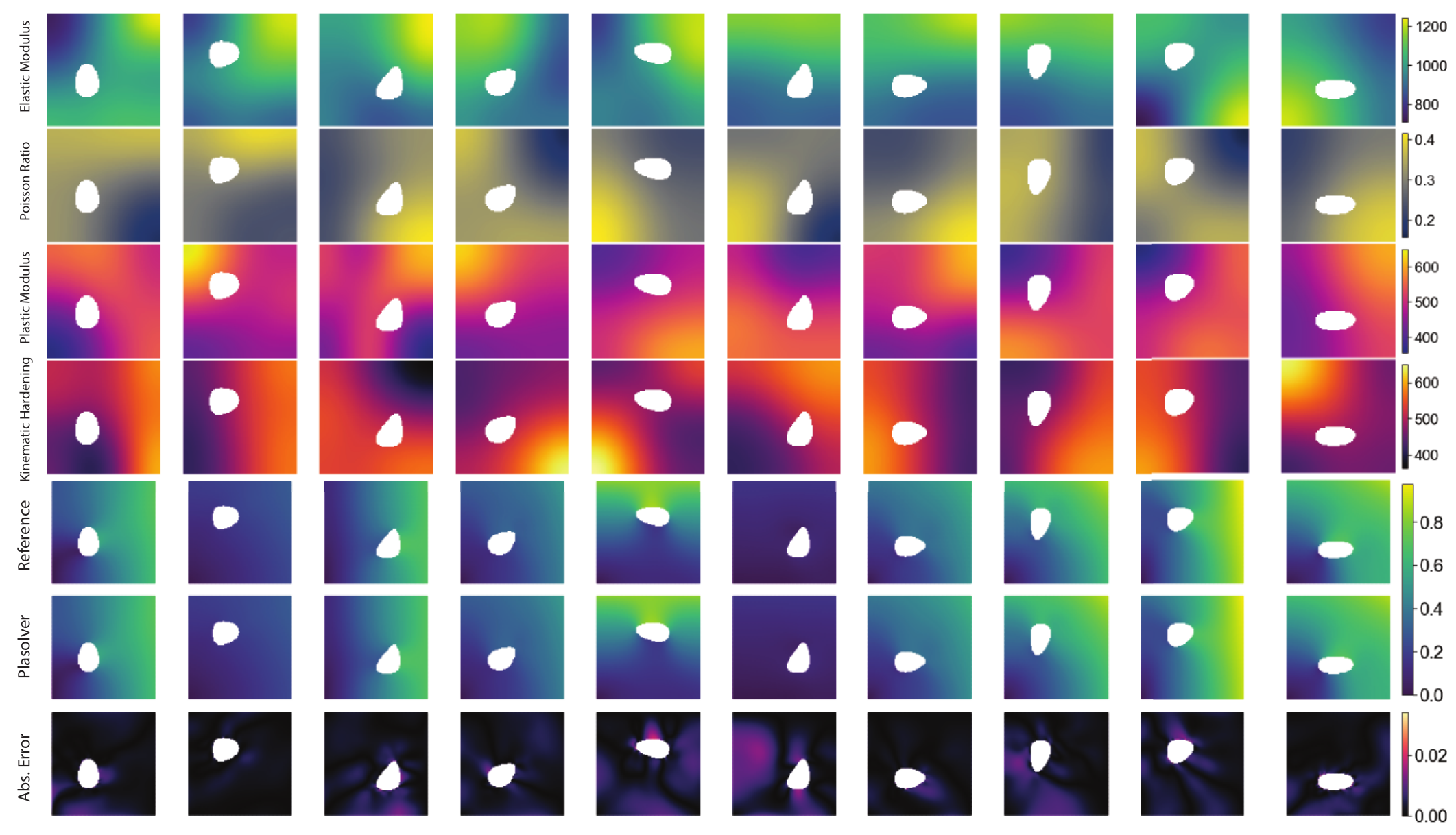}
		\par\end{centering}
	\caption{Comparison of the displacement-magnitude fields predicted by the pretrained Plasolver model and the finite element reference solutions for different materials, geometries, and loading paths at the intermediate load step, namely the 20th step.
		\label{fig:Plasolver_material_more_contourf}}
\end{figure}

\section{Supplementary code}
The code of this work will be available at \url{https://github.com/yizheng-wang/Research-on-Solving-Partial-Differential-Equations-of-Solid-Mechanics-Based-on-PINN} after accepted.

\bibliographystyle{elsarticle-num}
\addcontentsline{toc}{section}{\refname}\bibliography{reference.bib}

\end{document}